\documentclass[aps,prd,twocolumn,superscriptaddress,nofootinbib,10pt]{revtex4-1}

\pdfoutput=1
\usepackage[utf8]{inputenc}
\usepackage{graphicx,amsmath,amsfonts,amssymb,aas_macros,slashed}
\usepackage{bbold}
\usepackage{datetime}
\usepackage{numprint}
\usepackage{enumitem}
\usepackage{graphicx,color,amsmath}
\usepackage{hyperref}

\allowdisplaybreaks

\newcommand{\GeV}{\ensuremath{\mathrm{GeV}}}
\newcommand{\TeV}{\ensuremath{\mathrm{TeV}}}

\usepackage{pgfplots}
\usepackage{bm}
\usepackage{mathtools}

\makeatletter
\newcommand{\thickhline}{%
    \noalign {\ifnum 0=`}\fi \hrule height 1pt
    \futurelet \reserved@a \@xhline
}
\newcolumntype{"}{@{\hskip\tabcolsep\vrule width 1pt\hskip\tabcolsep}}
\makeatother

\newcommand{\sdark}{s_{\chi\bar\chi}} %

\usetikzlibrary{positioning,arrows,patterns}
\usetikzlibrary{decorations.markings}
\usetikzlibrary{calc}

\tikzset{
    photon/.style={decorate, decoration={snake}, draw=black},
    vector/.style={decorate, decoration={snake}, draw},
	provector/.style={decorate, decoration={snake,amplitude=2.5pt}, draw},
	antivector/.style={decorate, decoration={snake,amplitude=-2.5pt}, draw},
    fermion/.style={draw=black, postaction={decorate},
        decoration={markings,mark=at position .55 with {\arrow[draw=black]{>}}}},
    fermionbar/.style={draw=black, postaction={decorate},
        decoration={markings,mark=at position .55 with {\arrow[draw=black]{<}}}},
    fermionnoarrow/.style={draw=black},
    gluon/.style={decorate, draw=black,
        decoration={coil,amplitude=4pt, segment length=5pt}},
    scalar/.style={dashed,draw=black, postaction={decorate},
        decoration={markings,mark=at position .55 with {\arrow[draw=black]{>}}}},
    scalarbar/.style={dashed,draw=black, postaction={decorate},
        decoration={markings,mark=at position .55 with {\arrow[draw=black]{<}}}},
    scalartwo/.style={dotted,draw=black, postaction={decorate},
        decoration={markings,mark=at position .55 with {\arrow[draw=black]{>}}}},
    scalartwobar/.style={dotted,draw=black, postaction={decorate},
        decoration={markings,mark=at position .55 with {\arrow[draw=black]{<}}}},
    scalarnoarrow/.style={dashed,draw=black},
    electron/.style={draw=black, postaction={decorate},
        decoration={markings,mark=at position .55 with {\arrow[draw=black]{>}}}},
	bigvector/.style={decorate, decoration={snake,amplitude=4pt}, draw},
    vertex/.style={draw,shape=circle,fill=black,minimum size=1pt,inner sep=0pt},
    fermion2/.style={double, draw=black, postaction={decorate},
		decoration={markings,mark=at position .55 with {\arrow[draw=black]{>}}}},
    momentum/.style={draw=black,line width=0.15mm, postaction={decorate},
        decoration={markings,mark=at position 1 with {\arrow[draw=black]{>}}}}
}

\begin{document}

\title{Dark sector-photon interactions in proton-beam experiments}

\author{Xiaoyong Chu}
\email{xiaoyong.chu@oeaw.ac.at}
\affiliation{Institute of High Energy Physics, Austrian Academy of Sciences, Nikolsdorfergasse 18, 1050 Vienna, Austria}
\author{Jui-Lin Kuo}
\email{jui-lin.kuo@oeaw.ac.at}
\affiliation{Institute of High Energy Physics, Austrian Academy of Sciences, Nikolsdorfergasse 18, 1050 Vienna, Austria}
\author{Josef Pradler}
\email{josef.pradler@oeaw.ac.at}
\affiliation{Institute of High Energy Physics, Austrian Academy of Sciences, Nikolsdorfergasse 18, 1050 Vienna, Austria}

\begin{abstract}
  We consider electromagnetically neutral dark states that couple to
  the photon through higher dimensional effective operators, such as
  electric and magnetic dipole moment, anapole moment and charge
  radius operators.
  We investigate the possibility of probing the existence of such dark
  states, taking a Dirac fermion $\chi$ as an example, at several
  representative proton-beam experiments.
  As no positive signal has been reported, we obtain upper limits (or
  projected sensitivities) on the corresponding electromagnetic form
  factors for dark states lighter than several GeV. We demonstrate
  that while the current limits from proton-beam experiments are at
  most comparable with those from high-energy electron colliders,
  future experiments, such as DUNE and SHiP, will be able to improve
  the sensitivities to electric and magnetic dipole moment
  interactions, owing to their high intensity.
\end{abstract}

\maketitle

\section{Introduction}
\label{sec:introduction}

The operation and development of high-intensity proton facilities are 
the backbone of the world-wide short- and long-baseline neutrino
program.  The collisions of high-energy proton beams on fixed targets
deliver the neutrino fluxes that are registered in near [$\mathcal{O}$(10-100
m)] and far [$\mathcal{O}$(100-1000 km)] detectors through charged and neutral
current interactions. In addition to mapping out the still elusive
neutrino sector of the Standard Model (SM), the near detectors
increasingly serve a second purpose: they become instruments to test
new physics beyond SM. Dark sector particles with masses in the
GeV-range and below can be produced and lead to observable signals in
many previous, existing and upcoming neutrino experiments, such as
LSND~\cite{Athanassopoulos:1996ds},
MiniBooNE~\cite{Aguilar-Arevalo:2018gpe},
COHERENT~\cite{Akimov:2017ade}, DUNE~\cite{Acciarri:2015uup}, among
others~\cite{Huber:2004ug}.  This dual purpose is further supported by
dedicated experiments that aim to probe dark sector states, such as
the proposed SHiP experiment or various beam-dump searches in the
past; for an overview see~\cite{Alexander:2016aln} and references
therein.

Among the prime dark-sector physics cases are the searches for
kinetically mixed dark photons and for new particles that are
connected to SM through the latter as a mediator, see
\textit{e.g.}~\cite{Essig:2013lka} and references therein. A scenario
that has received comparatively less attention is that some dark state
$\chi$ shares a \textit{direct} coupling to the SM photon.  Here, an
ample possibility (that in fact may find itself in both worlds above)
is a milli-charged state that carries a fraction~$\epsilon$ of the
electromagnetic unit of charge~$e$. The scenario has been scrutinized
not only at the intensity frontier,
\textit{e.g.}~\cite{Golowich:1986tj,Prinz:1998ua, Izaguirre:2013uxa,
  Soper:2014ska, Berlin:2018bsc, Magill:2018tbb, Liang:2019zkb}, but
also in terms of their cosmology and astrophysical
implications, \textit{e.g.}~\cite{Davidson:2000hf,Dubovsky:2003yn,McDermott:2010pa,Cline:2012is,Dolgov:2013una,Vogel:2013raa,Dvorkin:2013cea,Ali-Haimoud:2015pwa,Kamada:2016qjo,Agrawal:2017pnb,Munoz:2018pzp,Berlin:2018sjs,Barkana:2018cct,Chang:2018rso,Kovetz:2018zan,Xu:2018efh,Slatyer:2018aqg}

However, even if dark states are perfectly electrically neutral,
higher-dimensional \textit{effective} photon couplings are still
possible.  At mass-dimension five, magnetic- or electric dipole
moments (MDM or EDM), and at mass-dimension six an anapole moment (AM)
or charge radius interaction (CR) are possible for a Dirac fermion
$\chi$. The possibility that these interactions become the defining
feature of dark matter (DM) candidates has been explored
in~\cite{Pospelov:2000bq,Sigurdson:2004zp,Ho:2012bg}, and further
phenomenological studies were presented
in~\cite{Schmidt:2012yg,Kopp:2014tsa,Ibarra:2015fqa,Sandick:2016zut,Kavanagh:2018xeh,Trickle:2019ovy}.

The interest in light dark states not only concerns DM detection, but
also generally aims to test the presence of new sub-GeV particles in
nature. Taking this broader point of view, in~\cite{Chu:2018qrm} we
studied in-depth the possibility that $\chi$ particles---not
necessarily the main component of DM---carry electromagnetic (EM) form
factor interactions. In particular, we studied $\chi$ pair-production
in electron beams on fixed targets at NA64~\cite{Banerjee:2017hhz},
LDMX~\cite{Akesson:2018vlm}, BDX~\cite{Battaglieri:2016ggd}, and
mQ~\cite{Prinz:1998ua}, in $e^+e^-$ colliders
BaBar~\cite{Aubert:2001tu} and Belle-II~\cite{Abe:2010gxa} and in
proton-proton collisions at LHC~\cite{Evans:2008zzb}. In addition,
flavor constraints from rare meson decays such as
$K^{\pm}\to \pi^{\pm} + {\rm
  inv.}$~\cite{Aubert:2003yh,Bird:2004ts,Anisimovsky:2004hr,Artamonov:2009sz,Tanabashi:2018oca}
and precision observables such as ($g-2$) of the
muon~\cite{Jegerlehner:2009ry,Bennett:2006fi,Saito:2012zz,Grange:2015fou}
or constraints on the running of the fine-structure
constant~\cite{Sirlin:1980nh} were worked out. These studies of
electromagnetic moment dark states in the MeV-GeV mass bracket were
then further complemented by a detailed astrophysical study of stellar
cooling constraints, once the mass drops below the MeV
case~\cite{Chu:2019rok}; see also related~\cite{Chang:2019xva}.

In this work we bring the above phenomenological studies to a closure
by working out the current limits on and detection prospects of 
electromagnetically interacting $\chi$ particles utilizing
high-intensity proton beams. Concretely, we set limits from
LSND~\cite{Athanassopoulos:1996ds},
MiniBooNE-DM~\cite{Aguilar-Arevalo:2018wea},
CHARM-II~\cite{DeWinter:1989zg,Vilain:1994qy}, and
E613~\cite{Ball:1981nu} and forecast the sensitivity of
SHiP~\cite{Anelli:2015pba} and
DUNE~\cite{Acciarri:2015uup,Abi:2017aow}.  In this process we take
into account the most important $\chi$ pair-producing reactions from
the Drell-Yan (DY) process, as well as from scalar- and vector-meson
decays. With a detailed prediction of the energy- and
angular-differential flux at hand, we subsequently derive the
observable signals from $\chi$-electron scattering and $\chi$-nucleon
deep inelastic scattering.

The paper is organized as follows: in Sec.~\ref{Sec:emdm} the
electromagnetic form-factor interactions of $\chi$ are introduced.
Sec.~\ref{Sec:production} contains a detailed discussion of dark state
production in proton-beam experiments, and Sec.~\ref{sec:detection} calculates the
generic signal-generation in the detectors. We 
list parameters of the considered experiments in Sec.~\ref{sec:experiments} and 
present the derived results in the parameter space of dark state
coupling vs.~mass in Sec.~\ref{sec:result}. We conclude in Sec.~\ref{sec:conclusions} and
provide further details on our calculations in several appendices. 

\section{Electromagnetic Form Factors}
\label{Sec:emdm}
We consider a neutral Dirac fermion $\chi$ as the dark state which
interacts with the photon, $A_\mu$, or its field strength tensor,
$F_{\mu\nu}$.
At mass dimension-5, the Lagrangian reads
\begin{equation}
	\label{Eq:Lagrangian_dim5}
	\mathcal{L}_\chi^{\text{dim-}5} = \dfrac{1}{2}\mu_\chi\, \bar{\chi} \sigma^{\mu\nu} \chi \, F_{\mu\nu} + \dfrac{i}{2} d_\chi\, \bar{\chi} \sigma^{\mu\nu} \gamma^5 \chi \, F_{\mu\nu}\,,
\end{equation}
where $\mu_\chi$ and $d_\chi$ are the MDM and EDM couplings, expressed in
units of the Bohr magneton $\mu_B \equiv e/(2m_e)$ below;
$m_e$ is the electron mass and
$\sigma^{\mu\nu} \equiv i [\gamma^\mu, \gamma^\nu]/2$.
At mass dimension-6, the Lagrangian reads 
\begin{equation}
	\label{Eq:Lagrangian_dim6}
	\mathcal{L}_\chi^{\text{dim-}6} = - a_\chi \,\bar{\chi} \gamma^\mu \gamma^5 \chi \, \partial^\nu F_{\mu\nu} + b_\chi \,\bar{\chi} \gamma^\mu \chi \, \partial^\nu F_{\mu\nu}\,, 
\end{equation}
in which $a_\chi$ and $b_\chi$ are the AM and CR couplings.
We take the values of all the couplings as real numbers.

The possible UV completion for the effective interaction in
Eqs.~\eqref{Eq:Lagrangian_dim5} and \eqref{Eq:Lagrangian_dim6} could
come from the compositeness of the dark
states~\cite{Bagnasco:1993st,Foadi:2008qv,Antipin:2015xia} or
radiatively, from extra U(1) charged particles at high energy scales.
For simplicity, we assume the effective operator approach holds for
the beam energy of those experiments considered throughout the paper.
The above interactions are then assembled in the matrix element of the
dark current,
\begin{equation}
	\langle \chi (p_f)|J_\chi^\mu (0)| \chi (p_i) \rangle = \bar{u} (p_f) \Gamma_\chi^\mu (q) u(p_i)\,,
\end{equation}
where $p_{i,f}$ and $q = p_i - p_f$ are both four-momenta.
The vertex factor is 
\begin{equation}
	\Gamma_\chi^\mu (q) = i \sigma^{\mu\nu} q_\nu (\mu_\chi +i d_\chi \gamma^5) + (q^2 \gamma^\mu - q^\mu \slashed q) (b_\chi - a_\chi \gamma^5)\,.
\end{equation}
The independence of momentum-transfer in the couplings follows from the assumption that
the UV completion scale is much higher than the center-of-mass energies
considered in this work.

\section{Production of dark states}
\label{Sec:production}

 \begin{table*}[t]
 \centering
 \begin{tabular}{cccccccc}
	 \toprule
  $E_{\rm beam}$ $\setminus$ {\rm meson}& $\pi^0$& $\eta$& $\eta'$ &$\rho$&$\omega$& $\phi$ & $J/\Psi$\\
  \hline
 $ 8.9\,{\rm GeV}$ (MiniBooNE-DM) & $8.6 \times 10^{-1}$  &$8.2 \times 10^{-2}$ & $4.9 \times 10^{-3}$ & $6.9\times 10^{-2}$ & $7.4 \times 10^{-2} $& $1.1 \times 10^{-4}$& 0\\
 $120\,{\rm GeV}$ (DUNE)& $2.9$  & $3.2 \times 10^{-1} $& $3.4 \times 10^{-2}$ & $3.7 \times 10^{-1}$ & $3.7 \times 10^{-1}$ & $1.1 \times 10^{-2}$ & $5.0 \times 10^{-7}$ \\
 $400/450\,{\rm GeV}$ (SHiP, E613/CHARM II)& $4.1$  & $4.6 \times 10^{-1}$ & $5.1 \times 10^{-2}$ & $5.4 \times 10^{-1}$ & $5.4 \times 10^{-1}$  & $1.9\times 10^{-2}$ & $8.0 \times 10^{-6}$\\
 \botrule
 \end{tabular}
 \caption{The number of mesons produced per POT from a
   \texttt{PYTHIA\,8.2} simulation.  }
  \label{tab:meson_production}
 \end{table*}

 At proton-beam experiments, dark states coupled to the photon can be
 produced via prompt processes (\textit{e.g.}~DY process or
 proton bremsstrahlung) and secondary processes (\textit{e.g.}~in
 meson decays or secondary collisions).  In this section, we discuss
 these production processes and provide the calculations of dominant
 channels. Numerical results, taking the SHiP experiment as an
 example, are shown in Fig.~\ref{fig:acceptance_SHiP}. The relative
 importance of the individual contributions does not change
 significantly from experiment to experiment.

\subsection{Drell-Yan Production}
\label{Sec:Drell-Yan}

Dark states with effective couplings to the photon can be
pair-produced directly through quark-antiquark annihilation.
To correctly estimate the $\chi$ production from proton-proton
collision, we utilize the event
generator~\texttt{MadGraph\,5}~\cite{Alwall:2014hca}, to obtain the
energy spectrum and angular distribution of dark states per collision,
denoted as $d^2\hat N^\text{DY}_\chi /(dE_\chi d\cos \theta_\chi)$, as
a function of $\chi$ energy $E_\chi$ and the angle between their
momentum and the beam axis, $\theta_\chi$.

We then take the thick target limit, and calculate the total yield of
dark states from the DY process via
\begin{equation}
	{d^2 N^\text{DY}_\chi \over dE_\chi d\cos \theta_\chi }= {\rm POT}\times {A^{\alpha_1 -\alpha_2}} \times {d^2\hat N^\text{DY}_\chi \over dE_\chi d\cos \theta_\chi}, 
\end{equation}
where the proton on target (POT) number is known for each experiment  and
$A$ is the atomic mass number of the target; $\alpha_1$, $\alpha_2$
are scaling-indices induced by scattering off a nucleus instead of a
proton for the DY cross section, and the total scattering cross
section, respectively. DY processes can be treated as incoherent and
thus $\alpha_1\simeq 1$. The value of $\alpha_2$, for inclusive
proton-nucleus scattering, typically of the order $\mathcal{O} (0.8)$,
depends on the exact target material, and only mildly affects the final
results~\cite{Alekhin:2015byh}. Here we take
$\alpha_2 = 0.9,\,0.88,\,0.8,\,0.71,\,0.6,$ for graphite, beryllium,
iron, molybdenum, and tungsten, respectively.

\subsection{Meson Decay}
\label{Sec:meson_decay}

Another important process is the secondary production of a $\chi$-pair
in the decays of scalar/vector mesons through an off-shell photon. Here we
consider the scalar mesons $\pi^0$, $\eta$ and $\eta'$, as well as
vector mesons $\rho$, $\omega$, $\phi$ and $J/\Psi$.

Typically, if the decays of meson into dark states are kinematically allowed,
they tend to dominate the  production rate.  For
example,~\cite{Harnik:2019zee} shows that the production of
milli-charged particles from meson decay is several orders of
magnitude larger than that from DY. Among them, the $\pi^0$ decay
contribution is the most important.
However, this picture changes when one considers higher-dimensional
operators. This is because the decay rate of light mesons into
$\chi$-pairs will receive additional suppression from their masses,
as shown below.
 
For scalar mesons, the dominant decay channel producing dark states is
a three-body decay with final states $\gamma \chi \bar{\chi}$.
By factorizing out the dark current part, we infer that%
 \begin{equation}
\label{Eq:Brsmchi}
    {\rm Br}({\rm sm} \rightarrow \gamma \chi \bar{\chi})=\dfrac{\Gamma_{{\rm sm} \rightarrow \gamma \chi \bar{\chi}}}{\Gamma_{{\rm sm} \rightarrow \gamma \gamma}} \times {\rm Br}({\rm sm} \rightarrow \gamma \gamma),
\end{equation}
where the subscript ``${\rm sm}$" denotes ``scalar meson''.
The branching ratios, ${\rm Br}({\rm sm} \rightarrow \gamma \gamma)$,
are taken from the PDG~\cite{Tanabashi:2018oca}.
It is worthwhile pointing out that in this step we neglect the mild
$q^2$-dependence induced by the meson transition form factors
$F_{{\rm sm}\gamma \gamma^{*}}(q^2,0)$. Such approximation is
particularly justified for the lighter mesons: the photon virtuality
is limited by kinematics, $q^2\leq m_{\rm sm}^2 $ and corrections
enter at the level of $q^2/m_{\rho}^2$ where $m_{\rho}$ is the
$\rho$-meson mass; see
e.g. \cite{Hoferichter:2014vra,Escribano:2015nra,Husek:2018inh} and
Fig.~\ref{fig:diffBr} in  App.~\ref{App:mesondecay}. 
To calculate $\Gamma_{{\rm sm} \rightarrow \gamma \chi \bar{\chi}}$
and thus the ratio of the two channels, we follow our previous
methodology in~\cite{Chu:2018qrm,Chu:2019rok} and provide the
corresponding expressions in App.~\ref{App:mesondecay}.

A vector meson, in turn, can decay into a $\chi$ pair
directly. Thus we compute the branching ratio
${\rm Br} ({\rm vm } \rightarrow \chi \bar{\chi})$, where the
subscript ``${\rm vm}$'' denotes ``vector meson''. For two-body
decays, one can separate the decay amplitude and phase space factors
to obtain
\begin{equation}
      {\rm Br} ({\rm vm } \rightarrow \chi \bar{\chi}) = {\rm Br} ({\rm vm } \rightarrow e^- e^+) \dfrac{f_{\chi} (m_{\rm vm}^2)}{f_e (m_{\rm vm}^2)} \sqrt{\dfrac{m_{\rm vm}^2 - 4 m_\chi^2}{m_{\rm vm}^2 - 4 m_e^2}},
\end{equation}
where the last two factors count the differences induced by the
interaction type and the phase space, respectively. The expression of
$f({m_{\rm vm}^2})$ for each interaction type is given in App.~\ref{App:mesondecay}, and has previously been derived in
\cite{Chu:2018qrm,Chu:2019rok}. In contrast to the (milli-)charged
case ($f_e$), the function $f_\chi$ heavily relies on the meson mass for
higher-dimensional operators, and the $\chi$ production rate becomes
enhanced for heavier meson decay; for more details see the
appendix. %

To calculate the $\chi$ production rate, the energy and angular
distributions of the produced scalar/vector mesons are
required. However, the latter are still poorly
understood.
One reasonable method is to estimate the normalized neutral meson
distribution using charged meson distributions. Taking the neutral
pion $\pi^0$ as an example, we follow~\cite{deNiverville:2016rqh} and
stipulate that $N_{\pi^0} \sim (N_{\pi^-} + N_{\pi^+}) /2$ and write
the $\pi^0$ distribution as
\begin{equation}
	\begin{split}
	\dfrac{d^2  N_{\pi^0}}{dE_{\pi^0} d\cos\theta_{\pi^0}} \simeq  \dfrac{1}{2} \left(\dfrac{d^2  N_{\pi^+}}{dE_{\pi^+} d\cos\theta_{\pi^+}} + \dfrac{d^2  N_{\pi^-}}{dE_{\pi^-} d\cos\theta_{\pi^-}} \right),
	\end{split}
\end{equation}
where $E_{\pi^{0,-,+}}$ is the energy of the pion and
$\theta_{\pi^{0,-,+}}$ is its respective emission angle relative to
the beam axis.
For charged meson distributions, we follow the literature and use the
Burman-Smith parameterization~\cite{Burman:1989ds} for sub-GeV kinetic
energy proton beams such as LSND, use the Sanford-Wang
distribution~\cite{AguilarArevalo:2008yp} for moderate beam energies
(several GeV) such as MiniBooNE, and use the so-called BMPT
distribution~\cite{Bonesini:2001iz} for larger beam energies (from
tens of GeV to hundreds of GeV), such as for DUNE and SHiP.

\begin{figure*}[t]
\centering
\includegraphics[width=\columnwidth]{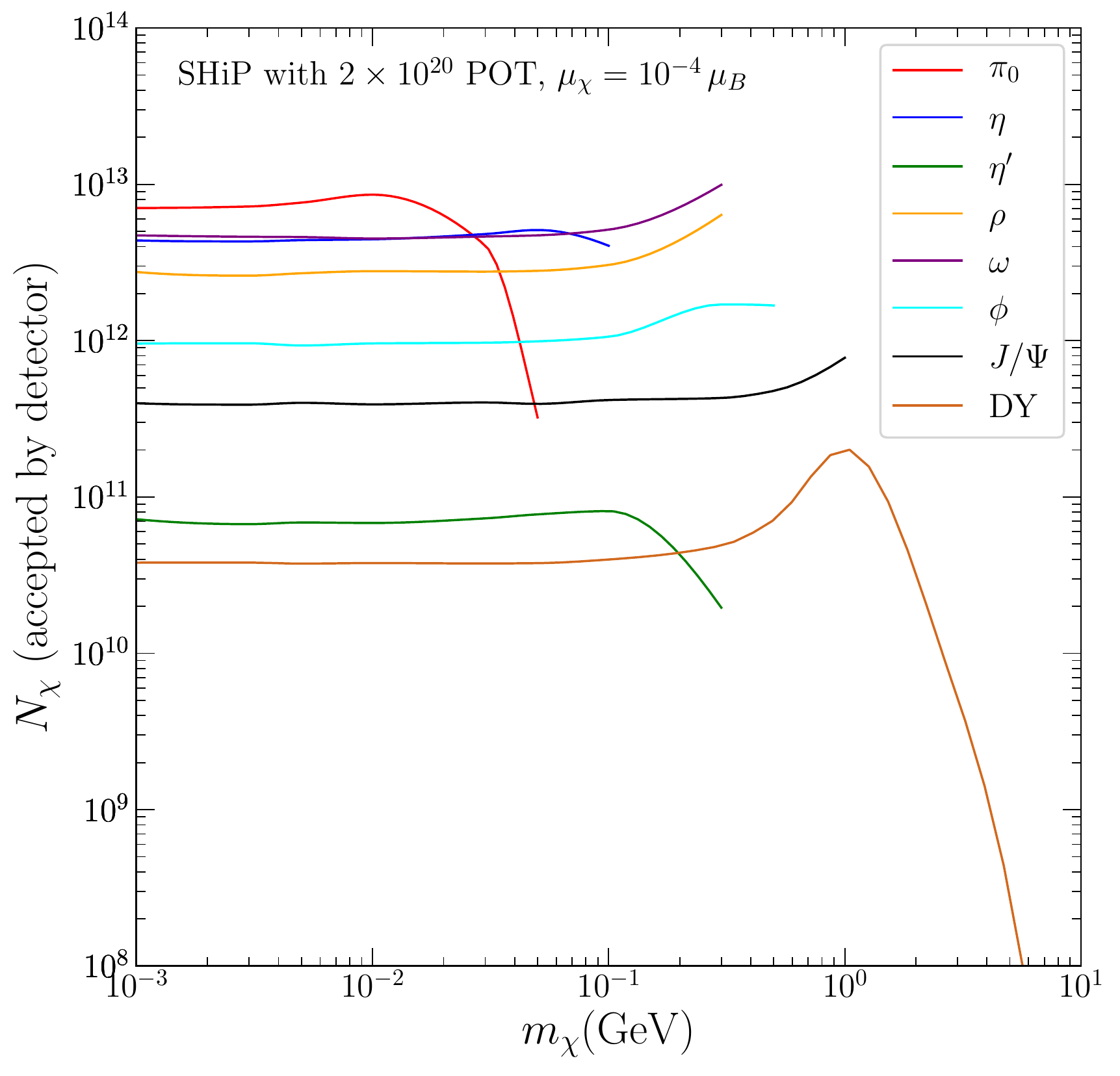}
\includegraphics[width=\columnwidth]{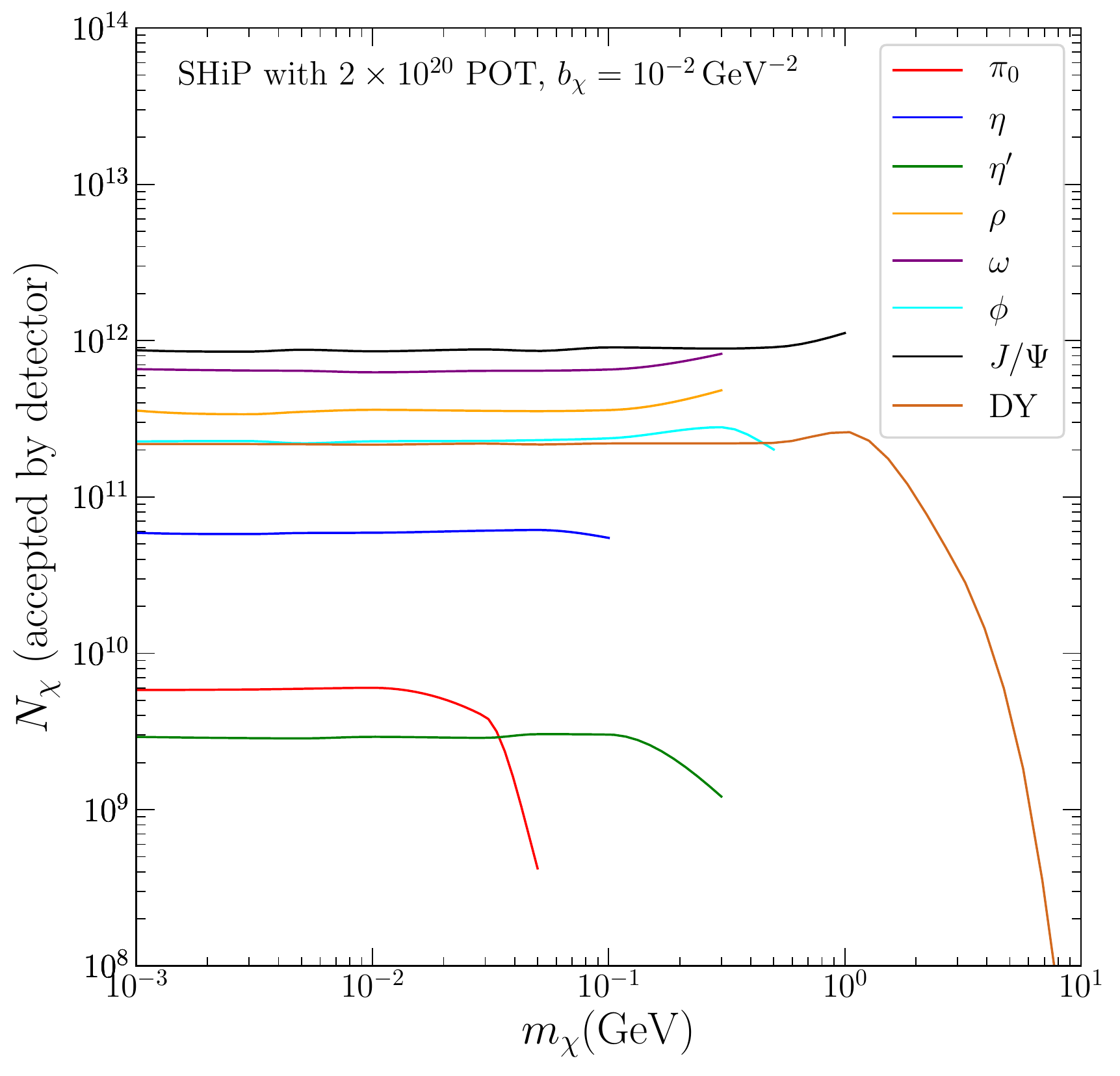}
\caption{The number of produced dark states reaching the SHiP detector
  and produced by a 400\,GeV proton beam, broken down into the
  individual contributing channels, for dimension-5 (left panel) and
  dimension-6 (right panel) operators. Here we only select 2 operators
  (MDM and CR) for demonstration.}
\label{fig:acceptance_SHiP}
\end{figure*}

Besides the normalized meson distribution discussed above, we also
require the total number of produced mesons in each experiment. For
this, we use
\texttt{PYTHIA\,8.2}~\cite{Sjostrand:2006za,Sjostrand:2014zea} to
simulate $pp$ collisions, and list the average number of mesons
produced per POT for each experiment in
Tab.~\ref{tab:meson_production}.  We assume that these meson
production rates per POT remain the same for $pN$ collisions;
for the latter, current  detailed simulations yield differing results, see,
\textit{e.g.},~\cite{Darme:2020ral} for a recent
discussion.\footnote{Although photo-production of light scalar mesons
  is known to scale as $A^{2/3}$~\cite{Krusche:2004xz} and the
  scaling-index for inclusive $pN$ scattering is about
  $\mathcal{O} (0.8)$ as mentioned above, effects of showers and the
  nuclear medium require dedicated simulations/measurements. }  The
meson multiplicities of our Tab.~\ref{tab:meson_production} lie within
the range of their adopted values in previous works,
e.g. \cite{Batell:2014yra, deNiverville:2016dkn,
  CERN-SHiP-NOTE-2016-004, deNiverville:2018dbu,
  DeRomeri:2019kic,Dobrich:2019dxc, Darme:2020ral}, and we estimate
the uncertainties only affect the final bounds by a factor of $1.2$ at
most. Finally, we have also extracted the information on their
momentum and angular distributions from \texttt{PYTHIA\,8.2}, which is
consistent with the fitted distributions mentioned
above~\cite{Dobrich:2019dxc}.

For the final distribution function of $\chi$ particles from meson
decay in the lab frame we find
\begin{equation}
	\label{Eq:spectrum_dist_chi}
	\begin{split}
\dfrac{d^2 N_\chi}{dE_\chi d\cos\theta_\chi} &= \sum_{m=\pi^0,..}\int  { d\cos\theta^* d\phi^* \over 4\pi} dE_\chi^* \, \dfrac{d \hat N^m_\chi}{dE_\chi^* } \\
& \times \dfrac{d^2 N_m}{dE_m d\cos\theta_m}\left| \dfrac{\partial (E_m, \cos\theta_m)}{\partial (E_\chi , \cos\theta_\chi)} \right|,%
        \end{split}
      \end{equation}
      where $E_\chi^*$, $\theta^*$ and $\phi^*$ are defined in the
      meson rest frame and denote respectively the $\chi$ energy, as
      well as the polar and azimuthal angles of the $\chi$ momentum
      w.r.t.~the direction of the boosted meson.
      In contrast, $E_\chi$ and $\theta_\chi$ are defined in lab
      frame, and represent the energy of $\chi$ and the angle of the
      $\chi$ momentum w.r.t.~the beam axis.  At last, $E_m$ and
      $\theta_m$, the energy of the meson and the angle of the meson
      momentum w.r.t.~the beam axis in lab frame, are functions of
      $\theta^*$, $\phi^*$, $E_\chi^*$, $E_\chi$ and
      $\theta_\chi$.
The dark state spectrum from each meson decay, $d \hat N^m_\chi/dE_\chi^*$, is defined as 
\begin{equation}
	\dfrac{d \hat N^m_\chi}{dE_\chi^*} \equiv  2  \dfrac{d {\rm Br}_\chi}{dE_\chi^*},
\end{equation}
where the factor $2$ accounts for the pair production of dark states and ${\rm Br}_\chi$ is the aforementioned ${\rm Br} ({\rm sm } \rightarrow \gamma\chi \bar{\chi})$ or  ${\rm Br} ({\rm vm } \rightarrow \chi \bar{\chi})$, depending on the spin of the meson. Their exact expressions are given in App.~\ref{App:mesondecay}.  Note that to obtain Eq.~\eqref{Eq:spectrum_dist_chi} we have used the fact that the meson decay at rest is isotropic.

In practice, we perform Monte Carlo simulations to numerically obtain
the distribution function of $\chi$ from meson decay, instead of
integrating Eq.~\eqref{Eq:spectrum_dist_chi} directly, as the latter
is prohibitively time-consuming.

\subsection{Other production mechanisms}
\label{Sec:pN_bremsstrahlung}

Here we discuss additional channels of $\chi$-pair
production. Prominently, proton-nucleus bremsstrahlung contributes to
the production of $\chi$ particles. The process can \textit{e.g.}~be
estimated using the Fermi-Weizs\"{a}cker-Williams
method~\cite{Fermi1924,vonWeizsacker:1934nji,Williams1934}, as has
been done in~\cite{Blumlein:2013cua,
  deNiverville:2016rqh, Feng:2017uoz, Tsai:2019mtm}.  However, for the
higher-dimensional interactions studied here, the production of
$\chi$-pairs through bremsstrahlung is generally dominated by the
contribution of the vector meson resonance at
$s_{\chi\bar\chi} \simeq
m^2_{\rho,\omega}$~\cite{Faessler:2009tn}. Since we have already taken
into account the resonant contribution through the vector meson decay
processes above, we will not consider the proton bremsstrahlung any
further; thereby we also avoid any double-counting.

Another source of $\chi$-pair production is the capture of pions onto
nuclei or protons via $p\pi^- \to n \gamma^{*} \to n \chi\bar\chi$. This process will
mostly result in low-energy $\chi$-particles and is not considered
further here.  At last, secondary collisions, \textit{e.g.}~between
secondary electrons/photons and the target, should not appreciably
contribute to the $\chi$ yield in our framework. We always neglect the
latter contributions in this work.

\section{Detection of dark states}
\label{sec:detection}

The dark states, produced in proton-nucleus collisions, travel
relativistically through the shield into the downstream detector,
leading to observable signals. In this work, we focus on their elastic
scattering with electrons in the detector (LSND, MiniBooNE-DM, CHARM
II, DUNE, SHiP) and hadronic shower signals caused by nuclear deep
inelastic scattering (DIS) in
E613. %

For simplicity, we will approximate the detector-shapes as cylinders
with a constant transverse cross-sectional area and a certain
depth. Thus, the geometric acceptance of the dark states is determined
by the target-detector distance and an effective size.
For the nearly spherical detector in MiniBooNE-DM, we take the
geometry into account in deriving the signal rate.

\begin{figure*}[t]
\centering
\includegraphics[width=2\columnwidth]{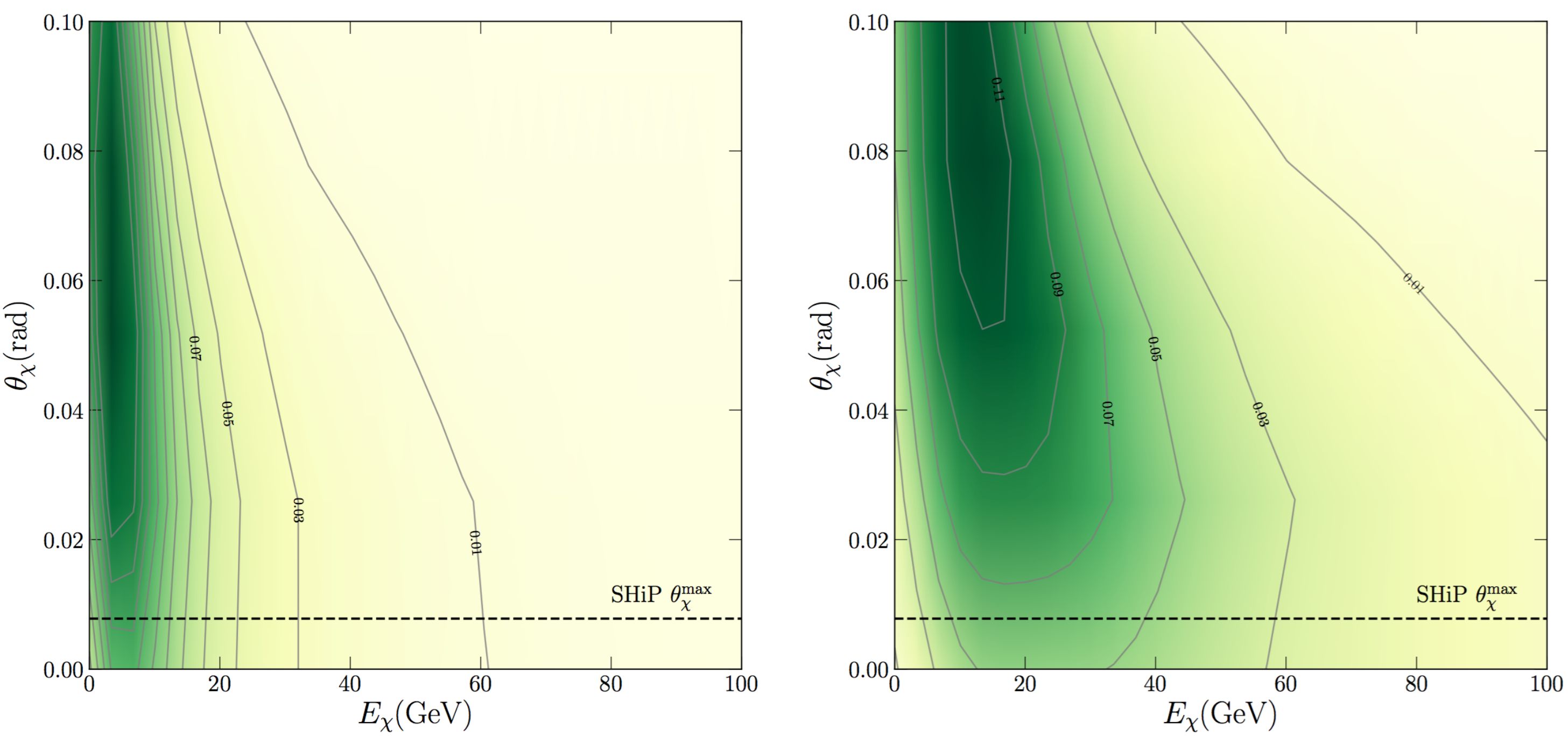}
\caption{Normalized energy and angular distribution of $\chi$
  particles, $d^2\hat N_\chi/ (dE_\chi d\theta_\chi)$, for SHiP with
  400\,GeV incident protons. Here we only select 2 operators, MDM
  (left) and CR (right) and $m_\chi = 1\,$MeV for
  demonstration.
}
\label{fig:distSHiP}
\end{figure*}

\subsection{Scattering on electrons}
\label{Sec:elastic_scattering}

When entering the detector, $\chi$ particles may scatter with
electrons and cause detectable recoil signals.
Following~\cite{Chu:2018qrm}, the master formula to calculate the
number of signal events reads
\begin{equation}
  \label{eq:master-det}
	\begin{split}
	N_{\rm sig}^{(e)} & = n_e  
	 \int_{E_{R}^{\rm min}}^{E_{R}^{\rm max}} dE_R \int_{E_\chi^{\rm min}} dE_\chi \,  L_{\rm det}  \\ & \quad \times  \epsilon_{\rm eff} \dfrac{dN_\chi}{dE_\chi} \dfrac{d\sigma_{\chi e}}{dE_R} \,,%
	\end{split}
\end{equation}
where $n_e$ is the electron number density of the target,
$L_{\rm det}$ is the depth of the detector, $E_R$ is the electron recoil energy
with respective experimental threshold and cutoff energies
$E_R^{\rm min}$ and $E_R^{\rm max}$, $E_\chi$ is the initial $\chi$
energy in lab frame, and $ \epsilon_{\rm eff}$ is the detection
efficiency.
The minimal energy of dark states $E_\chi^{\rm min}$ can be expressed
in terms of $E_R$ as
\begin{equation}
	E_\chi^{\rm min} = \dfrac{E_R}{2} + \dfrac{\sqrt{m_i (E_R +2m_i)(E_R m_i + 2m_\chi^2)}}{2m_i},
\end{equation}
where $m_i$ is target mass, \textit{i.e.}~the electron mass in this
case. %
The differential scattering cross section, $d\sigma_{\chi e}/{dE_R}$,
is found in App.~E of~\cite{Chu:2018qrm}.

\begin{figure*}[t]
\centering
\includegraphics[width=2\columnwidth]{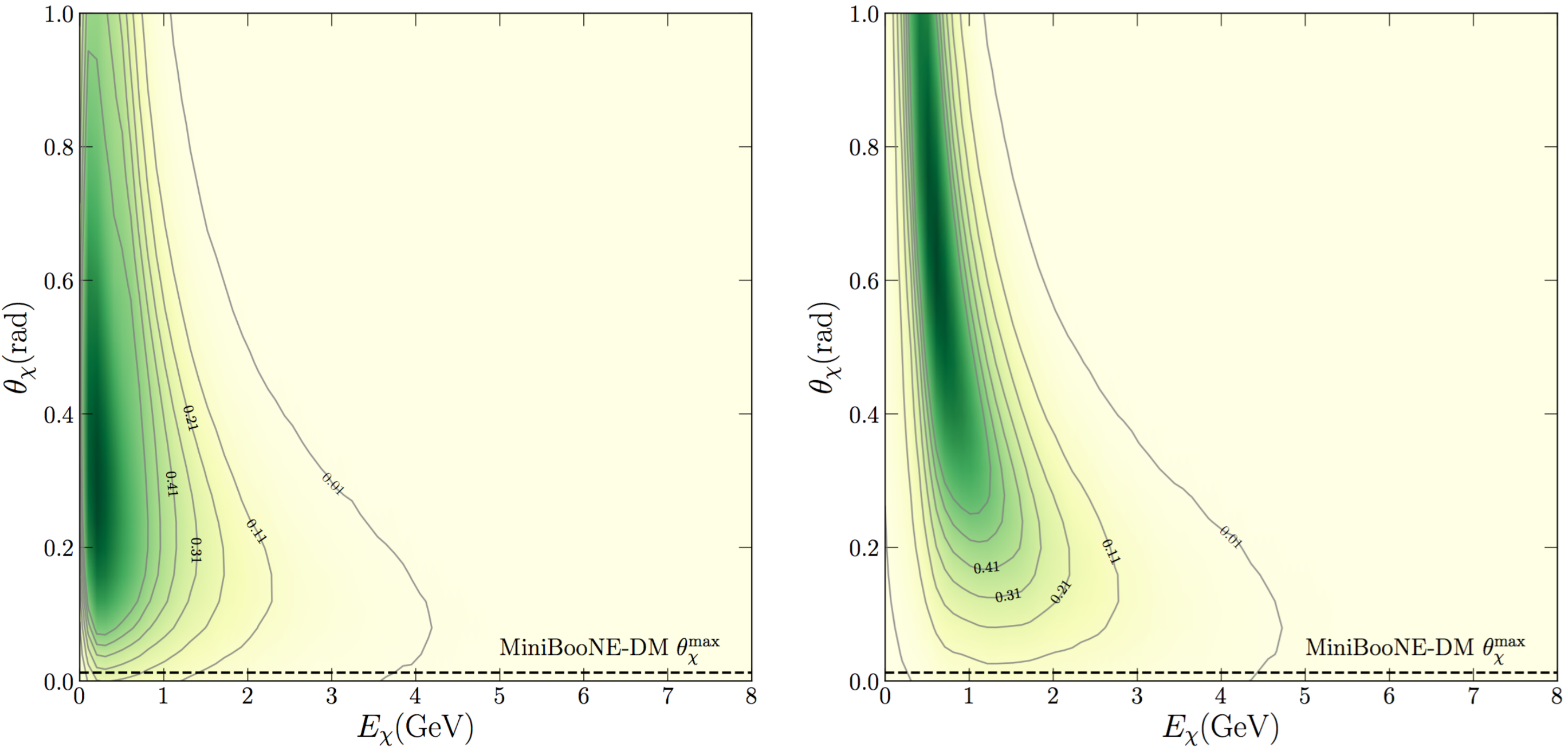}
\caption{Same as Fig.~\ref{fig:distSHiP} for MiniBooNE-DM with 8\,GeV
  incident protons: MDM (left) and CR (right) and $m_\chi = 1\,$MeV.
}
\label{fig:distMiniBooNE}
\end{figure*}

The spectrum of dark states that have entered the detector,
$dN_\chi / dE_\chi$, is obtained by summing up all production
processes in the previous section, and applying the detector geometric
cut,
\begin{equation}
  \label{eq:dn-det}
	\dfrac{dN_\chi}{dE_\chi} = \int^{1}_{\cos \theta_\chi^{\rm max}} d\cos \theta_{\chi} \,\dfrac{d^2 N_\chi}{dE_\chi d\cos\theta_\chi} . 
\end{equation}
The maximum opening angle $\theta_\chi^{\rm max}$ is obtained from the target-detector distance
and the effective size of the detector. This is illustrated in
Fig.~\ref{fig:distSHiP} for the SHiP experiment (400\,GeV proton) and
Fig.~\ref{fig:distMiniBooNE} for the MiniBooNE-DM experiment (8\,GeV
proton), where only $\chi$ particles below the horizontal dashed line
($\theta_\chi \le \theta_\chi^{\rm max}$)  enter the detector.
For the purpose of illustration, the two figures give the contours of
${d^2 \hat N_\chi}/{dE_\chi d\theta_\chi}$, normalized as per $\chi$
particle via
\begin{equation}
	\dfrac{d^2 \hat N_\chi}{dE_\chi d\theta_\chi} \equiv -{\sin\theta_\chi \over N_\chi}\dfrac{d^2  N_\chi}{dE_\chi d\cos\theta_\chi} \,,
\end{equation}
which is obviously independent of the values of form factor couplings.

As shown by the figures, only about $0.1$--$10^{-5}$ of the total
number of $\chi$ particles produced reach the detectors, and this
strongly suppresses the number of final events at low energy
experiments, such as at MiniBooNE-DM.\footnote{This is also one of the
  motivations for off-axis detectors in proton-beam experiments; see
  \textit{e.g.}~\cite{Coloma:2015pih,Frugiuele:2017zvx,deGouvea:2018cfv}. }
Moreover, such reduction becomes more severe for dark particles
generated from heavy meson decay, and is largely insensitive to
$m_\chi$ for $\chi$ particles from DY processes. Besides, for
higher-dimensional operators, a preference for more energetic $\chi$
particles can also be observed by comparing the left and right panels
in Fig.~\ref{fig:distSHiP} (also in
Fig.~\ref{fig:distMiniBooNE}). This is due to their different
energy-dependence in the production rate, and will be further
discussed in Sec.~\ref{subsec:compare}.

Several experiments also make cuts on the electron recoil angle,
$\theta_R$, in order to  reduce  backgrounds.
From kinematics, the recoil energy $E_R$ can be expressed in terms of $E_\chi$ and $\theta_R$ as 
\begin{equation}
	E_R =  \dfrac{4 m_e (E_\chi+m_e)^2}{- q_\chi^2 \cos(2\theta_R) + q_\chi^2 + 4 E_\chi m_e +2 (m_e^2 + m_\chi^2)} -m_e\,,
\end{equation}
where $q_\chi^2  \equiv E_\chi^2 - m_\chi^2$.
Hence, we take the cuts on $\theta_R$ as a further requirement
on the boundaries of $E_R$, where $\theta_R^{\rm min}$
($\theta_R^{\rm max}$) give upper (lower) limits on $E_R$.
 
For the spherical detector in the MiniBooNE-DM experiment, we use 
an incoming angle-dependent depth for the detector, which reads
\begin{equation} 
	L_{\rm det} (\theta_\chi)= 2 \sqrt{R^2 - D^2 \sin^2 \theta_\chi}\,,
\end{equation}
where the radius of the detector $R=6.1\,{\rm m }$, and the distance
between the collision point and the detector center
$D = 490\,{\rm m}$.

Finally, we note in passing that while we focus on electron recoil
signals, which in general provide better bounds, the expressions above
are easily generalized to describe nucleon recoils.

\subsection{Hadronic showers}

The dark states may also cause hadronic showers, which is relevant for
E613. %
Following \cite{Soper:2014ska} we consider the deep inelastic
scattering of $\chi$ with nucleons as the energy deposition process,
while neglecting any coherence effects since the typical momentum
transfer is larger than the QCD confinement scale.
It is worth pointing out that we do not consider the possibility of
multiple scatterings in the detector, since the coupling between
the $\chi$ particle and the photon is assumed to be weak; see
Sec.~\ref{Sec:mean_free_path}.

To derive the expected number of signal events, we first compute the
differential cross section of $\chi$-$N$ deep inelastic scattering.
The 4-momentum of $\chi$ before (after) scattering is denoted as $p_\chi$ ($p_\chi'$). The momentum transfer carried by the intermediate photon is defined as $q = p_\chi - p_\chi'$, which is spacelike.  
Following the DIS formalism for leptons, we introduce the Bjorken variable $x \equiv Q^2/(2m_N \nu)$, with $m_N \simeq m_p$ being the nucleon mass, $Q^2 \equiv  -q^2 > 0$ and $\nu$ being the energy transfer in the rest frame of the nucleons.
The differential cross section is then written as
\begin{equation}
\label{Eq:dxsecDIS}
\dfrac{d^2\sigma_{\chi N}}{d\nu dQ^2} = \dfrac{\alpha}{4m_N (E_\chi^2 -m_\chi^2)} \dfrac{L_{\mu\nu} W^{\mu\nu}}{Q^4}\,, 
\end{equation}
where the dark current $L^{\mu\nu}$ can be written in terms of the vertex factors of Sec.~\ref{Sec:emdm},
\begin{equation}
	L_{\mu\nu} = \dfrac{1}{2} {\rm Tr}[(\slashed p_\chi + m_\chi) \Gamma_{\mu} (q)(\slashed p_\chi ' + m_\chi) \Gamma_{\nu} (-q)]\,,
\end{equation}
with the factor $1/2$ coming from average over initial state
$\chi$-spins. 
The hadronic tensor $W^{\mu\nu}$ may be expressed as 
\begin{equation}
	W^{\mu\nu} = \left( - g^{\mu\nu} + \dfrac{q^\mu q^\nu}{q^2}\right) F_1 (x, Q^2) + a^{\mu\nu} F_2 (x, Q^2)\,,
\end{equation}
in which 
\begin{equation}
	a^{\mu\nu}  \equiv \dfrac{1}{p_N \cdot q} \left( p_N^\mu -\dfrac{p_N \cdot q}{q^2} q^\mu \right) \left( p_N^\nu -\dfrac{p_N \cdot q}{q^2} q^\nu \right)\,,
\end{equation}
with $p_N$ being the 4-momentum of the nucleon before the scattering.
We adopt the results for the two structure functions as 
\begin{equation}
	F_1 = \dfrac{1}{2x} \sum\limits_q  e_q^2 \,x f_{q} (x,Q^2), ~~F_2 = 2xF_1\,,
\end{equation} 
where $e_q$ is the charge of quarks in unit of electron charge.
We sum over flavors of light quarks/antiquarks,
$q = u, \bar{u}, d, \bar{d}, s, \bar{s}$, and use the values of parton
distribution function $ x f_{q} (x,Q^2)$ averaged over nucleons for
each corresponding nucleus from~\cite{Hirai:2007sx}.\footnote{Such
  parameterization is numerically equivalent to the one of
  \cite{Soper:2014ska} in the limit of $\nu^2 \gg Q^2$, which is the
  case in E613. }
The expected number of signal events is given by
\begin{equation}\label{stat:bounds}
N_{\rm sig} = n_N L_{\rm det} \epsilon_{\rm eff}
		\int  dE_\chi \int d\nu  dQ^2  \, \dfrac{dN_\chi}{dE_\chi} \,\dfrac{d^2\sigma_{\chi N}}{d\nu dQ^2}\, ,
\end{equation}
where $n_N$ is the number density of nucleons in the detector.  The
integration boundaries for $\nu$ and $Q^2$ are derived from kinematics
as
\begin{equation}
	E_{\rm cut} < \nu < E_\chi - m_\chi \, ,
\end{equation}
where $E_{\rm cut}$ is the experiment-specific threshold energy. The squared momentum transfer $Q^2$ lies in the range
\begin{equation}
		 2(E_\chi^2 - E_\chi \nu -  m_\chi^2 ) \mp 2\sqrt{E_\chi^2-m_\chi^2 }\sqrt{(E_\chi-\nu)^2 -m_\chi^2}  \, .
\end{equation}
Finally, there is the general requirement $x<1$.

\subsection{Mean-free-path of dark states}
\label{Sec:mean_free_path}

As already mentioned above, our calculations are based on the
assumption that $\chi$ particles travel freely, both in the shield
and in the detector. This may be validated by estimating the mean free
path of $\chi$, using transport cross section $\sigma^T_{\chi p}$ of
$\chi$-proton scatterings,
\begin{equation}
	\lambda_\chi \sim (n_p \sigma^T_{\chi p})^{-1} \,,
\end{equation}
where $n_{p}$ is the proton number density.
The transport cross section is used as it removes the influence of
soft scatterings that would not attenuate the flux of dark particles.

To obtain an estimate, we use the elastic scattering processes for
which the formul\ae\ can be found in App.~E of
\cite{Chu:2018qrm}. Here we take the typical distance between the
collision point and the detector to be 100\,m and the dump/shield mass
density to be 10\,g/cm$^3$. By requiring $\lambda_\chi \ge 100\,$m,
one can see that these proton-beam experiments are sensitive to the EM
form factor parameters
\begin{equation}
	\mu_\chi,d_\chi \le 0.005\,\mu_B\,, \text{~~and~~~}a_\chi,b_\chi \le 0.1 \text{GeV}^{-2}\,,
\end{equation}
for sub-GeV $\chi$ particles with $E_\chi = 5$\,GeV. As parameters
larger than these values above are already excluded by other probes,
we may always assume that $\chi$ particles scatter at best once inside
the entire experimental setup.

\section{Experiments}
\label{sec:experiments}

 \begin{table*}[t]
 \centering
 \begin{tabular}{lcc@{\hskip 0.3cm}lcccl}
	 \toprule
   Experiments        & POT ($10^{20}$) & $|\theta_\chi^{\rm max}|$ & Signal process and cuts                                                             & $N_{\rm bkg}$          & $\epsilon_{\rm eff}$ & on/off axis & Reference                                           \\
 \hline
  LSND               & 1800            & -                         & e-recoil ($E_R \in   [18,52]\,{\rm MeV}$, $\theta_R \leq \pi/2$)                  & $N_{\rm sig} \leq 110$ & 0.16                 & $31^\circ$  & \cite{Athanassopoulos:1996ds, deNiverville:2016dkn} \\
 MiniBooNE-DM        & 1.86            & $12.4\,{\rm mrad}$        & e-recoil ($E_R \in  [75,850]\,{\rm MeV}$, $\theta_R \leq 140\,{\rm mrad}$)        & 0                      & 0.2                  & $0^\circ$   & \cite{Aguilar-Arevalo:2018wea}                      \\
 CHARM II            & 0.25            & $2.1 \,{\rm mrad}$        & e-recoil ($E_R \in  [3,24]\,{\rm GeV}$, $E_R\theta_R^2 \leq 3\,{\rm MeV}$)        & 5429                   & $\sim 1$             & $0^\circ$   & \cite{DeWinter:1989zg, Vilain:1994qy}               \\
  DUNE (10 yr) & 11/yr           & $3.4\,{\rm mrad}$         & e-recoil ($E_R \in   [0.6,15]\,{\rm GeV}$, $E_R\theta_R^2 \leq 1\,{\rm MeV}$) & 8930/yr                & 0.5                  & $0^\circ$   & \cite{Brown:2018rcz,Hostert:2019iia}                \\
 SHiP                & 2               & $7.8\,{\rm mrad}$         & e-recoil ($E_R \in   [1,20]\,{\rm GeV}$, $\theta_R \in   [10,20]\,{\rm mrad}$)    & 846                    & $\sim 1$             & $0^\circ$   & \cite{Anelli:2015pba,Buonocore:2018xjk}             \\
   E613              & 0.0018          & 12.8\,mrad                & had.~shower ($E_N^{\rm dep} \geq 20\,{\rm GeV}$ per event)                      & $N_{\rm sig} \leq 180$ & $\sim 1$             & $0^\circ$   & \cite{Soper:2014ska,Ball:1981nu}                    \\
 \botrule 
 \end{tabular}
 \caption{Summary of key
   parameters from each experiment.  Here $\theta_\chi^{\rm max}$ is
   the maximal angle between $\chi$'s momentum and the beam axis in
   order for $\chi$ to pass through the detector, $E_R$ is the recoil
   energy of the target, $\theta_R$ is the recoil angle of the target
   with respect to the $\chi$ momentum and $\epsilon_{\rm eff}$ is the
   detection efficiency of considered signal.
 }
  \label{tab:experiment}
 \end{table*}

 In this section, we briefly review the relevant details of each
 experiment under consideration. In order to derive the ensuing
 90\%~C.L.~limits, we require that the number of events generated by the dark
 states, 
\begin{equation}
	N_{\rm sig} \leq {\rm Max}[0, N_{\rm obs} - N_{\rm bkg}] + 1.28 \sqrt{N_{\rm obs}}\,,
      \end{equation}
       where $N_{\rm obs}$ is the number of actual observed
      events and $N_{\rm bkg}$ is the expected number of background
      events.  When making forecasts for future experiments, we assume
      $N_{\rm obs} = N_{\rm bkg}$.  The standard criterion
      $N_{\rm sig} \leq 2.3$ is adopted if no events were observed.
      For each experiment, the summary of relevant parameters can be
      found in Table.~\ref{tab:experiment}.

\subsection{LSND}

At the Liquid Scintillator Neutrino Detector (LSND) experiment, a
proton beam of $800\,{\rm MeV}$ kinetic energy was conducted onto
water or a high-$Z$ target such as
copper~\cite{Athanassopoulos:1996ds}.  The detector was located at a distance of
$35\,{\rm m}$ from the beam dump, with an off-axis angle of
$31^\circ$, and an active volume comprised of an $8.3\,{\rm m}$ long
cylinder with a diameter of $5.7\,{\rm m}$, filled with 167 tonnes of
mineral oil ${\rm CH}_2$~\cite{Mills:2001wvh}.

Due to the low beam energy, we consider $\pi^0$ decay as the only
$\chi$ production channel in LSND as other heavier mesons decay and
DY channels are suppressed. As it is difficult to generate the
total production rate of $\pi^0$ in \texttt{PYTHIA\,8.2} at such low
energy, we instead estimate it via the ratio
$(\sigma_{pp\to X+\pi^0}+2\sigma_{pp\to X+2\pi^0})/\sigma_{pp}$, which
measurements put at a value of approximately~0.1~\cite{Shimizu:1982dx,
  Achilli:2011sw}.  Under the assumption that this ratio remains
unchanged for proton-nuclear scattering, we adopt the value
0.1$\pi^0$/POT as our fiducial value in the calculation. This is close
to the production rate of positively charged mesons in LSND, about
0.08$\pi^+$/POT~\cite{Allen:1989dt}, as well as the value used in
COHERENT experiment, 0.09$\pi^{0}$/POT~\cite{Akimov:2019xdj}.

In the MDM case with $m_{\chi} \ll m_{\pi}$, the $\chi$ flux entering
the detector is then approximately
\begin{equation}
	\Phi_\chi \simeq 2.2 \times 10^{5}\,{\rm cm}^{-2} \left(\dfrac{\mu_\chi}{2 \times 10^{-5}\,\mu_B}\right)^2\,, 
\end{equation}
yielding the constraint $\mu_\chi \leq 2 \times 10^{-5}\,\mu_B$. This can be rescaled to compare with the LSND results~\cite{Auerbach:2001wg}, which estimates that the $\nu_e$ flux  entering the detector, $\Phi_{\nu_e}$, is about $1.2\times 10^{14}\,{\rm cm}^{-2}$, leading to a bound on $\nu_e$'s MDM at $\mu_{\nu_e} \le 10^{-9}\,\mu_B$~\cite{Auerbach:2001wg}. 
One can see the equality  
\begin{equation}
\left( \Phi_\chi \times \mu_\chi^2 \right) |_{\mu_\chi = 2 \times 10^{-5}\mu_B}  \simeq    \Phi_{\nu_e} \times (10^{-9}\,\mu_B)^2   \,, \end{equation}
is approximately satisfied, suggesting that our treatment of the detector works well.

\subsection{MiniBooNE-DM}

The Booster Neutrino Experiment, MiniBooNE, operates at the Fermi
National Accelerator~\cite{AguilarArevalo:2008qa}.
The Booster delivers a proton beam with kinetic energy
$E_{\rm beam} = 8\,{\rm GeV}$ ($\sqrt{s} \sim 4.3\,{\rm GeV}$) on a
beryllium ($A_{\rm Be} = 9$) target.
The center of the spherical on-axis detector is placed $490\,{\rm m }$
downstream from the beam dump with a diameter of $12.2\,{\rm m}$
filled with 818 tonnes of mineral oil ${\rm C}_n {\rm H}_{2n+2}$
($n\sim 20$).
In practice, we are more interested in the off-target mode of
MiniBooNE, where the proton beam hits directly the steel beam dump,
with an ensuing smaller high-energy neutrino background.  This is
referred to as MiniBooNE-DM, which has data with $1.86 \times 10^{20}$
POT~\cite{Aguilar-Arevalo:2018wea}.  By only focusing on electrons
with extremely small recoil angles, the background was effectively
reduced to zero in this off-target
mode~\cite{Aguilar-Arevalo:2018wea}. That is, we derive the 90\%
C.L. limits on the couplings of dark states to the photon by requiring
$N_\text{sig}\le 2.3$.

It is well known that in the on-target mode with
$1.3\times 10^{21}$\,POT, MiniBooNE reported a significant excess of
electron-like events~\cite{Aguilar-Arevalo:2018gpe}. In addition, the
background event of a single electron recoil is estimated to be about
one hundred, after the same cuts as
above~\cite{Dharmapalan:2012xp}. Substituting these values into
Eq.~\eqref{stat:bounds} in turn suggests that the on-target mode
should lead to slightly weaker limits than those from MiniBooNE-DM,
despite its larger POT number.

\subsection{CHARM II}

CERN High energy AcceleRator Mixed field facility II
(CHARM II) was a fixed-target experiment designed for a
precision measurement of the weak angle. It utilized a 450\,GeV proton
beam on a Be target, and collected data with
$2.5\times 10^{19}$\,POT during 1987-1991~\cite{Vilain:1994qy}. The
main detector is a 692~t glass calorimeter (SiO$_2$, on average
$\langle A\rangle \simeq 20.7$ per nucleus), and has an active area
of $3.7\times 3.7$\,m$^2$, about 870\,m away from the target along the beam
axis~\cite{DeWinter:1989zg}.  In this study, we focus on the single
electron recoil signals, as the detector has an almost 100\%
efficiency to record electromagnetic showers for recoil energy
$E_R\in [3,\,24]$\,GeV.

To estimate the number of background events, we take 
$N_{\rm obs} = 5429$ reported in \cite{Vilain:1994qy}, largely induced by electron
scattering with energetic $\nu_{\mu} + \bar \nu_{\mu}$ particles. This
estimation is conservative, as CHARM II was able to determine the
value of the Weinberg angle with the uncertainty below several
percents.

\subsection{DUNE}

The Deep Underground Neutrino Experiment (DUNE) is proposed to be
performed at the Long-Baseline Neutrino Facility (LBNF), and can be
used to probe light dark particles~\cite{Acciarri:2015uup,
  Abi:2017aow}. At DUNE, a graphite ($A_{\rm C} = 12 $) target is hit by a
proton beam with an initial energy $E_{\rm beam} = 120\,{\rm GeV}$.
The near detector (75~t fiducial mass) will be placed $574\,{\rm m }$
downstream from the target. It is on-axis and a parallelepiped with a
size $4\times 3 \times 5\,{\rm m^3}$ and we use $5$~m as its effective
depth~\cite{Harnik:2019zee}.
The detector is filled with liquid Argon (LAr).

We take a 10-year run of the DUNE experiment, with a total POT of
$1.1\times 10^{22}$. The observable signals we consider for DUNE are
single electron events caused by $\chi$-$e$ scatterings. The detection
efficiency is assumed to be $\epsilon_{\rm eff} = 0.5$ for the LAr
time projection chamber. Following \cite{Hostert:2019iia,
  DeRomeri:2019kic}, we require the cut on the electron recoil angle
to satisfy $E_R \theta_R^2 \le 1$\,MeV, which significantly reduces
the number of background events from charged-current $\nu_e$-$n$ scattering; see
Tab.~\ref{tab:experiment} for details of the parameters.

\subsection{SHiP}

A fixed-target facility to Search for Hidden Particles (SHiP) is
proposed at the CERN super proton synchrotron (SPS)
accelerator~\cite{Anelli:2015pba}.
At the SPS facility, a proton beam with
$E_{\rm beam} = 400\,{\rm GeV}$ ($\sqrt{s} \sim 27.4 \, {\rm GeV}$) is
deployed to collide with the titanium-zirconium doped molybdenum target
($A_{\rm Mo}= 95.95$).
An emulsion cloud chamber detector will be located $56.5 \, {\rm m }$
downstream from the target, and it will be filled with layers of nuclear emulsion films. Following the latest SHiP
report~\cite{Ahdida:2704147}, the size of the detector ($\sim$ 8
tonnes) is set to be $80 \times 80 \times 100\,{\rm cm^3}$. We assume
a 100\% detection efficiency for simplicity.\footnote{A unity
  efficiency was also used in~\cite{Buonocore:2018xjk,
    Jodlowski:2019ycu}.}

The detection process we consider for SHiP is also $\chi$-$e$
scatterings. With $2\times 10^{20}$ POT after 5-years of operation the
number of background events is estimated to be $846$, which is
dominated by $\nu_e$ quasi-elastic scattering with a soft final state
proton~\cite{Ahdida:2704147}. %

\subsection{E613}
\label{subsec:E613}

E613 was a beam dump experiment at Fermilab, set up to study neutrino
production, with a $400\,{\rm GeV}$ proton beam hitting a tungsten
target~\cite{Ball:1981nu}. The detector, $55.8\,{\rm m}$ away from the
target, consisted of 200 tonnes lead plus liquid scintillator. Its
size was $1.5\times 3 \times 3 \,{\rm m}^3$, with a mass density of
about $10\, {\rm g/cm}^3$. In order to compare with the previous
results~\cite{Soper:2014ska, Mohanty:2015koa}, we only consider a
circular region of the detector with a radius of 0.75\,m along the
beam axis. Moreover, for nucleon-recoil events in E613, the energy
deposit needs to be larger than 20\,GeV, in order to be
recorded. We require the number of
such events to be below 180 during its $1.8 \times 10^{17}$\,POT run
to obtain the constraints.

We assume a thick target so that each incident proton scatters
once. This is different from the treatment by \cite{Soper:2014ska,
  Mohanty:2015koa}, which estimated the number of scatter events per
POT following the scaling
\begin{equation}\label{eq:soper}
	L_\text{T} \times  n_T \,\sigma_{p T}   \,
\end{equation}
with $L_\text{T}$ ($n_T$) being the total length (the nucleon number
density) of the target, and $ \sigma_{p T}$ the scattering cross
section between proton and target. For E613, where $L_\text{T}$ is
much larger than the mean-free-path of a 400\,GeV proton (a few cm in
tungsten), Eq.~\eqref{eq:soper} significantly over-estimates the total
number of produced $\chi$ particles. As a result, our limits are 
weaker than those derived in~\cite{Mohanty:2015koa}. We revise the
previous results in the next section.

\begin{figure*}[!ht]
\centering
\includegraphics[width=\columnwidth]{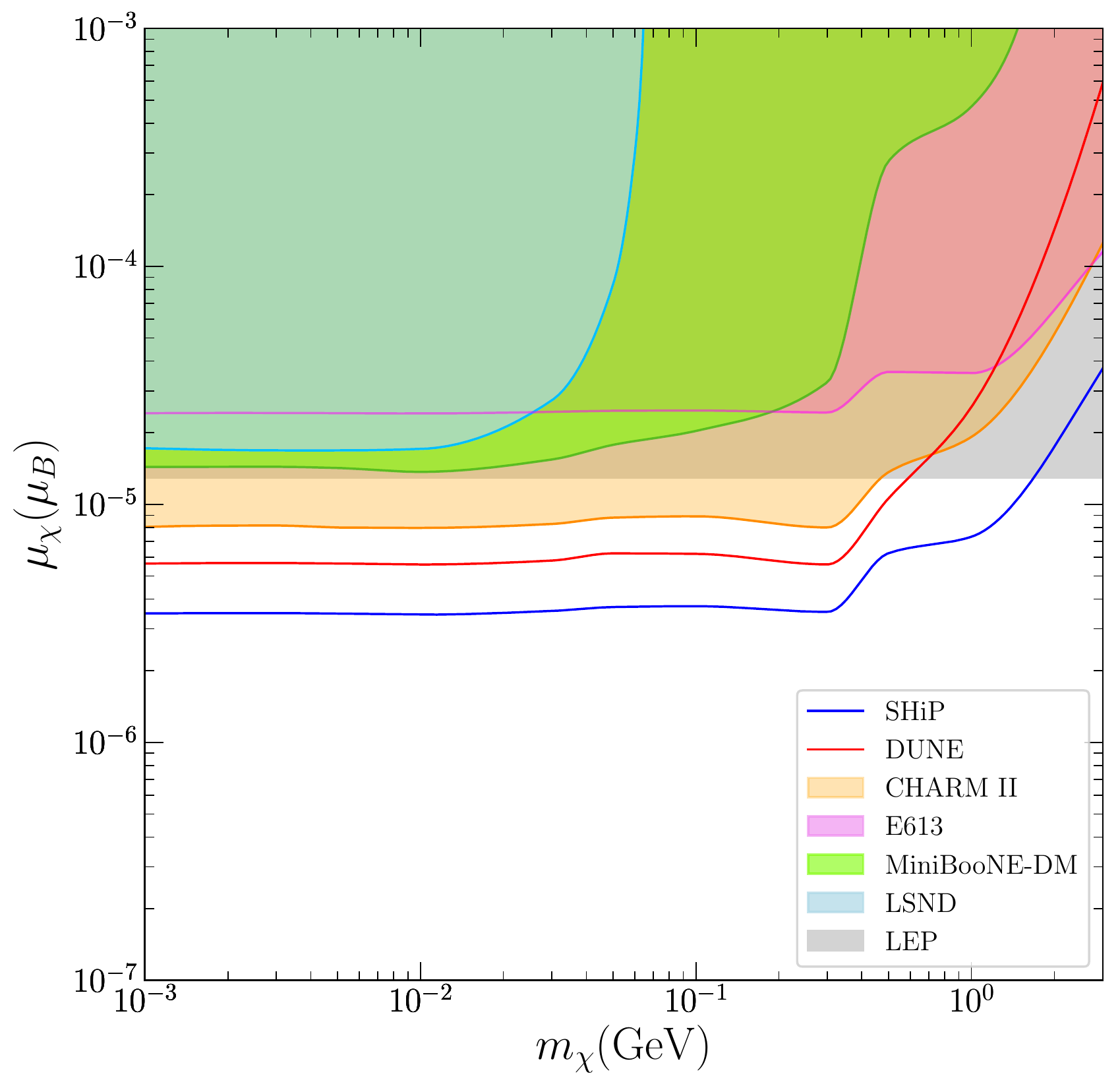}
\includegraphics[width=\columnwidth]{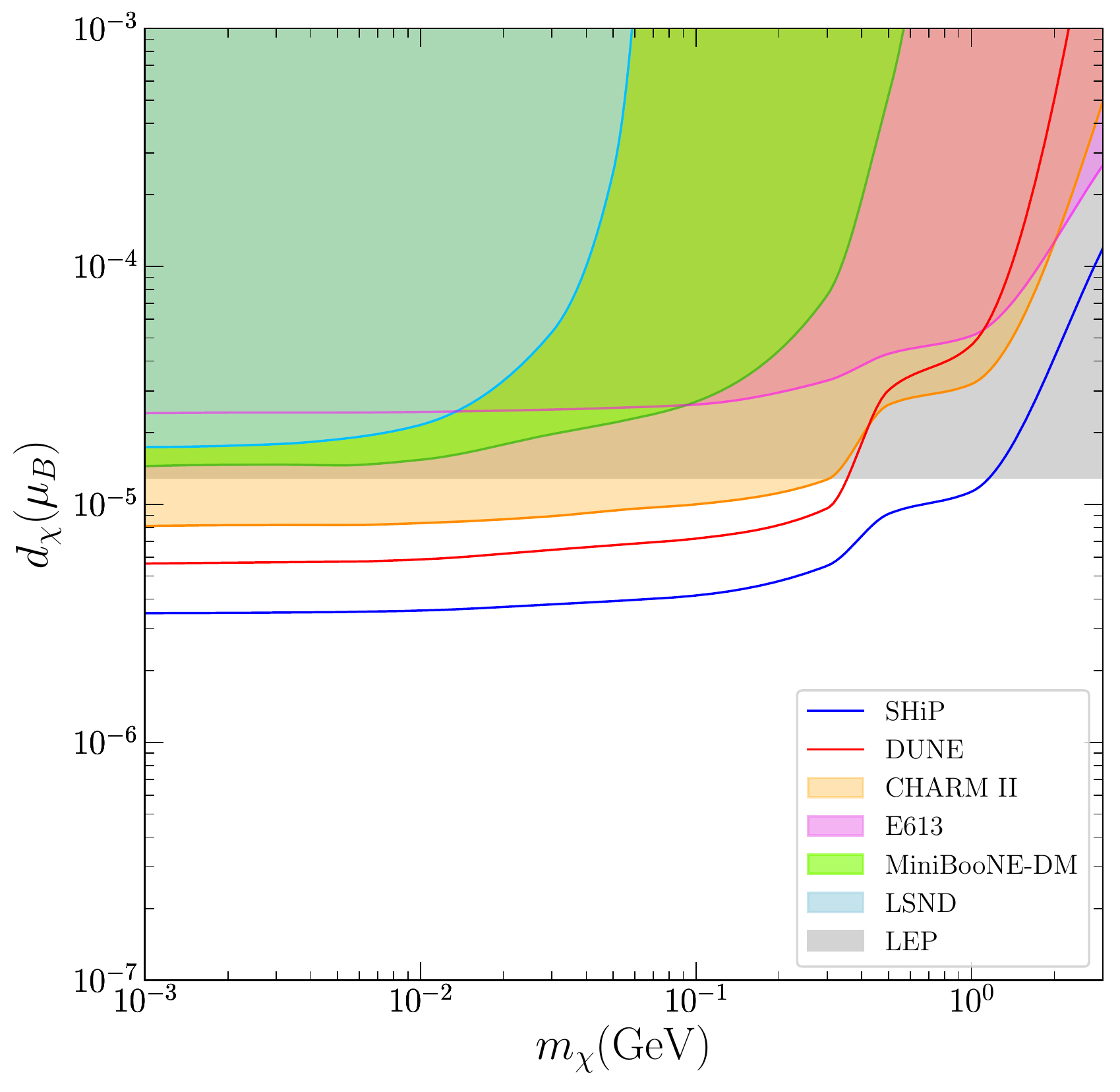}
\caption{Summary of 90\% C.L. excluded regions on the EM form factors
  for the dim-5 operators MDM (left) and EDM (right). Shaded regions
  are excluded; projected sensitivities from future experiments are
  shown as solid lines. The  LEP bound is taken from~\cite{Chu:2018qrm}.}
  \label{fig:landscapeDim5}
\end{figure*}

\begin{figure}[!ht]
\centering
\includegraphics[width=\columnwidth]{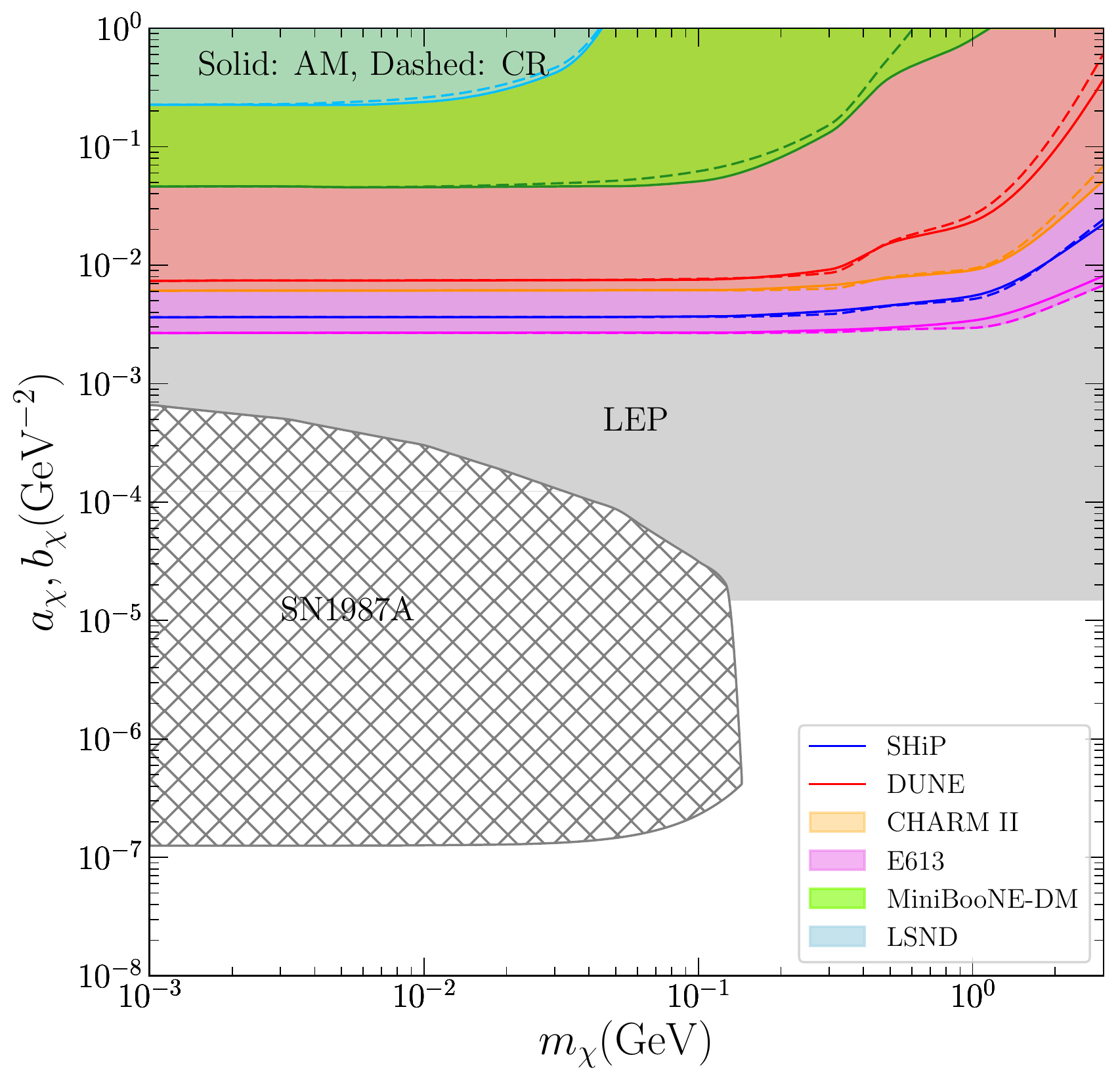}
\caption{Same as Fig.~\ref{fig:landscapeDim5} but for dim-6
  operators (AM/CR). The SN1987A  bound is  taken from~\cite{Chu:2019rok}.
}
  \label{fig:landscapeDim6}
\end{figure}

\subsection{Other experiments}

There also exist many other proton-beam experiments which adopt
similar setups to those we have studied above, such as COHERENT with a
1\,GeV proton beam~\cite{Akimov:2017ade}, JSNS$^2$ with a 3\,GeV
proton beam~\cite{Ajimura:2017fld}, NO$\nu$A with a 120\,GeV proton
beam~\cite{Adamson:2017qqn}, as well as WA66 with a 400\,GeV proton beam~\cite{Talebzadeh:1987rq}.
Nevertheless, %
these experiments are in general not expected to provide noticeably
stronger (projected) bounds than those obtained above~(see
\textit{e.g.}~\cite{CooperSarkar:1991xz, Ge:2017mcq,
  deNiverville:2018dbu,Jordan:2018gcd,Buonocore:2019esg,
  Akimov:2019xdj}), and are thus not further studied in this work.

A different new bound on light dark states comes from the NA62
experiment, which has recently improved the constraint on
 $\pi^0\to \gamma+\text{inv.}$ by three orders of
magnitude~\cite{CortinaGil:2019nuo}. This puts upper bounds on the
MDM/EDM interactions of our interest as
\begin{equation}
\label{Eq:NA62}
	\mu_\chi (d_\chi) \lesssim 2.4 \times 10^{-4}\, \mu_B,
\end{equation}
for $m_\chi \ll m_\pi/2$. They are weaker than the bounds obtained
above, and become even weaker for higher-dimensional operators,
\textit{i.e.}~the AM/CR interactions.

High-energy colliders become more important for $\chi$ particles
heavier than pions. For instance, at LHC, the upgrade of the MoEDAL
experiment will be equipped with three deep liquid scintillator
layers~\cite{Pinfold:2019nqj}. In addition, there will be the milliQan
detector~\cite{Haas:2014dda,Ball:2016zrp} which will be composed of
three stacks of plastic scintillators. Both experiments are designed
to be sensitive to milli-charged dark particles, of which the
scattering cross section with electron/nucleus is dramatically
enhanced at low momentum-transfer. As suggested in \cite{Sher:2017wya,
  Frank:2019pgk}, such experiments will constrain the EDM form factor
of dark states, where there also exists an enhancement---although
milder---in low momentum-transfer $\chi$-$e$\,($\chi$-$N$) region of
elastic scattering. Moreover, proposed future colliders, such as
HL-LHC and ILC, will be able to further improve the experimental
sensitivity on all the EM form factors studied here; see
e.g. \cite{Kadota:2014mea, Primulando:2015lfa,
  Alves:2017uls}.  %

\section{Results}
\label{sec:result}

In this section, we first compare the production efficiency of  
various production channels, and then summarize our bounds on the EM
form factors of dark
states.

\subsection{Comparison of Production Channels}
\label{subsec:compare}

In contrast to dark state-photon interactions through milli-charge,
higher-dimensional operators are considered in this work. Therefore,
dimensional analysis demands an extra energy scale $E$ to compensate
for the presence of the dimensionful coupling ($E$ for dimension-5
operators and $E^2$ for dimension-6 operators) in cross sections and
branching ratios, in comparison to the dimension-4 case. This typically
suppresses the yield of dark states.

For DY, the relevant energy scale is of the order of the $pp$
collision energy, $\sqrt{s}$.  We can then infer that for dimension-5
(dimension-6) operators the resulting cross section will contain a
dimensionless factor $\mu_\chi^2 s$ and $d_\chi^2 s$ ($ a_\chi^2 s^2$
and $b_\chi^2 s^2$).%
\footnote{The use of effective operators is justified when these
  products do not exceed unity. This is not guaranteed in the top
  portions of Figs.~\ref{fig:landscapeDim5} and
  \ref{fig:landscapeDim6}, but we expect that the region remains
  excluded by associated LEP bounds that resolve the UV particle
  content. We leave a derivation of such UV-dependent high-energy collider
  constraints for dedicated future work.}
Thus, the cross sections involving dimension-5 and 6 operators are 
further suppressed relative to dimension-4 interactions for  
$d_\chi^{-1},\mu_\chi^{-1} \gg \sqrt{s}$ and
$a_\chi^{-1}, b_\chi^{-1} \gg s$, which incidentally are also required
for the treatment of Eqs.~\eqref{Eq:Lagrangian_dim5} and
\eqref{Eq:Lagrangian_dim6} as effective operators.  
As a result, the DY process gains in relevance relative to the meson decay in the
production of $\chi$ particles, especially for dimension-6 operators
as for the latter, the relevant energy scale is roughly the meson
mass.

In addition, because of the mass-scaling, the relative importance of
decaying meson contributions is also modified. The branching ratios
into $\chi$-pairs from light mesons become suppressed.
Therefore, we can see that although heavier mesons are produced at
lower rates, as shown in Tab.~\ref{tab:meson_production}, the final
yields of dark states from their decay are comparable to (dominate
over) those from light mesons for dimension-5 (dimension-6) operators.

The $\chi$ production rate of each channel, after applying the
geometric cut, is demonstrated in Fig.~\ref{fig:acceptance_SHiP}. One
can see that due to the reasons above, the overall pattern in our
$\chi$ production rate becomes very different from those of
milli-charged particles (see \textit{e.g.}~\cite{Harnik:2019zee}) and
dark photons (see \textit{e.g.}~\cite{deNiverville:2016rqh}), where
light meson decay is the most important production channel unless it is
kinematically suppressed.\footnote{We have checked that our code
  reproduces Fig.~2 of \cite{Harnik:2019zee} when switching the
  effective operators to the milli-charged interaction.}

\subsection{Constraints}
The 90\% C.L.~constraints on the EM form factors derived above are
shown by the colored regions in Fig.~\ref{fig:landscapeDim5} (MDM and
EDM) and Fig.~\ref{fig:landscapeDim6} (AM and CR), together with our
previous constraints (gray regions)~\cite{Chu:2018qrm,Chu:2019rok}.
As explained above, the strengths of higher-dimensional interactions
are energy-sensitive, and constraints derived from current proton-beam
experiments, with $\sqrt{s}$ below several to tens of GeV, turn out
not to be competitive with the constraint from LEP~\cite{Chu:2018qrm}.
For dimension-5 operators, future experiments such as DUNE (10-year)
and SHiP will improve the sensitivity by a factor of 2--3,
and become stronger than LEP for $m_\chi < 1\,{\rm GeV}$ due to their
high intensity. It is worth pointing out that the astrophysical bound
from SN1987A constrains the MeV-region below
$10^{-8}\mu_B$~\cite{Chu:2019rok}, well below the current and
projected experimental sensitivity.

For dimension-6 operators, the production and detection rates of light
dark states are even more sensitive to the center-of-mass energy,
suggesting it is unlikely for low-energy experiments to play any role
in the foreseeable future. In E613 the initial energy of $\chi$ needs
to be above 20\,GeV to trigger an observable signal, but such large
$E_\chi$ also enhances the $\chi$-proton scattering, making it
difficult for $\chi$ particles to travel through the shield unless
$a_\chi,b_\chi \ll 10^{-2}$\,GeV$^{-2}$.\footnote{In this region, the
  validity of the use of effective operators is also in question.}
Thus, future high energy colliders have better potential to probe
dimension-6 dark state interactions.

\begin{figure}[tb]
\centering
\includegraphics[width=\columnwidth]{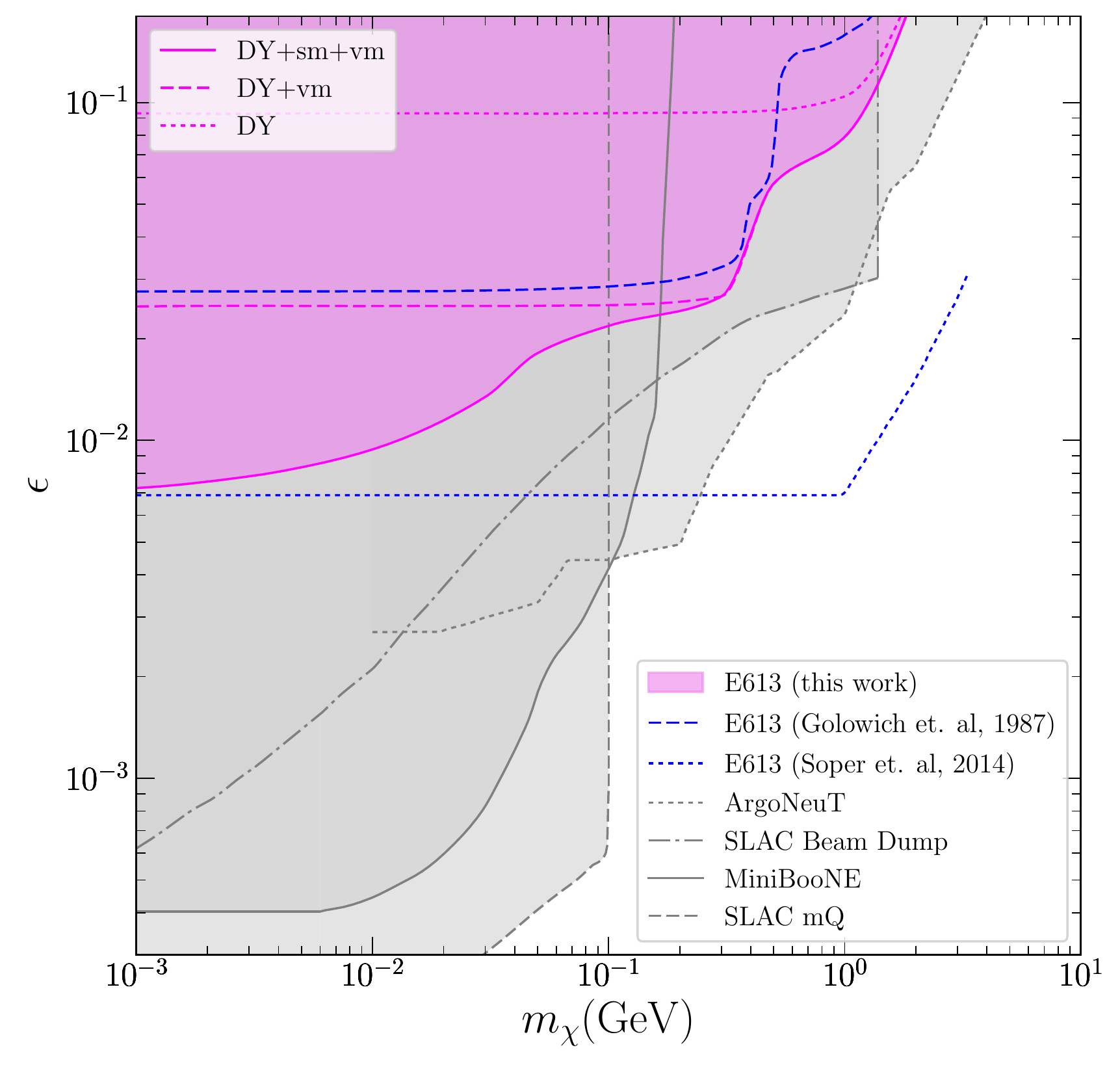}
\caption{Revised upper bounds from E613 on milli-charged dark states
  from DY production only (pink dotted line), DY + vector meson decay
  (pink dashed) and DY + vector/scalar meson decay (pink solid). This
  corrects a previously derived limit from DY production (blue
  dotted)~\cite{Soper:2014ska} and improves previous work utilizing
  DY + vector meson decay only (blue
  dashed)~\cite{Golowich:1986tj}. Other bounds shown are from the ArgoNeuT~\cite{Acciarri:2019jly} (dotted grey),
  MiniBooNE~\cite{Magill:2018tbb} (solid gray), SLAC beam dump
  (dash-dotted gray) and mQ~\cite{Prinz:1998ua} (dashed gray)
  experiments. See~\cite{Liang:2019zkb,
    Magill:2018tbb} for the sensitivity reach of Babar, BESIII, and
  future experiments. }
  \label{fig:milli-charged}
\end{figure}

At last, due to the consideration given in Sec.~\ref{subsec:E613}, we
also revise the E613 bound on milli-charged particles from
\cite{Soper:2014ska}, although it has been surpassed by bounds derived
from later experiments\,\cite{Prinz:1998ua, Magill:2018tbb,
  Liu:2018jdi, Gninenko:2018ter, Chu:2018qrm}.  Our derivation also
improves w.r.t.~a much earlier work~\cite{Golowich:1986tj}, by adding
the production through decays of scalar mesons and by imposing the
BMPT distribution for mesons. As shown in
Fig.~\ref{fig:milli-charged}, if only DY processes are taken into
account, our bound is weaker than that from \cite{Soper:2014ska} by  
about a factor of~7. By adding contributions from vector meson decay,
the bound becomes stronger, approximately in agreement with
\cite{Golowich:1986tj} (dashed lines). Our final exclusion limit,
taking into account all these contributions, is shown as the pink
shaded region in the figure.

\section{Conclusions}
\label{sec:conclusions}

In this work we study the production and detection of neutral
fermionic dark states $\chi$ that carry EM form factors in proton-beam
experiments.  We consider the production of $\chi\bar\chi$-pairs in
the collision of high-intensity protons on nuclear targets through
prompt Drell-Yan scattering and in secondary meson decays. The
detectable signals considered are single electron recoil events at LSND
and MiniBooNE-DM, CHARM II, as well as at the proposed DUNE and SHiP
experiments, and hadronic showers caused by nuclear deep inelastic
scattering at E613.

Owing to the higher dimensionality of the considered operators
(dimension 5 and 6), the relative importance of production channels is
biased towards processes with larger intrinsic energy. As a
consequence, Drell-Yan production and production in \textit{heavy}
meson decays gain prominence when compared to the milli-charged and
dark photon cases, for which pion decays dominate the dark state yield.

We compute in detail the energy and angular distribution of the
produced dark state flux and set the strongest constraints on the
existence of $\chi$-particles with MDM and EDM interactions in the
MeV-GeV mass bracket, excluding dimensionful coefficients
$\mu_{\chi}, d_{\chi} \gtrsim 8\times 10^{-6} \mu_B$, corresponding to
an effective scale $\Lambda_5 < 0.4\,\TeV$. For the dimension-6 AM and
CR interactions, we find
$a_{\chi}, b_{\chi} \gtrsim 3\times 10^{-3} \,\GeV^{-2}$ are excluded,
pointing towards a comparably lower effective scale of
$\Lambda_6 < 20\,\GeV$. In the latter case, the constraint is
superseded by LEP. Finally, as a by-product of our study, we also
revise previously obtained proton-beam dump bounds on milli-charged
particles.

With a strong connection to the neutrino program, proton beam
experiments constitute an active and diverse field, with a number of
new experiments proposed such as SHiP and DUNE. However, because the
interactions considered here are higher-dimensional, we find that the
prospects of significantly improving the direct sensitivity on EM form
factor couplings rather hinges on the future of high-energy collider
experiments and their ability to produce collisions with an ever
increased center-of-mass energy.

\vspace{.3cm}

\paragraph*{Acknowledgments}
We thank Giacomo Marocco, Samuel McDermott, and 
Subir Sarkar for useful discussions.  
The authors are supported by the New Frontiers program
  of the Austrian Academy of Sciences. JLK is supported by the Austrian
  Science Fund FWF under the Doctoral Program W1252-N27 Particles and
  Interactions. We acknowledge the use of computer packages for
  algebraic calculations~\cite{Mertig:1990an,Shtabovenko:2016sxi}.
\vspace{1.cm}

\appendix

\section{Decay rates of scalar mesons}
\label{App:mesondecay}

\begin{figure*}[tb]
\centering
\includegraphics[width=\columnwidth]{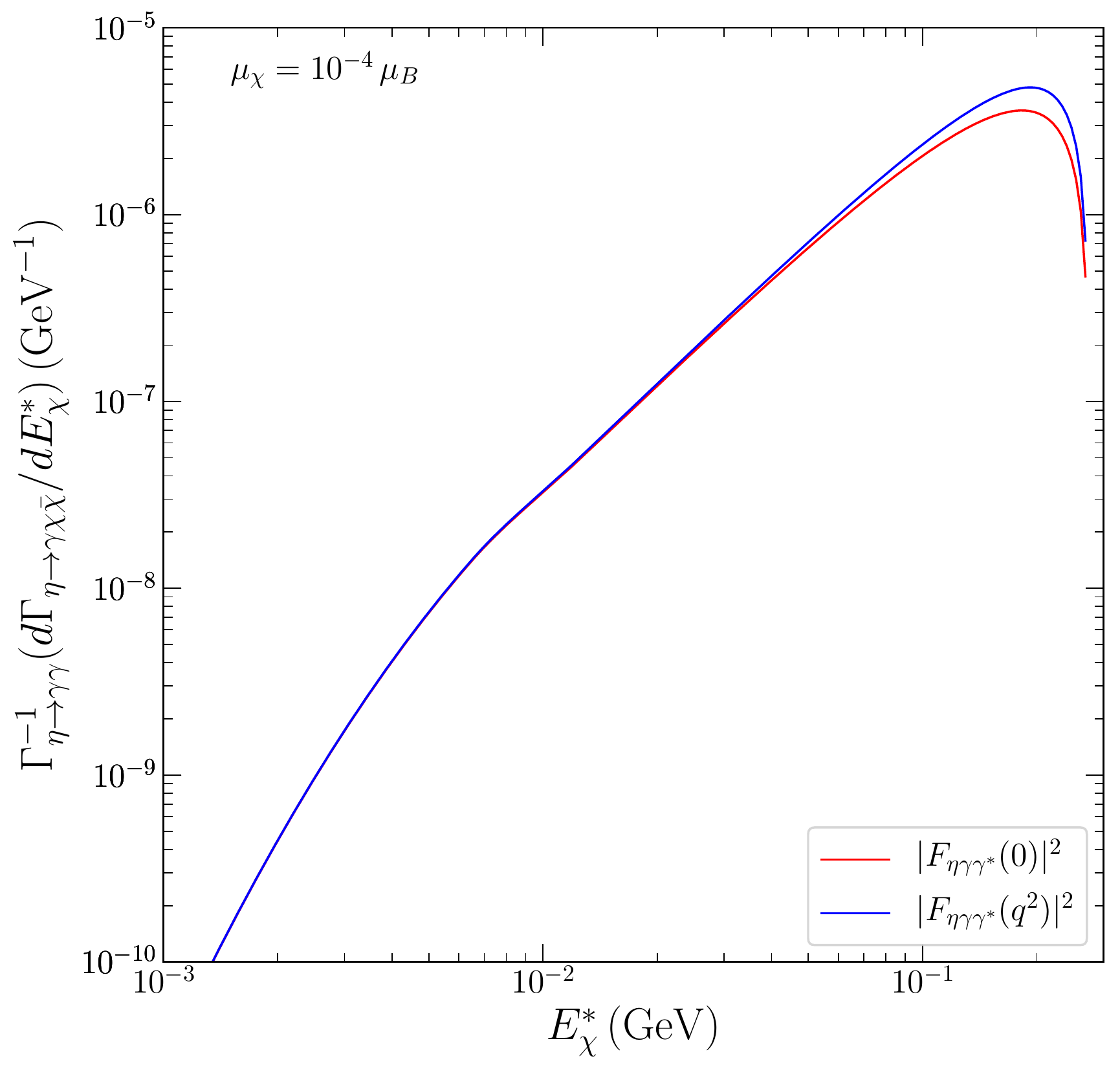}
\includegraphics[width=\columnwidth]{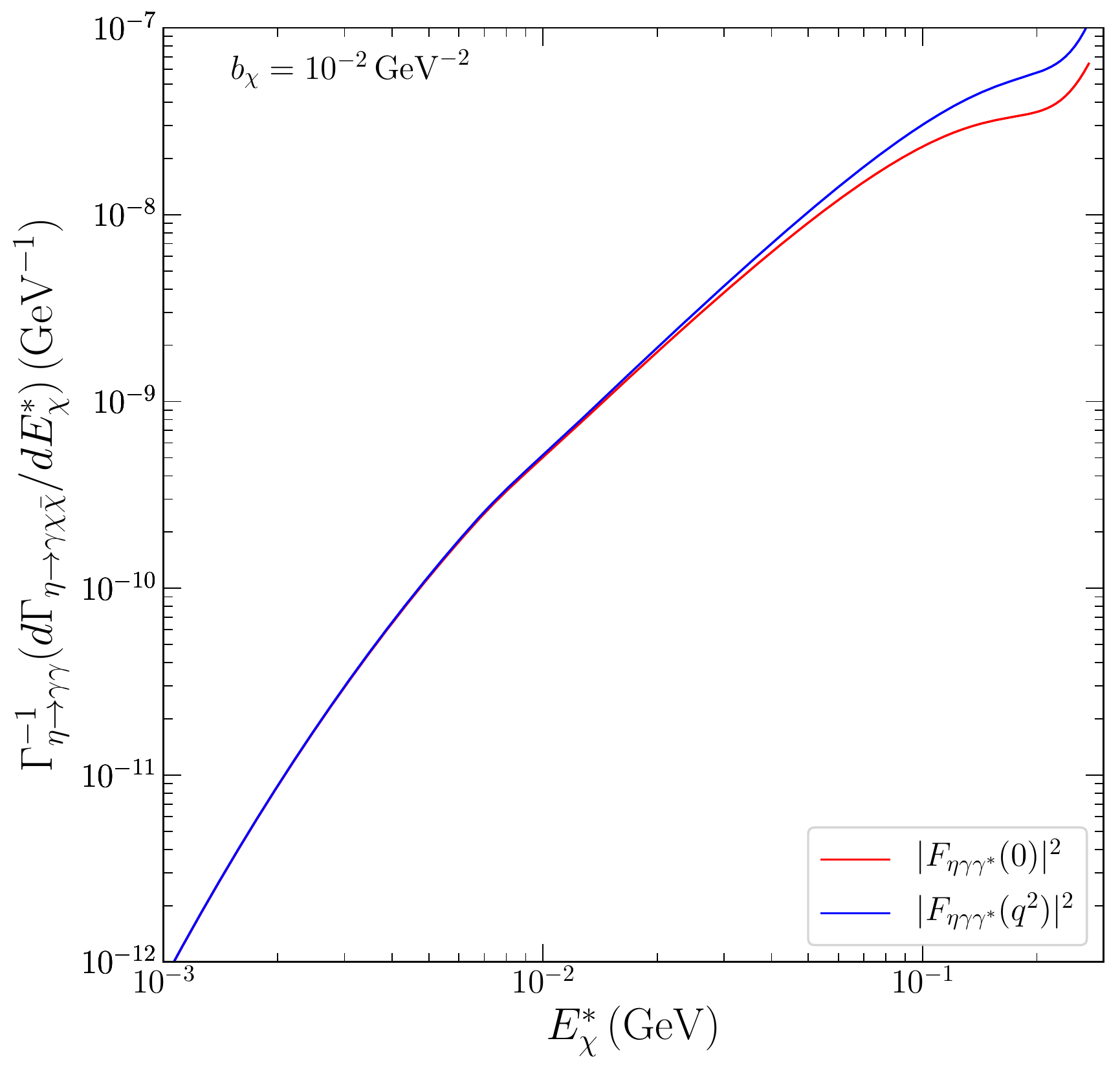}
\caption{Comparison of differential decay rate of $\eta$ meson into dark states  without\,(red) or with\,(blue)  meson transition form factor for  MDM (left) and CR (right) with $m_\chi =1$\,MeV. The decay rates are normalized with $\Gamma^{-1}_{\eta \rightarrow \gamma\gamma}$.
} 
  \label{fig:diffBr}
\end{figure*}

The decay rate of scalar mesons into a photon plus a $\chi$-pair,
$ \Gamma_\chi\equiv \Gamma_{{\rm sm} \rightarrow \gamma\chi\bar{\chi}}
$, is given by
\begin{equation}
\label{Eq:sm_total_drate}
  	\Gamma_\chi = \int^{m^2_\text{sm}}_{4m_\chi^2}d\sdark \, \Gamma_{{\rm sm}  \rightarrow \gamma \gamma^*}(\sdark) \dfrac{f_\chi(\sdark)}{16\pi^2 \sdark^2}\sqrt{1-{4m_\chi^2\over \sdark}}\, ,
      \end{equation}
      where $\Gamma_{{\rm sm} \rightarrow \gamma \gamma^*}$ is the
      decay rate with an off-shell photon,
\begin{equation}
	\Gamma_{{\rm sm}  \rightarrow \gamma \gamma^*} (\sdark)= \dfrac{\alpha^2 (m_{\rm sm}^2 - \sdark)^3}{32 \pi^3 m_{\rm sm}^3 F_{\rm sm}^2}\,,
\end{equation}
with $F_{\rm sm}$ being the decay constant of the meson. Since we are
allowed to neglect the momentum-dependence of the EM transition form
factors of the scalar mesons (shown below), $F_{\rm sm}$ will drop out
in the branching ratio, Eq.~\eqref{Eq:Brsmchi}.

In Eq.~\eqref{Eq:sm_total_drate},  the function
$f_\chi (\sdark)$ is defined by the phase space integral of the squared amplitude  of $\gamma^* \to \chi\bar\chi $
      \begin{align*}
         \int &  \frac{d^3p_{\chi}}{(2\pi)^3 2E_{\chi}} \frac{d^3p_{\bar\chi}}{(2\pi)^3 2E_{\bar\chi}} (2\pi)^4
        \delta^4 (q -p_\chi -p_{\bar\chi}) L^{\nu\sigma}    \\
       & = \dfrac{1}{16\pi} \sqrt{1-\dfrac{4 m_\chi^2}{\sdark}} f(\sdark) \left(-g^{\nu\sigma} + \dfrac{q^{\nu}q^{\sigma}}{\sdark} \right)\,
      \end{align*}
      with  $q^2 = \sdark$. 
      The explicit expressions of $f_\chi (\sdark)$ were already
      obtained in~\cite{Chu:2018qrm}, and are also listed below:
      \begin{eqnarray}
   {\rm mQ}: f_\chi (\sdark) &=& \dfrac{16\pi \alpha }{3}\epsilon^2 \sdark \left(1+\dfrac{2m_\chi^2}{\sdark}\right), \\
	{\rm MDM}: f_\chi (\sdark) &=& \dfrac{2}{3}\mu_\chi^2 \sdark^2 \left(1+\dfrac{8m_\chi^2}{\sdark}\right), \\
   {\rm EDM}: f_\chi (\sdark) &=&\dfrac{2}{3} d_\chi^2 \sdark^2 \left(1-\dfrac{4m_\chi^2}{\sdark}\right), \\
{\rm AM}: f_\chi (\sdark) &=&\dfrac{4}{3} a_\chi^2 \sdark^3 \left(1-\dfrac{4m_\chi^2}{\sdark}\right), \\
{\rm CR}: f_\chi (\sdark) &=&\dfrac{4}{3} b_\chi^2 \sdark^3 \left(1+\dfrac{2m_\chi^2}{\sdark}\right)\,.
\end{eqnarray}
The expression for $f_e (m_{\rm vm}^2)$, used for vector meson decay  in
Sec.~\ref{Sec:meson_decay}, is then given by
\begin{equation}
   f_e (m_{\rm vm}^2) =  \dfrac{16\pi \alpha}{3} m_{\rm vm}^2  \left(1+ \dfrac{2m_e^2}{m_{\rm vm}^2} \right).
\end{equation}

To infer the energy spectrum of $\chi$ from scalar meson decay, we also need
to know the differential decay rate $d\Gamma_\chi /d E^*_\chi$ in the
rest frame of the meson ($\vec p_1 =0$). To this end, 
 we first compute the amplitude of the process
${\rm sm} (p_1) \rightarrow \gamma (p_2) + \chi (p_3) +\bar{\chi}
(p_4)$.  %
and define the two Lorentz-invariant variables
$s_{23} = (p_2 + p_3)^2$ and $s_{42} = (p_4 + p_2)^2$, so that
$\sdark$ becomes $\sdark = m_{\rm sm}^2 + 2m_\chi^2 -s_{23} -s_{42}$.
The corresponding squared amplitudes, summed over the spin of final states for each
EM form factor are obtained as follows:
\begin{widetext}
\begin{eqnarray}
&\,&\,\,\,\,\,\,
\begin{split}
 {\rm mQ}: \sum\limits_{\rm spins} |\mathcal{M}|^2 =&\,\, \dfrac{\alpha^2 \epsilon^2}{\pi^2 F_{\rm sm}^2 (2m_\chi^2+m_{\rm sm}^2 -s_{23} -s_{42} )^2} \times \lbrace 12m_\chi^6 +2m_\chi^4 [m_{\rm sm}^2 - 7 (s_{23} + s_{42})] \\
& \,+ m_\chi^2 [-2m_{\rm sm}^2 (s_{23}+ s_{42} )+6 s_{23}^2 +8 s_{23} s_{42}  +6 s_{42}^2 ] + (s_{23}^2 +s_{42}^2) (m_{\rm sm}^2 - s_{23} - s_{42} ) \rbrace, 
\end{split}
\\
&\,&
\begin{split}
 {\rm MDM}: \sum\limits_{\rm spins} |\mathcal{M}|^2 =&\,\, \dfrac{2\alpha^2 \mu_\chi^2}{\pi^2 F_{\rm sm}^2 (2m_\chi^2+m_{\rm sm}^2 -s_{23} -s_{42} )} \times \lbrace 6m_\chi^6 +m_\chi^4 [m_{\rm sm}^2 - 7 (s_{23} + s_{42})] \\
& \,+ m_\chi^2 [2 (s_{23}^2 + 3 s_{23} s_{42} + s_{42}^2)-m_{\rm sm}^2 (s_{23}+ s_{42} )] -s_{23} s_{42} (-m_{\rm sm}^2 + s_{23} + s_{42} ) \rbrace, 
\end{split}
\\
&\,&
\begin{split}
{\rm EDM}: \sum\limits_{\rm spins} |\mathcal{M}|^2 = &\,\,  \dfrac{-2\alpha^2 d_\chi^2}{\pi^2 F_{\rm sm}^2 (2m_\chi^2+m_{\rm sm}^2 -s_{23} -s_{42} )} \times \lbrace 2 m_\chi^6 -m_\chi^4 (m_{\rm sm}^2 + s_{23} + s_{42} ) \\
& \,\, + m_\chi^2 [m_{\rm sm}^2 (s_{23} + s_{42}) -2 s_{23} s_{42} ] + s_{23} s_{42} (-m_{\rm sm}^2 +s_{23} + s_{42}) \rbrace,
\end{split}
\\
&\,&\,\,\,\,
\begin{split}
{\rm AM}: \sum\limits_{\rm spins} |\mathcal{M}|^2 = &\,\, \dfrac{-\alpha^2 a_\chi^2}{\pi^2 F_{\rm sm}^2} \times \lbrace 4 m_\chi^6 - 2m_\chi^4 (m_{\rm sm}^2 + s_{23} +s_{42} ) + 2m_\chi^2 [m_{\rm sm}^2 ( s_{23} +s_{42} )- s_{23}^2 - s_{42}^2 ] \\
&\,\, - (s_{23}^2 + s_{42}^2 )(m_{\rm sm}^2 - s_{23} - s_{42} ) \rbrace,
\end{split}
\\
&\,&\,\,\,\,\,
\begin{split}
{\rm CR}: \sum\limits_{\rm spins} |\mathcal{M}|^2 = &\,\, \dfrac{\alpha^2 b_\chi^2 }{\pi^2 F_{\rm sm}^2} \times \lbrace 12 m_\chi^ 6 + 2m_\chi^4 [m_{\rm sm}^2  -7 (s_{23} + s_{42}) ] -m_\chi^2 [2 m_{\rm sm}^2 ( s_{23} + s_{42} ) - 6 s_{23}^2  - 8 s_{23} s_{42} - 6 s_{42}^2 ] \\
&\,\, +(s_{23}^2 + s_{42}^2 ) ( m_{\rm sm}^2 - s_{23} - s_{42}) \rbrace .
\end{split}
\end{eqnarray}
\end{widetext}
Then the Dalitz plot allows us to express the differential decay rate
in the rest frame of the meson as
\begin{equation}
\dfrac{d \Gamma_\chi}{d E_\chi^* } = \dfrac{1}{128\pi^3 m_{\rm sm}^2}  \int  ds_{23} \,   \sum\limits_{\rm spins} |\mathcal{M}|^2\, , 
\end{equation}
where the integration boundaries of $s_{23}$ are given by
\begin{equation}
\begin{split}
    [s_{23}]^{\pm} &= \,  \dfrac{1}{2 s_{42}} \Big[ (m_\chi^2 - s_{42}) (m_\chi^2  -m_{\rm sm}^2 +s_{42} ) + 2m_\chi^2  s_{42} \\
& \left. \mp (m_\chi^2 - s_{42} ) \sqrt{m_\chi^4 - 2 m_\chi^2 (m_{\rm sm}^2 + s_{42} )+ (m_{\rm sm}^2 -s_{42})^2} \right]. 
\end{split}
\end{equation}
Here, $s_{42}$ is  
\begin{equation}
s_{42} =  m_\chi^2 + m_{\rm sm}^2 - 2E_\chi^* m_{\rm sm}.
\end{equation}
The allowed kinematic range of $E_\chi^*$ is $m_\chi \leq E_\chi^* \leq m_{\rm sm} /2$. 
At last we  arrive at the differential branching ratio via
\begin{equation}
	 \dfrac{d {\rm Br}_\chi}{dE_\chi^*} = {\rm Br}_{{\rm sm}\to \gamma \gamma }\times \dfrac{1}{\Gamma_{ {\rm sm}\to \gamma \gamma } }\dfrac{d\Gamma_\chi}{dE_\chi^*}\,,
\end{equation}
from which one can directly see that the meson decay constant,
$F_{\rm sm}$, cancels out in the ratio of $\Gamma_\chi $ and
$\Gamma_{ {\rm sm}\to \gamma \gamma }$.

At the end of this section, we comment on the assumption of using
constant transition form factors for scalar meson decay. Vector meson
dominance suggests that the assumption holds well for
$m_{\rm sm}^2\ll m_\rho^2$, which is the case for $\pi^0$ decays. For
the heavier scalar mesons considered in this work, $\eta/\eta'$, we
have numerically evaluated the differential decay rate using the EM
transition form factor. For the $\eta$ meson~\cite{Husek:2019axa}, the
results are given in Fig.~\ref{fig:diffBr}, which shows that the shape
of $d{\rm Br}_\chi/dE_\chi^*$ is only affected mildly by the
(kinematically limited) virtuality of the intermediate photon. In the
total decay rate for $m_\chi = 1\,(100)$\,MeV, ${\rm Br}_\chi$ would increase by a factor of
1.3\,(1.7)  in the case of MDM/EDM, and by
a factor of 1.8\,(1.9) in the case of
AM/CR. Hence, neglecting the momentum-dependence of the transition
form factors leads to slightly weaker bounds, and is hence
conservative.

\section{$\chi$ distribution from meson decay}

In this section, the derivation of Eq.~\eqref{Eq:spectrum_dist_chi}
is provided.
In general, the number of $\chi$ particles produced from a certain meson
distribution is given by

\begin{eqnarray}
\label{Eq:Nchi_general}
      N_\chi &=&  \int dE_m \, d\cos\theta_m \, \dfrac{d^2 N_m}{dE_m \, d\cos\theta_m} \times 2 \rm Br_\chi \notag \\
         &=& \int dE_m \, d\cos\theta_m \,  \dfrac{d^2 N_m}{dE_m \, d\cos\theta_m} \notag\\
       &  & \times \int dE_\chi^* \,d\cos\theta^* d\phi^*   \dfrac{d^3 \hat{N}_\chi}{dE_\chi^* \, d\cos\theta^* d\phi^*},
\end{eqnarray}
where $E_m$ and $\theta_m$ are the energy of meson and angle between
the meson momentum and the beam axis in the lab frame, respectively.
The energy of $\chi$ in the rest frame of the meson is denoted by
$E_\chi^*$. Finally, $\theta^*$, $\phi^*$ are the polar and azimuthal
angles of the $\chi$ momentum in the rest frame of the meson
w.r.t.~the lab-frame meson momentum.

In practice, we are interested in the distribution of $\chi$ particles
in terms of $E_\chi$ and $\theta_\chi$, which are the energy of $\chi$
and the polar angle of the $\chi$ momentum w.r.t.~the beam axis in the lab
frame.  A Lorentz transformation allows to express the last quantities
as functions of $E_m$, $\cos\theta_m$, $E_\chi^*$, $\cos\theta^*$ and
$\phi^*$. Then, by inserting the two delta-functions
\begin{align*}
  &\int dE_\chi \, \delta \left[E_\chi - E_\chi (E_m, \cos\theta_m, E_\chi^*, \cos\theta^*,\phi^*)\right]
  = 1\,, \\
 & \int d\cos\theta_\chi \, \delta \left[\cos\theta_\chi - \cos\theta_\chi
  (E_m, \cos\theta_m, E_\chi^*, \cos\theta^*,\phi^*)\right]  = 1, 
\end{align*}
into Eq.~\eqref{Eq:Nchi_general} and using the fact that the decay is
isotropic in meson rest frame, we arrive at
\begin{eqnarray}
N_\chi &=& \int \dfrac{d\cos\theta^* d\phi^*}{4\pi} dE_\chi^*\, dE_m \, d\cos\theta_m \, dE_\chi \, d\cos\theta_\chi \,\dfrac{d \hat{N}_\chi}{dE_\chi^*}\notag \\
&\times &  \dfrac{d^2 N_m}{dE_m \, d\cos\theta_m}  \delta \left[E_\chi - E_\chi (E_m, \cos\theta_m, E_\chi^*, \cos\theta^*,\phi^*)\right] \notag \\  
&\times & \delta \left[\cos\theta_\chi - \cos\theta_\chi (E_m, \cos\theta_m, E_\chi^*, \cos\theta^*,\phi^*)\right] \, .
\end{eqnarray}

Next we use the two delta functions to perform the integrals over $E_m$ and $\theta_m$ leading to,
\begin{equation}
\begin{split}
    N_\chi &= \int \dfrac{d\cos\theta^* d\phi^*}{4\pi} dE_\chi^* \, dE_\chi \,d\cos\theta_\chi  \\
 &\times \dfrac{d \hat{N}_\chi}{dE_\chi^*}\, \dfrac{d^2 N_m}{dE_m \, d\cos\theta_m}  \left|\dfrac{\partial (E_m, \cos\theta_m)}{\partial (E_\chi , \cos\theta_\chi)} \right|. 
\end{split}
\end{equation}
where the last factor $|\partial(\dots)/\partial(\dots)|$ is the Jacobian of the variable transformation. 

In the end, the distribution function of $\chi$ particles from meson decay in the lab frame in terms of $E_\chi$ and $\theta_\chi$ reads
\begin{equation}
\begin{split}
	   \dfrac{d^2 N_\chi}{dE_\chi d\cos\theta_\chi} &= \int \dfrac{d\cos\theta^* d\phi^*}{4\pi} dE_\chi^* \dfrac{d \hat{N}_\chi}{dE_\chi^*} \\
&\times  \dfrac{d^2 N_m}{dE_m \, d\cos\theta_m} \left| \dfrac{\partial (E_m, \cos\theta_m)}{\partial (E_\chi , \cos\theta_\chi)} \right|\,,
\end{split}
\end{equation}
Summing up the contribution from each meson, we retrieve
Eq.~\eqref{Eq:spectrum_dist_chi} of the main text.

\quad

\section{$L_{\mu\nu} W^{\mu\nu}$ in DIS cross section}
\label{App:xsecdis}

The DIS differential cross section, given in Eq.~\eqref{Eq:dxsecDIS},
contains the contraction of dark and hadronic matrix element
$L_{\mu\nu} W^{\mu\nu}$.
In the following we list $L_{\mu\nu} W^{\mu\nu}$ for each EM form factor interaction: 
\begin{widetext}
\begin{eqnarray}
 {\rm mQ}: L_{\mu\nu} W^{\mu\nu} &=&  4 \pi \alpha \epsilon^2 \left\lbrace 2 F_1 (Q^2 - 2m_\chi^2 ) - \dfrac{m_N  F_2}{\nu} [4 E_\chi (\nu - E_\chi) +Q^2] \right\rbrace, 
\\
 {\rm MDM}: L_{\mu\nu} W^{\mu\nu} &=& \mu_\chi^2 \left\lbrace Q^2 F_1 (8m_\chi^2 -Q^2)  + \dfrac{m_N F_2}{\nu} [4E_\chi^2 Q^2 - 4 E_\chi \nu Q^2  -4 m_\chi^2  (\nu^2 +Q^2 ) +\nu^2 Q^2]  \right\rbrace, 
\\
{\rm EDM}: L_{\mu\nu} W^{\mu\nu} &=& d_\chi^2 \left\lbrace -Q^2 F_1 (4 m_\chi^2 +Q^2) + \dfrac{m_N F_2 }{\nu} Q^2 (\nu -2 E_\chi)^2 \right\rbrace ,
\\
{\rm AM}: L_{\mu\nu} W^{\mu\nu} &=& a_\chi^2 \left\lbrace 2Q^4 F_1 (4 m_\chi^2 +Q^2 ) - \dfrac{m_N F_2 }{\nu} Q^2 [-4 E_\chi^2 Q^2 + 4 E_\chi \nu Q^2 + 4m_\chi^2 (\nu^2 + Q^2 ) +Q^4 ] \right\rbrace,
\\
{\rm CR}: L_{\mu\nu} W^{\mu\nu} &=& b_\chi^2 \left\lbrace 2Q^4 F_1 (Q^2 -2 m_\chi^2 ) - \dfrac{m_N F_2 }{\nu} Q^4 [4 E_\chi (\nu -E_\chi) 
+Q^2 ]\right\rbrace .
\end{eqnarray}
\end{widetext}

\bibliography{refs}

\begin{thebibliography}{135}%
\makeatletter
\providecommand \@ifxundefined [1]{%
 \@ifx{#1\undefined}
}%
\providecommand \@ifnum [1]{%
 \ifnum #1\expandafter \@firstoftwo
 \else \expandafter \@secondoftwo
 \fi
}%
\providecommand \@ifx [1]{%
 \ifx #1\expandafter \@firstoftwo
 \else \expandafter \@secondoftwo
 \fi
}%
\providecommand \natexlab [1]{#1}%
\providecommand \enquote  [1]{``#1''}%
\providecommand \bibnamefont  [1]{#1}%
\providecommand \bibfnamefont [1]{#1}%
\providecommand \citenamefont [1]{#1}%
\providecommand \href@noop [0]{\@secondoftwo}%
\providecommand \href [0]{\begingroup \@sanitize@url \@href}%
\providecommand \@href[1]{\@@startlink{#1}\@@href}%
\providecommand \@@href[1]{\endgroup#1\@@endlink}%
\providecommand \@sanitize@url [0]{\catcode `\\12\catcode `\$12\catcode
  `\&12\catcode `\#12\catcode `\^12\catcode `\_12\catcode `\%12\relax}%
\providecommand \@@startlink[1]{}%
\providecommand \@@endlink[0]{}%
\providecommand \url  [0]{\begingroup\@sanitize@url \@url }%
\providecommand \@url [1]{\endgroup\@href {#1}{\urlprefix }}%
\providecommand \urlprefix  [0]{URL }%
\providecommand \Eprint [0]{\href }%
\providecommand \doibase [0]{http://dx.doi.org/}%
\providecommand \selectlanguage [0]{\@gobble}%
\providecommand \bibinfo  [0]{\@secondoftwo}%
\providecommand \bibfield  [0]{\@secondoftwo}%
\providecommand \translation [1]{[#1]}%
\providecommand \BibitemOpen [0]{}%
\providecommand \bibitemStop [0]{}%
\providecommand \bibitemNoStop [0]{.\EOS\space}%
\providecommand \EOS [0]{\spacefactor3000\relax}%
\providecommand \BibitemShut  [1]{\csname bibitem#1\endcsname}%
\let\auto@bib@innerbib\@empty
\bibitem [{\citenamefont {Athanassopoulos}\ \emph {et~al.}(1997)\citenamefont
  {Athanassopoulos} \emph {et~al.}}]{Athanassopoulos:1996ds}%
  \BibitemOpen
  \bibfield  {author} {\bibinfo {author} {\bibfnamefont {C.}~\bibnamefont
  {Athanassopoulos}} \emph {et~al.} (\bibinfo {collaboration} {LSND}),\ }\href
  {\doibase 10.1016/S0168-9002(96)01155-2} {\bibfield  {journal} {\bibinfo
  {journal} {Nucl. Instrum. Meth.}\ }\textbf {\bibinfo {volume} {A388}},\
  \bibinfo {pages} {149} (\bibinfo {year} {1997})},\ \Eprint
  {http://arxiv.org/abs/nucl-ex/9605002} {arXiv:nucl-ex/9605002 [nucl-ex]}
  \BibitemShut {NoStop}%
\bibitem [{\citenamefont {Aguilar-Arevalo}\ \emph
  {et~al.}(2018{\natexlab{a}})\citenamefont {Aguilar-Arevalo} \emph
  {et~al.}}]{Aguilar-Arevalo:2018gpe}%
  \BibitemOpen
  \bibfield  {author} {\bibinfo {author} {\bibfnamefont {A.~A.}\ \bibnamefont
  {Aguilar-Arevalo}} \emph {et~al.} (\bibinfo {collaboration} {MiniBooNE}),\
  }\href@noop {} {\  (\bibinfo {year} {2018}{\natexlab{a}})},\ \Eprint
  {http://arxiv.org/abs/1805.12028} {arXiv:1805.12028 [hep-ex]} \BibitemShut
  {NoStop}%
\bibitem [{\citenamefont {Akimov}\ \emph {et~al.}(2017)\citenamefont {Akimov}
  \emph {et~al.}}]{Akimov:2017ade}%
  \BibitemOpen
  \bibfield  {author} {\bibinfo {author} {\bibfnamefont {D.}~\bibnamefont
  {Akimov}} \emph {et~al.} (\bibinfo {collaboration} {COHERENT}),\ }\href
  {\doibase 10.1126/science.aao0990} {\bibfield  {journal} {\bibinfo  {journal}
  {Science}\ }\textbf {\bibinfo {volume} {357}},\ \bibinfo {pages} {1123}
  (\bibinfo {year} {2017})},\ \Eprint {http://arxiv.org/abs/1708.01294}
  {arXiv:1708.01294 [nucl-ex]} \BibitemShut {NoStop}%
\bibitem [{\citenamefont {Acciarri}\ \emph {et~al.}(2015)\citenamefont
  {Acciarri} \emph {et~al.}}]{Acciarri:2015uup}%
  \BibitemOpen
  \bibfield  {author} {\bibinfo {author} {\bibfnamefont {R.}~\bibnamefont
  {Acciarri}} \emph {et~al.} (\bibinfo {collaboration} {DUNE}),\ }\href@noop {}
  {\  (\bibinfo {year} {2015})},\ \Eprint {http://arxiv.org/abs/1512.06148}
  {arXiv:1512.06148 [physics.ins-det]} \BibitemShut {NoStop}%
\bibitem [{\citenamefont {Huber}\ \emph {et~al.}(2004)\citenamefont {Huber},
  \citenamefont {Lindner}, \citenamefont {Rolinec}, \citenamefont {Schwetz},\
  and\ \citenamefont {Winter}}]{Huber:2004ug}%
  \BibitemOpen
  \bibfield  {author} {\bibinfo {author} {\bibfnamefont {P.}~\bibnamefont
  {Huber}}, \bibinfo {author} {\bibfnamefont {M.}~\bibnamefont {Lindner}},
  \bibinfo {author} {\bibfnamefont {M.}~\bibnamefont {Rolinec}}, \bibinfo
  {author} {\bibfnamefont {T.}~\bibnamefont {Schwetz}}, \ and\ \bibinfo
  {author} {\bibfnamefont {W.}~\bibnamefont {Winter}},\ }\href {\doibase
  10.1103/PhysRevD.70.073014} {\bibfield  {journal} {\bibinfo  {journal} {Phys.
  Rev.}\ }\textbf {\bibinfo {volume} {D70}},\ \bibinfo {pages} {073014}
  (\bibinfo {year} {2004})},\ \Eprint {http://arxiv.org/abs/hep-ph/0403068}
  {arXiv:hep-ph/0403068 [hep-ph]} \BibitemShut {NoStop}%
\bibitem [{\citenamefont {Alexander}\ \emph {et~al.}(2016)\citenamefont
  {Alexander} \emph {et~al.}}]{Alexander:2016aln}%
  \BibitemOpen
  \bibfield  {author} {\bibinfo {author} {\bibfnamefont {J.}~\bibnamefont
  {Alexander}} \emph {et~al.}\ }(\bibinfo {year} {2016})\ \Eprint
  {http://arxiv.org/abs/1608.08632} {arXiv:1608.08632 [hep-ph]} \BibitemShut
  {NoStop}%
\bibitem [{\citenamefont {Essig}\ \emph {et~al.}(2013)\citenamefont {Essig}
  \emph {et~al.}}]{Essig:2013lka}%
  \BibitemOpen
  \bibfield  {author} {\bibinfo {author} {\bibfnamefont {R.}~\bibnamefont
  {Essig}} \emph {et~al.},\ }in\ \href
  {https://inspirehep.net/record/1263039/files/arXiv:1311.0029.pdf} {\emph
  {\bibinfo {booktitle} {{Proceedings, 2013 Community Summer Study on the
  Future of U.S. Particle Physics: Snowmass on the Mississippi (CSS2013):
  Minneapolis, MN, USA, July 29-August 6, 2013}}}}\ (\bibinfo {year} {2013})\
  \Eprint {http://arxiv.org/abs/1311.0029} {arXiv:1311.0029 [hep-ph]}
  \BibitemShut {NoStop}%
\bibitem [{\citenamefont {Golowich}\ and\ \citenamefont
  {Robinett}(1987)}]{Golowich:1986tj}%
  \BibitemOpen
  \bibfield  {author} {\bibinfo {author} {\bibfnamefont {E.}~\bibnamefont
  {Golowich}}\ and\ \bibinfo {author} {\bibfnamefont {R.~W.}\ \bibnamefont
  {Robinett}},\ }\href {\doibase 10.1103/PhysRevD.35.391} {\bibfield  {journal}
  {\bibinfo  {journal} {Phys. Rev.}\ }\textbf {\bibinfo {volume} {D35}},\
  \bibinfo {pages} {391} (\bibinfo {year} {1987})}\BibitemShut {NoStop}%
\bibitem [{\citenamefont {Prinz}\ \emph {et~al.}(1998)\citenamefont {Prinz}
  \emph {et~al.}}]{Prinz:1998ua}%
  \BibitemOpen
  \bibfield  {author} {\bibinfo {author} {\bibfnamefont {A.~A.}\ \bibnamefont
  {Prinz}} \emph {et~al.},\ }\href {\doibase 10.1103/PhysRevLett.81.1175}
  {\bibfield  {journal} {\bibinfo  {journal} {Phys. Rev. Lett.}\ }\textbf
  {\bibinfo {volume} {81}},\ \bibinfo {pages} {1175} (\bibinfo {year}
  {1998})},\ \Eprint {http://arxiv.org/abs/hep-ex/9804008}
  {arXiv:hep-ex/9804008 [hep-ex]} \BibitemShut {NoStop}%
\bibitem [{\citenamefont {Izaguirre}\ \emph {et~al.}(2013)\citenamefont
  {Izaguirre}, \citenamefont {Krnjaic}, \citenamefont {Schuster},\ and\
  \citenamefont {Toro}}]{Izaguirre:2013uxa}%
  \BibitemOpen
  \bibfield  {author} {\bibinfo {author} {\bibfnamefont {E.}~\bibnamefont
  {Izaguirre}}, \bibinfo {author} {\bibfnamefont {G.}~\bibnamefont {Krnjaic}},
  \bibinfo {author} {\bibfnamefont {P.}~\bibnamefont {Schuster}}, \ and\
  \bibinfo {author} {\bibfnamefont {N.}~\bibnamefont {Toro}},\ }\href {\doibase
  10.1103/PhysRevD.88.114015} {\bibfield  {journal} {\bibinfo  {journal} {Phys.
  Rev.}\ }\textbf {\bibinfo {volume} {D88}},\ \bibinfo {pages} {114015}
  (\bibinfo {year} {2013})},\ \Eprint {http://arxiv.org/abs/1307.6554}
  {arXiv:1307.6554 [hep-ph]} \BibitemShut {NoStop}%
\bibitem [{\citenamefont {Soper}\ \emph {et~al.}(2014)\citenamefont {Soper},
  \citenamefont {Spannowsky}, \citenamefont {Wallace},\ and\ \citenamefont
  {Tait}}]{Soper:2014ska}%
  \BibitemOpen
  \bibfield  {author} {\bibinfo {author} {\bibfnamefont {D.~E.}\ \bibnamefont
  {Soper}}, \bibinfo {author} {\bibfnamefont {M.}~\bibnamefont {Spannowsky}},
  \bibinfo {author} {\bibfnamefont {C.~J.}\ \bibnamefont {Wallace}}, \ and\
  \bibinfo {author} {\bibfnamefont {T.~M.~P.}\ \bibnamefont {Tait}},\ }\href
  {\doibase 10.1103/PhysRevD.90.115005} {\bibfield  {journal} {\bibinfo
  {journal} {Phys. Rev.}\ }\textbf {\bibinfo {volume} {D90}},\ \bibinfo {pages}
  {115005} (\bibinfo {year} {2014})},\ \Eprint {http://arxiv.org/abs/1407.2623}
  {arXiv:1407.2623 [hep-ph]} \BibitemShut {NoStop}%
\bibitem [{\citenamefont {Berlin}\ \emph {et~al.}(2019)\citenamefont {Berlin},
  \citenamefont {Blinov}, \citenamefont {Krnjaic}, \citenamefont {Schuster},\
  and\ \citenamefont {Toro}}]{Berlin:2018bsc}%
  \BibitemOpen
  \bibfield  {author} {\bibinfo {author} {\bibfnamefont {A.}~\bibnamefont
  {Berlin}}, \bibinfo {author} {\bibfnamefont {N.}~\bibnamefont {Blinov}},
  \bibinfo {author} {\bibfnamefont {G.}~\bibnamefont {Krnjaic}}, \bibinfo
  {author} {\bibfnamefont {P.}~\bibnamefont {Schuster}}, \ and\ \bibinfo
  {author} {\bibfnamefont {N.}~\bibnamefont {Toro}},\ }\href {\doibase
  10.1103/PhysRevD.99.075001} {\bibfield  {journal} {\bibinfo  {journal} {Phys.
  Rev.}\ }\textbf {\bibinfo {volume} {D99}},\ \bibinfo {pages} {075001}
  (\bibinfo {year} {2019})},\ \Eprint {http://arxiv.org/abs/1807.01730}
  {arXiv:1807.01730 [hep-ph]} \BibitemShut {NoStop}%
\bibitem [{\citenamefont {Magill}\ \emph {et~al.}(2019)\citenamefont {Magill},
  \citenamefont {Plestid}, \citenamefont {Pospelov},\ and\ \citenamefont
  {Tsai}}]{Magill:2018tbb}%
  \BibitemOpen
  \bibfield  {author} {\bibinfo {author} {\bibfnamefont {G.}~\bibnamefont
  {Magill}}, \bibinfo {author} {\bibfnamefont {R.}~\bibnamefont {Plestid}},
  \bibinfo {author} {\bibfnamefont {M.}~\bibnamefont {Pospelov}}, \ and\
  \bibinfo {author} {\bibfnamefont {Y.-D.}\ \bibnamefont {Tsai}},\ }\href
  {\doibase 10.1103/PhysRevLett.122.071801} {\bibfield  {journal} {\bibinfo
  {journal} {Phys. Rev. Lett.}\ }\textbf {\bibinfo {volume} {122}},\ \bibinfo
  {pages} {071801} (\bibinfo {year} {2019})},\ \Eprint
  {http://arxiv.org/abs/1806.03310} {arXiv:1806.03310 [hep-ph]} \BibitemShut
  {NoStop}%
\bibitem [{\citenamefont {Liang}\ \emph {et~al.}(2019)\citenamefont {Liang},
  \citenamefont {Liu}, \citenamefont {Ma},\ and\ \citenamefont
  {Zhang}}]{Liang:2019zkb}%
  \BibitemOpen
  \bibfield  {author} {\bibinfo {author} {\bibfnamefont {J.}~\bibnamefont
  {Liang}}, \bibinfo {author} {\bibfnamefont {Z.}~\bibnamefont {Liu}}, \bibinfo
  {author} {\bibfnamefont {Y.}~\bibnamefont {Ma}}, \ and\ \bibinfo {author}
  {\bibfnamefont {Y.}~\bibnamefont {Zhang}},\ }\href@noop {} {\  (\bibinfo
  {year} {2019})},\ \Eprint {http://arxiv.org/abs/1909.06847} {arXiv:1909.06847
  [hep-ph]} \BibitemShut {NoStop}%
\bibitem [{\citenamefont {Davidson}\ \emph {et~al.}(2000)\citenamefont
  {Davidson}, \citenamefont {Hannestad},\ and\ \citenamefont
  {Raffelt}}]{Davidson:2000hf}%
  \BibitemOpen
  \bibfield  {author} {\bibinfo {author} {\bibfnamefont {S.}~\bibnamefont
  {Davidson}}, \bibinfo {author} {\bibfnamefont {S.}~\bibnamefont {Hannestad}},
  \ and\ \bibinfo {author} {\bibfnamefont {G.}~\bibnamefont {Raffelt}},\ }\href
  {\doibase 10.1088/1126-6708/2000/05/003} {\bibfield  {journal} {\bibinfo
  {journal} {JHEP}\ }\textbf {\bibinfo {volume} {05}},\ \bibinfo {pages} {003}
  (\bibinfo {year} {2000})},\ \Eprint {http://arxiv.org/abs/hep-ph/0001179}
  {arXiv:hep-ph/0001179 [hep-ph]} \BibitemShut {NoStop}%
\bibitem [{\citenamefont {Dubovsky}\ \emph {et~al.}(2004)\citenamefont
  {Dubovsky}, \citenamefont {Gorbunov},\ and\ \citenamefont
  {Rubtsov}}]{Dubovsky:2003yn}%
  \BibitemOpen
  \bibfield  {author} {\bibinfo {author} {\bibfnamefont {S.~L.}\ \bibnamefont
  {Dubovsky}}, \bibinfo {author} {\bibfnamefont {D.~S.}\ \bibnamefont
  {Gorbunov}}, \ and\ \bibinfo {author} {\bibfnamefont {G.~I.}\ \bibnamefont
  {Rubtsov}},\ }\href {\doibase 10.1134/1.1675909} {\bibfield  {journal}
  {\bibinfo  {journal} {JETP Lett.}\ }\textbf {\bibinfo {volume} {79}},\
  \bibinfo {pages} {1} (\bibinfo {year} {2004})},\ \bibinfo {note} {[Pisma Zh.
  Eksp. Teor. Fiz.79,3(2004)]},\ \Eprint {http://arxiv.org/abs/hep-ph/0311189}
  {arXiv:hep-ph/0311189 [hep-ph]} \BibitemShut {NoStop}%
\bibitem [{\citenamefont {McDermott}\ \emph {et~al.}(2011)\citenamefont
  {McDermott}, \citenamefont {Yu},\ and\ \citenamefont
  {Zurek}}]{McDermott:2010pa}%
  \BibitemOpen
  \bibfield  {author} {\bibinfo {author} {\bibfnamefont {S.~D.}\ \bibnamefont
  {McDermott}}, \bibinfo {author} {\bibfnamefont {H.-B.}\ \bibnamefont {Yu}}, \
  and\ \bibinfo {author} {\bibfnamefont {K.~M.}\ \bibnamefont {Zurek}},\ }\href
  {\doibase 10.1103/PhysRevD.83.063509} {\bibfield  {journal} {\bibinfo
  {journal} {Phys. Rev.}\ }\textbf {\bibinfo {volume} {D83}},\ \bibinfo {pages}
  {063509} (\bibinfo {year} {2011})},\ \Eprint {http://arxiv.org/abs/1011.2907}
  {arXiv:1011.2907 [hep-ph]} \BibitemShut {NoStop}%
\bibitem [{\citenamefont {Cline}\ \emph {et~al.}(2012)\citenamefont {Cline},
  \citenamefont {Liu},\ and\ \citenamefont {Xue}}]{Cline:2012is}%
  \BibitemOpen
  \bibfield  {author} {\bibinfo {author} {\bibfnamefont {J.~M.}\ \bibnamefont
  {Cline}}, \bibinfo {author} {\bibfnamefont {Z.}~\bibnamefont {Liu}}, \ and\
  \bibinfo {author} {\bibfnamefont {W.}~\bibnamefont {Xue}},\ }\href {\doibase
  10.1103/PhysRevD.85.101302} {\bibfield  {journal} {\bibinfo  {journal} {Phys.
  Rev.}\ }\textbf {\bibinfo {volume} {D85}},\ \bibinfo {pages} {101302}
  (\bibinfo {year} {2012})},\ \Eprint {http://arxiv.org/abs/1201.4858}
  {arXiv:1201.4858 [hep-ph]} \BibitemShut {NoStop}%
\bibitem [{\citenamefont {Dolgov}\ \emph {et~al.}(2013)\citenamefont {Dolgov},
  \citenamefont {Dubovsky}, \citenamefont {Rubtsov},\ and\ \citenamefont
  {Tkachev}}]{Dolgov:2013una}%
  \BibitemOpen
  \bibfield  {author} {\bibinfo {author} {\bibfnamefont {A.~D.}\ \bibnamefont
  {Dolgov}}, \bibinfo {author} {\bibfnamefont {S.~L.}\ \bibnamefont
  {Dubovsky}}, \bibinfo {author} {\bibfnamefont {G.~I.}\ \bibnamefont
  {Rubtsov}}, \ and\ \bibinfo {author} {\bibfnamefont {I.~I.}\ \bibnamefont
  {Tkachev}},\ }\href {\doibase 10.1103/PhysRevD.88.117701} {\bibfield
  {journal} {\bibinfo  {journal} {Phys. Rev.}\ }\textbf {\bibinfo {volume}
  {D88}},\ \bibinfo {pages} {117701} (\bibinfo {year} {2013})},\ \Eprint
  {http://arxiv.org/abs/1310.2376} {arXiv:1310.2376 [hep-ph]} \BibitemShut
  {NoStop}%
\bibitem [{\citenamefont {Vogel}\ and\ \citenamefont
  {Redondo}(2014)}]{Vogel:2013raa}%
  \BibitemOpen
  \bibfield  {author} {\bibinfo {author} {\bibfnamefont {H.}~\bibnamefont
  {Vogel}}\ and\ \bibinfo {author} {\bibfnamefont {J.}~\bibnamefont
  {Redondo}},\ }\href {\doibase 10.1088/1475-7516/2014/02/029} {\bibfield
  {journal} {\bibinfo  {journal} {JCAP}\ }\textbf {\bibinfo {volume} {1402}},\
  \bibinfo {pages} {029} (\bibinfo {year} {2014})},\ \Eprint
  {http://arxiv.org/abs/1311.2600} {arXiv:1311.2600 [hep-ph]} \BibitemShut
  {NoStop}%
\bibitem [{\citenamefont {Dvorkin}\ \emph {et~al.}(2014)\citenamefont
  {Dvorkin}, \citenamefont {Blum},\ and\ \citenamefont
  {Kamionkowski}}]{Dvorkin:2013cea}%
  \BibitemOpen
  \bibfield  {author} {\bibinfo {author} {\bibfnamefont {C.}~\bibnamefont
  {Dvorkin}}, \bibinfo {author} {\bibfnamefont {K.}~\bibnamefont {Blum}}, \
  and\ \bibinfo {author} {\bibfnamefont {M.}~\bibnamefont {Kamionkowski}},\
  }\href {\doibase 10.1103/PhysRevD.89.023519} {\bibfield  {journal} {\bibinfo
  {journal} {Phys. Rev.}\ }\textbf {\bibinfo {volume} {D89}},\ \bibinfo {pages}
  {023519} (\bibinfo {year} {2014})},\ \Eprint {http://arxiv.org/abs/1311.2937}
  {arXiv:1311.2937 [astro-ph.CO]} \BibitemShut {NoStop}%
\bibitem [{\citenamefont {Ali-Haïmoud}\ \emph {et~al.}(2015)\citenamefont
  {Ali-Haïmoud}, \citenamefont {Chluba},\ and\ \citenamefont
  {Kamionkowski}}]{Ali-Haimoud:2015pwa}%
  \BibitemOpen
  \bibfield  {author} {\bibinfo {author} {\bibfnamefont {Y.}~\bibnamefont
  {Ali-Haïmoud}}, \bibinfo {author} {\bibfnamefont {J.}~\bibnamefont
  {Chluba}}, \ and\ \bibinfo {author} {\bibfnamefont {M.}~\bibnamefont
  {Kamionkowski}},\ }\href {\doibase 10.1103/PhysRevLett.115.071304} {\bibfield
   {journal} {\bibinfo  {journal} {Phys. Rev. Lett.}\ }\textbf {\bibinfo
  {volume} {115}},\ \bibinfo {pages} {071304} (\bibinfo {year} {2015})},\
  \Eprint {http://arxiv.org/abs/1506.04745} {arXiv:1506.04745 [astro-ph.CO]}
  \BibitemShut {NoStop}%
\bibitem [{\citenamefont {Kamada}\ \emph {et~al.}(2017)\citenamefont {Kamada},
  \citenamefont {Kohri}, \citenamefont {Takahashi},\ and\ \citenamefont
  {Yoshida}}]{Kamada:2016qjo}%
  \BibitemOpen
  \bibfield  {author} {\bibinfo {author} {\bibfnamefont {A.}~\bibnamefont
  {Kamada}}, \bibinfo {author} {\bibfnamefont {K.}~\bibnamefont {Kohri}},
  \bibinfo {author} {\bibfnamefont {T.}~\bibnamefont {Takahashi}}, \ and\
  \bibinfo {author} {\bibfnamefont {N.}~\bibnamefont {Yoshida}},\ }\href
  {\doibase 10.1103/PhysRevD.95.023502} {\bibfield  {journal} {\bibinfo
  {journal} {Phys. Rev.}\ }\textbf {\bibinfo {volume} {D95}},\ \bibinfo {pages}
  {023502} (\bibinfo {year} {2017})},\ \Eprint
  {http://arxiv.org/abs/1604.07926} {arXiv:1604.07926 [astro-ph.CO]}
  \BibitemShut {NoStop}%
\bibitem [{\citenamefont {Agrawal}\ and\ \citenamefont
  {Randall}(2017)}]{Agrawal:2017pnb}%
  \BibitemOpen
  \bibfield  {author} {\bibinfo {author} {\bibfnamefont {P.}~\bibnamefont
  {Agrawal}}\ and\ \bibinfo {author} {\bibfnamefont {L.}~\bibnamefont
  {Randall}},\ }\href {\doibase 10.1088/1475-7516/2017/12/019} {\bibfield
  {journal} {\bibinfo  {journal} {JCAP}\ }\textbf {\bibinfo {volume} {1712}},\
  \bibinfo {pages} {019} (\bibinfo {year} {2017})},\ \Eprint
  {http://arxiv.org/abs/1706.04195} {arXiv:1706.04195 [hep-ph]} \BibitemShut
  {NoStop}%
\bibitem [{\citenamefont {Muñoz}\ and\ \citenamefont
  {Loeb}(2018)}]{Munoz:2018pzp}%
  \BibitemOpen
  \bibfield  {author} {\bibinfo {author} {\bibfnamefont {J.~B.}\ \bibnamefont
  {Muñoz}}\ and\ \bibinfo {author} {\bibfnamefont {A.}~\bibnamefont {Loeb}},\
  }\href {\doibase 10.1038/s41586-018-0151-x} {\bibfield  {journal} {\bibinfo
  {journal} {Nature}\ }\textbf {\bibinfo {volume} {557}},\ \bibinfo {pages}
  {684} (\bibinfo {year} {2018})},\ \Eprint {http://arxiv.org/abs/1802.10094}
  {arXiv:1802.10094 [astro-ph.CO]} \BibitemShut {NoStop}%
\bibitem [{\citenamefont {Berlin}\ \emph {et~al.}(2018)\citenamefont {Berlin},
  \citenamefont {Hooper}, \citenamefont {Krnjaic},\ and\ \citenamefont
  {McDermott}}]{Berlin:2018sjs}%
  \BibitemOpen
  \bibfield  {author} {\bibinfo {author} {\bibfnamefont {A.}~\bibnamefont
  {Berlin}}, \bibinfo {author} {\bibfnamefont {D.}~\bibnamefont {Hooper}},
  \bibinfo {author} {\bibfnamefont {G.}~\bibnamefont {Krnjaic}}, \ and\
  \bibinfo {author} {\bibfnamefont {S.~D.}\ \bibnamefont {McDermott}},\ }\href
  {\doibase 10.1103/PhysRevLett.121.011102} {\bibfield  {journal} {\bibinfo
  {journal} {Phys. Rev. Lett.}\ }\textbf {\bibinfo {volume} {121}},\ \bibinfo
  {pages} {011102} (\bibinfo {year} {2018})},\ \Eprint
  {http://arxiv.org/abs/1803.02804} {arXiv:1803.02804 [hep-ph]} \BibitemShut
  {NoStop}%
\bibitem [{\citenamefont {Barkana}\ \emph {et~al.}(2018)\citenamefont
  {Barkana}, \citenamefont {Outmezguine}, \citenamefont {Redigolo},\ and\
  \citenamefont {Volansky}}]{Barkana:2018cct}%
  \BibitemOpen
  \bibfield  {author} {\bibinfo {author} {\bibfnamefont {R.}~\bibnamefont
  {Barkana}}, \bibinfo {author} {\bibfnamefont {N.~J.}\ \bibnamefont
  {Outmezguine}}, \bibinfo {author} {\bibfnamefont {D.}~\bibnamefont
  {Redigolo}}, \ and\ \bibinfo {author} {\bibfnamefont {T.}~\bibnamefont
  {Volansky}},\ }\href {\doibase 10.1103/PhysRevD.98.103005} {\bibfield
  {journal} {\bibinfo  {journal} {Phys. Rev.}\ }\textbf {\bibinfo {volume}
  {D98}},\ \bibinfo {pages} {103005} (\bibinfo {year} {2018})},\ \Eprint
  {http://arxiv.org/abs/1803.03091} {arXiv:1803.03091 [hep-ph]} \BibitemShut
  {NoStop}%
\bibitem [{\citenamefont {Chang}\ \emph {et~al.}(2018)\citenamefont {Chang},
  \citenamefont {Essig},\ and\ \citenamefont {McDermott}}]{Chang:2018rso}%
  \BibitemOpen
  \bibfield  {author} {\bibinfo {author} {\bibfnamefont {J.~H.}\ \bibnamefont
  {Chang}}, \bibinfo {author} {\bibfnamefont {R.}~\bibnamefont {Essig}}, \ and\
  \bibinfo {author} {\bibfnamefont {S.~D.}\ \bibnamefont {McDermott}},\ }\href
  {\doibase 10.1007/JHEP09(2018)051} {\bibfield  {journal} {\bibinfo  {journal}
  {JHEP}\ }\textbf {\bibinfo {volume} {09}},\ \bibinfo {pages} {051} (\bibinfo
  {year} {2018})},\ \Eprint {http://arxiv.org/abs/1803.00993} {arXiv:1803.00993
  [hep-ph]} \BibitemShut {NoStop}%
\bibitem [{\citenamefont {Kovetz}\ \emph {et~al.}(2018)\citenamefont {Kovetz},
  \citenamefont {Poulin}, \citenamefont {Gluscevic}, \citenamefont {Boddy},
  \citenamefont {Barkana},\ and\ \citenamefont
  {Kamionkowski}}]{Kovetz:2018zan}%
  \BibitemOpen
  \bibfield  {author} {\bibinfo {author} {\bibfnamefont {E.~D.}\ \bibnamefont
  {Kovetz}}, \bibinfo {author} {\bibfnamefont {V.}~\bibnamefont {Poulin}},
  \bibinfo {author} {\bibfnamefont {V.}~\bibnamefont {Gluscevic}}, \bibinfo
  {author} {\bibfnamefont {K.~K.}\ \bibnamefont {Boddy}}, \bibinfo {author}
  {\bibfnamefont {R.}~\bibnamefont {Barkana}}, \ and\ \bibinfo {author}
  {\bibfnamefont {M.}~\bibnamefont {Kamionkowski}},\ }\href {\doibase
  10.1103/PhysRevD.98.103529} {\bibfield  {journal} {\bibinfo  {journal} {Phys.
  Rev.}\ }\textbf {\bibinfo {volume} {D98}},\ \bibinfo {pages} {103529}
  (\bibinfo {year} {2018})},\ \Eprint {http://arxiv.org/abs/1807.11482}
  {arXiv:1807.11482 [astro-ph.CO]} \BibitemShut {NoStop}%
\bibitem [{\citenamefont {Xu}\ \emph {et~al.}(2018)\citenamefont {Xu},
  \citenamefont {Dvorkin},\ and\ \citenamefont {Chael}}]{Xu:2018efh}%
  \BibitemOpen
  \bibfield  {author} {\bibinfo {author} {\bibfnamefont {W.~L.}\ \bibnamefont
  {Xu}}, \bibinfo {author} {\bibfnamefont {C.}~\bibnamefont {Dvorkin}}, \ and\
  \bibinfo {author} {\bibfnamefont {A.}~\bibnamefont {Chael}},\ }\href
  {\doibase 10.1103/PhysRevD.97.103530} {\bibfield  {journal} {\bibinfo
  {journal} {Phys. Rev.}\ }\textbf {\bibinfo {volume} {D97}},\ \bibinfo {pages}
  {103530} (\bibinfo {year} {2018})},\ \Eprint
  {http://arxiv.org/abs/1802.06788} {arXiv:1802.06788 [astro-ph.CO]}
  \BibitemShut {NoStop}%
\bibitem [{\citenamefont {Slatyer}\ and\ \citenamefont
  {Wu}(2018)}]{Slatyer:2018aqg}%
  \BibitemOpen
  \bibfield  {author} {\bibinfo {author} {\bibfnamefont {T.~R.}\ \bibnamefont
  {Slatyer}}\ and\ \bibinfo {author} {\bibfnamefont {C.-L.}\ \bibnamefont
  {Wu}},\ }\href {\doibase 10.1103/PhysRevD.98.023013} {\bibfield  {journal}
  {\bibinfo  {journal} {Phys. Rev.}\ }\textbf {\bibinfo {volume} {D98}},\
  \bibinfo {pages} {023013} (\bibinfo {year} {2018})},\ \Eprint
  {http://arxiv.org/abs/1803.09734} {arXiv:1803.09734 [astro-ph.CO]}
  \BibitemShut {NoStop}%
\bibitem [{\citenamefont {Pospelov}\ and\ \citenamefont {ter
  Veldhuis}(2000)}]{Pospelov:2000bq}%
  \BibitemOpen
  \bibfield  {author} {\bibinfo {author} {\bibfnamefont {M.}~\bibnamefont
  {Pospelov}}\ and\ \bibinfo {author} {\bibfnamefont {T.}~\bibnamefont {ter
  Veldhuis}},\ }\href {\doibase 10.1016/S0370-2693(00)00358-0} {\bibfield
  {journal} {\bibinfo  {journal} {Phys. Lett.}\ }\textbf {\bibinfo {volume}
  {B480}},\ \bibinfo {pages} {181} (\bibinfo {year} {2000})},\ \Eprint
  {http://arxiv.org/abs/hep-ph/0003010} {arXiv:hep-ph/0003010 [hep-ph]}
  \BibitemShut {NoStop}%
\bibitem [{\citenamefont {Sigurdson}\ \emph {et~al.}(2004)\citenamefont
  {Sigurdson}, \citenamefont {Doran}, \citenamefont {Kurylov}, \citenamefont
  {Caldwell},\ and\ \citenamefont {Kamionkowski}}]{Sigurdson:2004zp}%
  \BibitemOpen
  \bibfield  {author} {\bibinfo {author} {\bibfnamefont {K.}~\bibnamefont
  {Sigurdson}}, \bibinfo {author} {\bibfnamefont {M.}~\bibnamefont {Doran}},
  \bibinfo {author} {\bibfnamefont {A.}~\bibnamefont {Kurylov}}, \bibinfo
  {author} {\bibfnamefont {R.~R.}\ \bibnamefont {Caldwell}}, \ and\ \bibinfo
  {author} {\bibfnamefont {M.}~\bibnamefont {Kamionkowski}},\ }\href {\doibase
  10.1103/PhysRevD.70.083501, 10.1103/PhysRevD.73.089903} {\bibfield  {journal}
  {\bibinfo  {journal} {Phys. Rev.}\ }\textbf {\bibinfo {volume} {D70}},\
  \bibinfo {pages} {083501} (\bibinfo {year} {2004})},\ \bibinfo {note}
  {[Erratum: Phys. Rev.D73,089903(2006)]},\ \Eprint
  {http://arxiv.org/abs/astro-ph/0406355} {arXiv:astro-ph/0406355 [astro-ph]}
  \BibitemShut {NoStop}%
\bibitem [{\citenamefont {Ho}\ and\ \citenamefont
  {Scherrer}(2013)}]{Ho:2012bg}%
  \BibitemOpen
  \bibfield  {author} {\bibinfo {author} {\bibfnamefont {C.~M.}\ \bibnamefont
  {Ho}}\ and\ \bibinfo {author} {\bibfnamefont {R.~J.}\ \bibnamefont
  {Scherrer}},\ }\href {\doibase 10.1016/j.physletb.2013.04.039} {\bibfield
  {journal} {\bibinfo  {journal} {Phys. Lett.}\ }\textbf {\bibinfo {volume}
  {B722}},\ \bibinfo {pages} {341} (\bibinfo {year} {2013})},\ \Eprint
  {http://arxiv.org/abs/1211.0503} {arXiv:1211.0503 [hep-ph]} \BibitemShut
  {NoStop}%
\bibitem [{\citenamefont {Schmidt}\ \emph {et~al.}(2012)\citenamefont
  {Schmidt}, \citenamefont {Schwetz},\ and\ \citenamefont
  {Toma}}]{Schmidt:2012yg}%
  \BibitemOpen
  \bibfield  {author} {\bibinfo {author} {\bibfnamefont {D.}~\bibnamefont
  {Schmidt}}, \bibinfo {author} {\bibfnamefont {T.}~\bibnamefont {Schwetz}}, \
  and\ \bibinfo {author} {\bibfnamefont {T.}~\bibnamefont {Toma}},\ }\href
  {\doibase 10.1103/PhysRevD.85.073009} {\bibfield  {journal} {\bibinfo
  {journal} {Phys. Rev.}\ }\textbf {\bibinfo {volume} {D85}},\ \bibinfo {pages}
  {073009} (\bibinfo {year} {2012})},\ \Eprint {http://arxiv.org/abs/1201.0906}
  {arXiv:1201.0906 [hep-ph]} \BibitemShut {NoStop}%
\bibitem [{\citenamefont {Kopp}\ \emph {et~al.}(2014)\citenamefont {Kopp},
  \citenamefont {Michaels},\ and\ \citenamefont {Smirnov}}]{Kopp:2014tsa}%
  \BibitemOpen
  \bibfield  {author} {\bibinfo {author} {\bibfnamefont {J.}~\bibnamefont
  {Kopp}}, \bibinfo {author} {\bibfnamefont {L.}~\bibnamefont {Michaels}}, \
  and\ \bibinfo {author} {\bibfnamefont {J.}~\bibnamefont {Smirnov}},\ }\href
  {\doibase 10.1088/1475-7516/2014/04/022} {\bibfield  {journal} {\bibinfo
  {journal} {JCAP}\ }\textbf {\bibinfo {volume} {1404}},\ \bibinfo {pages}
  {022} (\bibinfo {year} {2014})},\ \Eprint {http://arxiv.org/abs/1401.6457}
  {arXiv:1401.6457 [hep-ph]} \BibitemShut {NoStop}%
\bibitem [{\citenamefont {Ibarra}\ and\ \citenamefont
  {Wild}(2015)}]{Ibarra:2015fqa}%
  \BibitemOpen
  \bibfield  {author} {\bibinfo {author} {\bibfnamefont {A.}~\bibnamefont
  {Ibarra}}\ and\ \bibinfo {author} {\bibfnamefont {S.}~\bibnamefont {Wild}},\
  }\href {\doibase 10.1088/1475-7516/2015/05/047} {\bibfield  {journal}
  {\bibinfo  {journal} {JCAP}\ }\textbf {\bibinfo {volume} {1505}},\ \bibinfo
  {pages} {047} (\bibinfo {year} {2015})},\ \Eprint
  {http://arxiv.org/abs/1503.03382} {arXiv:1503.03382 [hep-ph]} \BibitemShut
  {NoStop}%
\bibitem [{\citenamefont {Sandick}\ \emph {et~al.}(2016)\citenamefont
  {Sandick}, \citenamefont {Sinha},\ and\ \citenamefont
  {Teng}}]{Sandick:2016zut}%
  \BibitemOpen
  \bibfield  {author} {\bibinfo {author} {\bibfnamefont {P.}~\bibnamefont
  {Sandick}}, \bibinfo {author} {\bibfnamefont {K.}~\bibnamefont {Sinha}}, \
  and\ \bibinfo {author} {\bibfnamefont {F.}~\bibnamefont {Teng}},\ }\href
  {\doibase 10.1007/JHEP10(2016)018} {\bibfield  {journal} {\bibinfo  {journal}
  {JHEP}\ }\textbf {\bibinfo {volume} {10}},\ \bibinfo {pages} {018} (\bibinfo
  {year} {2016})},\ \Eprint {http://arxiv.org/abs/1608.00642} {arXiv:1608.00642
  [hep-ph]} \BibitemShut {NoStop}%
\bibitem [{\citenamefont {Kavanagh}\ \emph {et~al.}(2019)\citenamefont
  {Kavanagh}, \citenamefont {Panci},\ and\ \citenamefont
  {Ziegler}}]{Kavanagh:2018xeh}%
  \BibitemOpen
  \bibfield  {author} {\bibinfo {author} {\bibfnamefont {B.~J.}\ \bibnamefont
  {Kavanagh}}, \bibinfo {author} {\bibfnamefont {P.}~\bibnamefont {Panci}}, \
  and\ \bibinfo {author} {\bibfnamefont {R.}~\bibnamefont {Ziegler}},\ }\href
  {\doibase 10.1007/JHEP04(2019)089} {\bibfield  {journal} {\bibinfo  {journal}
  {JHEP}\ }\textbf {\bibinfo {volume} {04}},\ \bibinfo {pages} {089} (\bibinfo
  {year} {2019})},\ \Eprint {http://arxiv.org/abs/1810.00033} {arXiv:1810.00033
  [hep-ph]} \BibitemShut {NoStop}%
\bibitem [{\citenamefont {Trickle}\ \emph {et~al.}(2019)\citenamefont
  {Trickle}, \citenamefont {Zhang},\ and\ \citenamefont
  {Zurek}}]{Trickle:2019ovy}%
  \BibitemOpen
  \bibfield  {author} {\bibinfo {author} {\bibfnamefont {T.}~\bibnamefont
  {Trickle}}, \bibinfo {author} {\bibfnamefont {Z.}~\bibnamefont {Zhang}}, \
  and\ \bibinfo {author} {\bibfnamefont {K.~M.}\ \bibnamefont {Zurek}},\
  }\href@noop {} {\  (\bibinfo {year} {2019})},\ \Eprint
  {http://arxiv.org/abs/1905.13744} {arXiv:1905.13744 [hep-ph]} \BibitemShut
  {NoStop}%
\bibitem [{\citenamefont {Chu}\ \emph {et~al.}(2019{\natexlab{a}})\citenamefont
  {Chu}, \citenamefont {Pradler},\ and\ \citenamefont
  {Semmelrock}}]{Chu:2018qrm}%
  \BibitemOpen
  \bibfield  {author} {\bibinfo {author} {\bibfnamefont {X.}~\bibnamefont
  {Chu}}, \bibinfo {author} {\bibfnamefont {J.}~\bibnamefont {Pradler}}, \ and\
  \bibinfo {author} {\bibfnamefont {L.}~\bibnamefont {Semmelrock}},\ }\href
  {\doibase 10.1103/PhysRevD.99.015040} {\bibfield  {journal} {\bibinfo
  {journal} {Phys. Rev.}\ }\textbf {\bibinfo {volume} {D99}},\ \bibinfo {pages}
  {015040} (\bibinfo {year} {2019}{\natexlab{a}})},\ \Eprint
  {http://arxiv.org/abs/1811.04095} {arXiv:1811.04095 [hep-ph]} \BibitemShut
  {NoStop}%
\bibitem [{\citenamefont {Banerjee}\ \emph {et~al.}(2017)\citenamefont
  {Banerjee} \emph {et~al.}}]{Banerjee:2017hhz}%
  \BibitemOpen
  \bibfield  {author} {\bibinfo {author} {\bibfnamefont {D.}~\bibnamefont
  {Banerjee}} \emph {et~al.} (\bibinfo {collaboration} {NA64}),\ }\href@noop {}
  {\  (\bibinfo {year} {2017})},\ \Eprint {http://arxiv.org/abs/1710.00971}
  {arXiv:1710.00971 [hep-ex]} \BibitemShut {NoStop}%
\bibitem [{\citenamefont {Åkesson}\ \emph {et~al.}(2018)\citenamefont
  {Åkesson} \emph {et~al.}}]{Akesson:2018vlm}%
  \BibitemOpen
  \bibfield  {author} {\bibinfo {author} {\bibfnamefont {T.}~\bibnamefont
  {Åkesson}} \emph {et~al.} (\bibinfo {collaboration} {LDMX}),\ }\href@noop {}
  {\  (\bibinfo {year} {2018})},\ \Eprint {http://arxiv.org/abs/1808.05219}
  {arXiv:1808.05219 [hep-ex]} \BibitemShut {NoStop}%
\bibitem [{\citenamefont {Battaglieri}\ \emph {et~al.}(2016)\citenamefont
  {Battaglieri} \emph {et~al.}}]{Battaglieri:2016ggd}%
  \BibitemOpen
  \bibfield  {author} {\bibinfo {author} {\bibfnamefont {M.}~\bibnamefont
  {Battaglieri}} \emph {et~al.} (\bibinfo {collaboration} {BDX}),\ }\href@noop
  {} {\  (\bibinfo {year} {2016})},\ \Eprint {http://arxiv.org/abs/1607.01390}
  {arXiv:1607.01390 [hep-ex]} \BibitemShut {NoStop}%
\bibitem [{\citenamefont {Aubert}\ \emph {et~al.}(2002)\citenamefont {Aubert}
  \emph {et~al.}}]{Aubert:2001tu}%
  \BibitemOpen
  \bibfield  {author} {\bibinfo {author} {\bibfnamefont {B.}~\bibnamefont
  {Aubert}} \emph {et~al.} (\bibinfo {collaboration} {BaBar}),\ }\href
  {\doibase 10.1016/S0168-9002(01)02012-5} {\bibfield  {journal} {\bibinfo
  {journal} {Nucl. Instrum. Meth.}\ }\textbf {\bibinfo {volume} {A479}},\
  \bibinfo {pages} {1} (\bibinfo {year} {2002})},\ \Eprint
  {http://arxiv.org/abs/hep-ex/0105044} {arXiv:hep-ex/0105044 [hep-ex]}
  \BibitemShut {NoStop}%
\bibitem [{\citenamefont {Abe}\ \emph {et~al.}(2010)\citenamefont {Abe} \emph
  {et~al.}}]{Abe:2010gxa}%
  \BibitemOpen
  \bibfield  {author} {\bibinfo {author} {\bibfnamefont {T.}~\bibnamefont
  {Abe}} \emph {et~al.} (\bibinfo {collaboration} {Belle-II}),\ }\href@noop {}
  {\  (\bibinfo {year} {2010})},\ \Eprint {http://arxiv.org/abs/1011.0352}
  {arXiv:1011.0352 [physics.ins-det]} \BibitemShut {NoStop}%
\bibitem [{\citenamefont {Evans}\ and\ \citenamefont
  {Bryant}(2008)}]{Evans:2008zzb}%
  \BibitemOpen
  \bibfield  {author} {\bibinfo {author} {\bibfnamefont {L.}~\bibnamefont
  {Evans}}\ and\ \bibinfo {author} {\bibfnamefont {P.}~\bibnamefont {Bryant}},\
  }\href {\doibase 10.1088/1748-0221/3/08/S08001} {\bibfield  {journal}
  {\bibinfo  {journal} {JINST}\ }\textbf {\bibinfo {volume} {3}},\ \bibinfo
  {pages} {S08001} (\bibinfo {year} {2008})}\BibitemShut {NoStop}%
\bibitem [{\citenamefont {Aubert}\ \emph {et~al.}(2003)\citenamefont {Aubert}
  \emph {et~al.}}]{Aubert:2003yh}%
  \BibitemOpen
  \bibfield  {author} {\bibinfo {author} {\bibfnamefont {B.}~\bibnamefont
  {Aubert}} \emph {et~al.} (\bibinfo {collaboration} {BaBar}),\ }in\ \href
  {https://oraweb.slac.stanford.edu/pls/slacquery/BABAR_DOCUMENTS.Search?P_SLAC_PUB=SLAC-PUB-9710}
  {\emph {\bibinfo {booktitle} {{38th Rencontres de Moriond on Electroweak
  Interactions and Unified Theories Les Arcs, France, March 15-22, 2003}}}}\
  (\bibinfo {year} {2003})\ \Eprint {http://arxiv.org/abs/hep-ex/0304020}
  {arXiv:hep-ex/0304020 [hep-ex]} \BibitemShut {NoStop}%
\bibitem [{\citenamefont {Bird}\ \emph {et~al.}(2004)\citenamefont {Bird},
  \citenamefont {Jackson}, \citenamefont {Kowalewski},\ and\ \citenamefont
  {Pospelov}}]{Bird:2004ts}%
  \BibitemOpen
  \bibfield  {author} {\bibinfo {author} {\bibfnamefont {C.}~\bibnamefont
  {Bird}}, \bibinfo {author} {\bibfnamefont {P.}~\bibnamefont {Jackson}},
  \bibinfo {author} {\bibfnamefont {R.~V.}\ \bibnamefont {Kowalewski}}, \ and\
  \bibinfo {author} {\bibfnamefont {M.}~\bibnamefont {Pospelov}},\ }\href
  {\doibase 10.1103/PhysRevLett.93.201803} {\bibfield  {journal} {\bibinfo
  {journal} {Phys. Rev. Lett.}\ }\textbf {\bibinfo {volume} {93}},\ \bibinfo
  {pages} {201803} (\bibinfo {year} {2004})},\ \Eprint
  {http://arxiv.org/abs/hep-ph/0401195} {arXiv:hep-ph/0401195 [hep-ph]}
  \BibitemShut {NoStop}%
\bibitem [{\citenamefont {Anisimovsky}\ \emph {et~al.}(2004)\citenamefont
  {Anisimovsky} \emph {et~al.}}]{Anisimovsky:2004hr}%
  \BibitemOpen
  \bibfield  {author} {\bibinfo {author} {\bibfnamefont {V.~V.}\ \bibnamefont
  {Anisimovsky}} \emph {et~al.} (\bibinfo {collaboration} {E949}),\ }\href
  {\doibase 10.1103/PhysRevLett.93.031801} {\bibfield  {journal} {\bibinfo
  {journal} {Phys. Rev. Lett.}\ }\textbf {\bibinfo {volume} {93}},\ \bibinfo
  {pages} {031801} (\bibinfo {year} {2004})},\ \Eprint
  {http://arxiv.org/abs/hep-ex/0403036} {arXiv:hep-ex/0403036 [hep-ex]}
  \BibitemShut {NoStop}%
\bibitem [{\citenamefont {Artamonov}\ \emph {et~al.}(2009)\citenamefont
  {Artamonov} \emph {et~al.}}]{Artamonov:2009sz}%
  \BibitemOpen
  \bibfield  {author} {\bibinfo {author} {\bibfnamefont {A.~V.}\ \bibnamefont
  {Artamonov}} \emph {et~al.} (\bibinfo {collaboration} {BNL-E949}),\ }\href
  {\doibase 10.1103/PhysRevD.79.092004} {\bibfield  {journal} {\bibinfo
  {journal} {Phys. Rev.}\ }\textbf {\bibinfo {volume} {D79}},\ \bibinfo {pages}
  {092004} (\bibinfo {year} {2009})},\ \Eprint {http://arxiv.org/abs/0903.0030}
  {arXiv:0903.0030 [hep-ex]} \BibitemShut {NoStop}%
\bibitem [{\citenamefont {Tanabashi}\ \emph {et~al.}(2018)\citenamefont
  {Tanabashi} \emph {et~al.}}]{Tanabashi:2018oca}%
  \BibitemOpen
  \bibfield  {author} {\bibinfo {author} {\bibfnamefont {M.}~\bibnamefont
  {Tanabashi}} \emph {et~al.} (\bibinfo {collaboration} {Particle Data
  Group}),\ }\href {\doibase 10.1103/PhysRevD.98.030001} {\bibfield  {journal}
  {\bibinfo  {journal} {Phys. Rev.}\ }\textbf {\bibinfo {volume} {D98}},\
  \bibinfo {pages} {030001} (\bibinfo {year} {2018})}\BibitemShut {NoStop}%
\bibitem [{\citenamefont {Jegerlehner}\ and\ \citenamefont
  {Nyffeler}(2009)}]{Jegerlehner:2009ry}%
  \BibitemOpen
  \bibfield  {author} {\bibinfo {author} {\bibfnamefont {F.}~\bibnamefont
  {Jegerlehner}}\ and\ \bibinfo {author} {\bibfnamefont {A.}~\bibnamefont
  {Nyffeler}},\ }\href {\doibase 10.1016/j.physrep.2009.04.003} {\bibfield
  {journal} {\bibinfo  {journal} {Phys. Rept.}\ }\textbf {\bibinfo {volume}
  {477}},\ \bibinfo {pages} {1} (\bibinfo {year} {2009})},\ \Eprint
  {http://arxiv.org/abs/0902.3360} {arXiv:0902.3360 [hep-ph]} \BibitemShut
  {NoStop}%
\bibitem [{\citenamefont {Bennett}\ \emph {et~al.}(2006)\citenamefont {Bennett}
  \emph {et~al.}}]{Bennett:2006fi}%
  \BibitemOpen
  \bibfield  {author} {\bibinfo {author} {\bibfnamefont {G.~W.}\ \bibnamefont
  {Bennett}} \emph {et~al.} (\bibinfo {collaboration} {Muon g-2}),\ }\href
  {\doibase 10.1103/PhysRevD.73.072003} {\bibfield  {journal} {\bibinfo
  {journal} {Phys. Rev.}\ }\textbf {\bibinfo {volume} {D73}},\ \bibinfo {pages}
  {072003} (\bibinfo {year} {2006})},\ \Eprint
  {http://arxiv.org/abs/hep-ex/0602035} {arXiv:hep-ex/0602035 [hep-ex]}
  \BibitemShut {NoStop}%
\bibitem [{\citenamefont {Saito}(2012)}]{Saito:2012zz}%
  \BibitemOpen
  \bibfield  {author} {\bibinfo {author} {\bibfnamefont {N.}~\bibnamefont
  {Saito}} (\bibinfo {collaboration} {J-PARC g-'2/EDM}),\ }\bibfield
  {booktitle} {\emph {\bibinfo {booktitle} {{Proceedings, International
  Workshop on Grand Unified Theories (GUT2012): Kyoto, Japan, March 15-17,
  2012}}},\ }\href {\doibase 10.1063/1.4742078} {\bibfield  {journal} {\bibinfo
   {journal} {AIP Conf. Proc.}\ }\textbf {\bibinfo {volume} {1467}},\ \bibinfo
  {pages} {45} (\bibinfo {year} {2012})}\BibitemShut {NoStop}%
\bibitem [{\citenamefont {Grange}\ \emph {et~al.}(2015)\citenamefont {Grange}
  \emph {et~al.}}]{Grange:2015fou}%
  \BibitemOpen
  \bibfield  {author} {\bibinfo {author} {\bibfnamefont {J.}~\bibnamefont
  {Grange}} \emph {et~al.} (\bibinfo {collaboration} {Muon g-2}),\ }\href@noop
  {} {\  (\bibinfo {year} {2015})},\ \Eprint {http://arxiv.org/abs/1501.06858}
  {arXiv:1501.06858 [physics.ins-det]} \BibitemShut {NoStop}%
\bibitem [{\citenamefont {Sirlin}(1980)}]{Sirlin:1980nh}%
  \BibitemOpen
  \bibfield  {author} {\bibinfo {author} {\bibfnamefont {A.}~\bibnamefont
  {Sirlin}},\ }\href {\doibase 10.1103/PhysRevD.22.971} {\bibfield  {journal}
  {\bibinfo  {journal} {Phys. Rev.}\ }\textbf {\bibinfo {volume} {D22}},\
  \bibinfo {pages} {971} (\bibinfo {year} {1980})}\BibitemShut {NoStop}%
\bibitem [{\citenamefont {Chu}\ \emph {et~al.}(2019{\natexlab{b}})\citenamefont
  {Chu}, \citenamefont {Kuo}, \citenamefont {Pradler},\ and\ \citenamefont
  {Semmelrock}}]{Chu:2019rok}%
  \BibitemOpen
  \bibfield  {author} {\bibinfo {author} {\bibfnamefont {X.}~\bibnamefont
  {Chu}}, \bibinfo {author} {\bibfnamefont {J.-L.}\ \bibnamefont {Kuo}},
  \bibinfo {author} {\bibfnamefont {J.}~\bibnamefont {Pradler}}, \ and\
  \bibinfo {author} {\bibfnamefont {L.}~\bibnamefont {Semmelrock}},\ }\href
  {\doibase 10.1103/PhysRevD.100.083002} {\bibfield  {journal} {\bibinfo
  {journal} {Phys. Rev.}\ }\textbf {\bibinfo {volume} {D100}},\ \bibinfo
  {pages} {083002} (\bibinfo {year} {2019}{\natexlab{b}})},\ \Eprint
  {http://arxiv.org/abs/1908.00553} {arXiv:1908.00553 [hep-ph]} \BibitemShut
  {NoStop}%
\bibitem [{\citenamefont {Chang}\ \emph {et~al.}(2019)\citenamefont {Chang},
  \citenamefont {Essig},\ and\ \citenamefont {Reinert}}]{Chang:2019xva}%
  \BibitemOpen
  \bibfield  {author} {\bibinfo {author} {\bibfnamefont {J.~H.}\ \bibnamefont
  {Chang}}, \bibinfo {author} {\bibfnamefont {R.}~\bibnamefont {Essig}}, \ and\
  \bibinfo {author} {\bibfnamefont {A.}~\bibnamefont {Reinert}},\ }\href@noop
  {} {\  (\bibinfo {year} {2019})},\ \Eprint {http://arxiv.org/abs/1911.03389}
  {arXiv:1911.03389 [hep-ph]} \BibitemShut {NoStop}%
\bibitem [{\citenamefont {Aguilar-Arevalo}\ \emph
  {et~al.}(2018{\natexlab{b}})\citenamefont {Aguilar-Arevalo} \emph
  {et~al.}}]{Aguilar-Arevalo:2018wea}%
  \BibitemOpen
  \bibfield  {author} {\bibinfo {author} {\bibfnamefont {A.~A.}\ \bibnamefont
  {Aguilar-Arevalo}} \emph {et~al.} (\bibinfo {collaboration} {MiniBooNE DM}),\
  }\href {\doibase 10.1103/PhysRevD.98.112004} {\bibfield  {journal} {\bibinfo
  {journal} {Phys. Rev.}\ }\textbf {\bibinfo {volume} {D98}},\ \bibinfo {pages}
  {112004} (\bibinfo {year} {2018}{\natexlab{b}})},\ \Eprint
  {http://arxiv.org/abs/1807.06137} {arXiv:1807.06137 [hep-ex]} \BibitemShut
  {NoStop}%
\bibitem [{\citenamefont {De~Winter}\ \emph {et~al.}(1989)\citenamefont
  {De~Winter} \emph {et~al.}}]{DeWinter:1989zg}%
  \BibitemOpen
  \bibfield  {author} {\bibinfo {author} {\bibfnamefont {K.}~\bibnamefont
  {De~Winter}} \emph {et~al.} (\bibinfo {collaboration} {CHARM-II}),\ }\href
  {\doibase 10.1016/0168-9002(89)91190-X} {\bibfield  {journal} {\bibinfo
  {journal} {Nucl. Instrum. Meth.}\ }\textbf {\bibinfo {volume} {A278}},\
  \bibinfo {pages} {670} (\bibinfo {year} {1989})}\BibitemShut {NoStop}%
\bibitem [{\citenamefont {Vilain}\ \emph {et~al.}(1994)\citenamefont {Vilain}
  \emph {et~al.}}]{Vilain:1994qy}%
  \BibitemOpen
  \bibfield  {author} {\bibinfo {author} {\bibfnamefont {P.}~\bibnamefont
  {Vilain}} \emph {et~al.} (\bibinfo {collaboration} {CHARM-II}),\ }\href
  {\doibase 10.1016/0370-2693(94)91421-4} {\bibfield  {journal} {\bibinfo
  {journal} {Phys. Lett.}\ }\textbf {\bibinfo {volume} {B335}},\ \bibinfo
  {pages} {246} (\bibinfo {year} {1994})}\BibitemShut {NoStop}%
\bibitem [{\citenamefont {Ball}\ \emph {et~al.}(1980)\citenamefont {Ball} \emph
  {et~al.}}]{Ball:1981nu}%
  \BibitemOpen
  \bibfield  {author} {\bibinfo {author} {\bibfnamefont {R.}~\bibnamefont
  {Ball}} \emph {et~al.},\ }\bibfield  {booktitle} {\emph {\bibinfo {booktitle}
  {{Proceedings: Mini-Conference and Workshop on Neutrino Mass, Telemark,
  Wisconsin, Oct 2-4, 1980}}},\ }\href@noop {} {\bibfield  {journal} {\bibinfo
  {journal} {eConf}\ }\textbf {\bibinfo {volume} {C801002}},\ \bibinfo {pages}
  {172} (\bibinfo {year} {1980})}\BibitemShut {NoStop}%
\bibitem [{\citenamefont {Anelli}\ \emph {et~al.}(2015)\citenamefont {Anelli}
  \emph {et~al.}}]{Anelli:2015pba}%
  \BibitemOpen
  \bibfield  {author} {\bibinfo {author} {\bibfnamefont {M.}~\bibnamefont
  {Anelli}} \emph {et~al.} (\bibinfo {collaboration} {SHiP}),\ }\href@noop {}
  {\  (\bibinfo {year} {2015})},\ \Eprint {http://arxiv.org/abs/1504.04956}
  {arXiv:1504.04956 [physics.ins-det]} \BibitemShut {NoStop}%
\bibitem [{\citenamefont {Abi}\ \emph {et~al.}(2017)\citenamefont {Abi} \emph
  {et~al.}}]{Abi:2017aow}%
  \BibitemOpen
  \bibfield  {author} {\bibinfo {author} {\bibfnamefont {B.}~\bibnamefont
  {Abi}} \emph {et~al.} (\bibinfo {collaboration} {DUNE}),\ }\href@noop {} {\
  (\bibinfo {year} {2017})},\ \Eprint {http://arxiv.org/abs/1706.07081}
  {arXiv:1706.07081 [physics.ins-det]} \BibitemShut {NoStop}%
\bibitem [{\citenamefont {Bagnasco}\ \emph {et~al.}(1994)\citenamefont
  {Bagnasco}, \citenamefont {Dine},\ and\ \citenamefont
  {Thomas}}]{Bagnasco:1993st}%
  \BibitemOpen
  \bibfield  {author} {\bibinfo {author} {\bibfnamefont {J.}~\bibnamefont
  {Bagnasco}}, \bibinfo {author} {\bibfnamefont {M.}~\bibnamefont {Dine}}, \
  and\ \bibinfo {author} {\bibfnamefont {S.~D.}\ \bibnamefont {Thomas}},\
  }\href {\doibase 10.1016/0370-2693(94)90830-3} {\bibfield  {journal}
  {\bibinfo  {journal} {Phys. Lett.}\ }\textbf {\bibinfo {volume} {B320}},\
  \bibinfo {pages} {99} (\bibinfo {year} {1994})},\ \Eprint
  {http://arxiv.org/abs/hep-ph/9310290} {arXiv:hep-ph/9310290 [hep-ph]}
  \BibitemShut {NoStop}%
\bibitem [{\citenamefont {Foadi}\ \emph {et~al.}(2009)\citenamefont {Foadi},
  \citenamefont {Frandsen},\ and\ \citenamefont {Sannino}}]{Foadi:2008qv}%
  \BibitemOpen
  \bibfield  {author} {\bibinfo {author} {\bibfnamefont {R.}~\bibnamefont
  {Foadi}}, \bibinfo {author} {\bibfnamefont {M.~T.}\ \bibnamefont {Frandsen}},
  \ and\ \bibinfo {author} {\bibfnamefont {F.}~\bibnamefont {Sannino}},\ }\href
  {\doibase 10.1103/PhysRevD.80.037702} {\bibfield  {journal} {\bibinfo
  {journal} {Phys. Rev.}\ }\textbf {\bibinfo {volume} {D80}},\ \bibinfo {pages}
  {037702} (\bibinfo {year} {2009})},\ \Eprint {http://arxiv.org/abs/0812.3406}
  {arXiv:0812.3406 [hep-ph]} \BibitemShut {NoStop}%
\bibitem [{\citenamefont {Antipin}\ \emph {et~al.}(2015)\citenamefont
  {Antipin}, \citenamefont {Redi}, \citenamefont {Strumia},\ and\ \citenamefont
  {Vigiani}}]{Antipin:2015xia}%
  \BibitemOpen
  \bibfield  {author} {\bibinfo {author} {\bibfnamefont {O.}~\bibnamefont
  {Antipin}}, \bibinfo {author} {\bibfnamefont {M.}~\bibnamefont {Redi}},
  \bibinfo {author} {\bibfnamefont {A.}~\bibnamefont {Strumia}}, \ and\
  \bibinfo {author} {\bibfnamefont {E.}~\bibnamefont {Vigiani}},\ }\href
  {\doibase 10.1007/JHEP07(2015)039} {\bibfield  {journal} {\bibinfo  {journal}
  {JHEP}\ }\textbf {\bibinfo {volume} {07}},\ \bibinfo {pages} {039} (\bibinfo
  {year} {2015})},\ \Eprint {http://arxiv.org/abs/1503.08749} {arXiv:1503.08749
  [hep-ph]} \BibitemShut {NoStop}%
\bibitem [{\citenamefont {Alwall}\ \emph {et~al.}(2014)\citenamefont {Alwall},
  \citenamefont {Frederix}, \citenamefont {Frixione}, \citenamefont {Hirschi},
  \citenamefont {Maltoni}, \citenamefont {Mattelaer}, \citenamefont {Shao},
  \citenamefont {Stelzer}, \citenamefont {Torrielli},\ and\ \citenamefont
  {Zaro}}]{Alwall:2014hca}%
  \BibitemOpen
  \bibfield  {author} {\bibinfo {author} {\bibfnamefont {J.}~\bibnamefont
  {Alwall}}, \bibinfo {author} {\bibfnamefont {R.}~\bibnamefont {Frederix}},
  \bibinfo {author} {\bibfnamefont {S.}~\bibnamefont {Frixione}}, \bibinfo
  {author} {\bibfnamefont {V.}~\bibnamefont {Hirschi}}, \bibinfo {author}
  {\bibfnamefont {F.}~\bibnamefont {Maltoni}}, \bibinfo {author} {\bibfnamefont
  {O.}~\bibnamefont {Mattelaer}}, \bibinfo {author} {\bibfnamefont {H.~S.}\
  \bibnamefont {Shao}}, \bibinfo {author} {\bibfnamefont {T.}~\bibnamefont
  {Stelzer}}, \bibinfo {author} {\bibfnamefont {P.}~\bibnamefont {Torrielli}},
  \ and\ \bibinfo {author} {\bibfnamefont {M.}~\bibnamefont {Zaro}},\ }\href
  {\doibase 10.1007/JHEP07(2014)079} {\bibfield  {journal} {\bibinfo  {journal}
  {JHEP}\ }\textbf {\bibinfo {volume} {07}},\ \bibinfo {pages} {079} (\bibinfo
  {year} {2014})},\ \Eprint {http://arxiv.org/abs/1405.0301} {arXiv:1405.0301
  [hep-ph]} \BibitemShut {NoStop}%
\bibitem [{\citenamefont {Alekhin}\ \emph {et~al.}(2016)\citenamefont {Alekhin}
  \emph {et~al.}}]{Alekhin:2015byh}%
  \BibitemOpen
  \bibfield  {author} {\bibinfo {author} {\bibfnamefont {S.}~\bibnamefont
  {Alekhin}} \emph {et~al.},\ }\href {\doibase 10.1088/0034-4885/79/12/124201}
  {\bibfield  {journal} {\bibinfo  {journal} {Rept. Prog. Phys.}\ }\textbf
  {\bibinfo {volume} {79}},\ \bibinfo {pages} {124201} (\bibinfo {year}
  {2016})},\ \Eprint {http://arxiv.org/abs/1504.04855} {arXiv:1504.04855
  [hep-ph]} \BibitemShut {NoStop}%
\bibitem [{\citenamefont {Harnik}\ \emph {et~al.}(2019)\citenamefont {Harnik},
  \citenamefont {Liu},\ and\ \citenamefont {Palamara}}]{Harnik:2019zee}%
  \BibitemOpen
  \bibfield  {author} {\bibinfo {author} {\bibfnamefont {R.}~\bibnamefont
  {Harnik}}, \bibinfo {author} {\bibfnamefont {Z.}~\bibnamefont {Liu}}, \ and\
  \bibinfo {author} {\bibfnamefont {O.}~\bibnamefont {Palamara}},\ }\href
  {\doibase 10.1007/JHEP07(2019)170} {\bibfield  {journal} {\bibinfo  {journal}
  {JHEP}\ }\textbf {\bibinfo {volume} {07}},\ \bibinfo {pages} {170} (\bibinfo
  {year} {2019})},\ \Eprint {http://arxiv.org/abs/1902.03246} {arXiv:1902.03246
  [hep-ph]} \BibitemShut {NoStop}%
\bibitem [{\citenamefont {Hoferichter}\ \emph {et~al.}(2014)\citenamefont
  {Hoferichter}, \citenamefont {Kubis}, \citenamefont {Leupold}, \citenamefont
  {Niecknig},\ and\ \citenamefont {Schneider}}]{Hoferichter:2014vra}%
  \BibitemOpen
  \bibfield  {author} {\bibinfo {author} {\bibfnamefont {M.}~\bibnamefont
  {Hoferichter}}, \bibinfo {author} {\bibfnamefont {B.}~\bibnamefont {Kubis}},
  \bibinfo {author} {\bibfnamefont {S.}~\bibnamefont {Leupold}}, \bibinfo
  {author} {\bibfnamefont {F.}~\bibnamefont {Niecknig}}, \ and\ \bibinfo
  {author} {\bibfnamefont {S.~P.}\ \bibnamefont {Schneider}},\ }\href {\doibase
  10.1140/epjc/s10052-014-3180-0} {\bibfield  {journal} {\bibinfo  {journal}
  {Eur. Phys. J.}\ }\textbf {\bibinfo {volume} {C74}},\ \bibinfo {pages} {3180}
  (\bibinfo {year} {2014})},\ \Eprint {http://arxiv.org/abs/1410.4691}
  {arXiv:1410.4691 [hep-ph]} \BibitemShut {NoStop}%
\bibitem [{\citenamefont {Escribano}\ \emph {et~al.}(2015)\citenamefont
  {Escribano}, \citenamefont {Masjuan},\ and\ \citenamefont
  {Sanchez-Puertas}}]{Escribano:2015nra}%
  \BibitemOpen
  \bibfield  {author} {\bibinfo {author} {\bibfnamefont {R.}~\bibnamefont
  {Escribano}}, \bibinfo {author} {\bibfnamefont {P.}~\bibnamefont {Masjuan}},
  \ and\ \bibinfo {author} {\bibfnamefont {P.}~\bibnamefont
  {Sanchez-Puertas}},\ }\href {\doibase 10.1140/epjc/s10052-015-3642-z}
  {\bibfield  {journal} {\bibinfo  {journal} {Eur. Phys. J.}\ }\textbf
  {\bibinfo {volume} {C75}},\ \bibinfo {pages} {414} (\bibinfo {year}
  {2015})},\ \Eprint {http://arxiv.org/abs/1504.07742} {arXiv:1504.07742
  [hep-ph]} \BibitemShut {NoStop}%
\bibitem [{\citenamefont {Husek}(2019{\natexlab{a}})}]{Husek:2018inh}%
  \BibitemOpen
  \bibfield  {author} {\bibinfo {author} {\bibfnamefont {T.}~\bibnamefont
  {Husek}},\ }\bibfield  {booktitle} {\emph {\bibinfo {booktitle}
  {{Proceedings, 15th International Workshop on Meson Physics (MESON 2018):
  Kraków, Poland, June 7-12, 2018}}},\ }\href {\doibase
  10.1051/epjconf/201919902015} {\bibfield  {journal} {\bibinfo  {journal} {EPJ
  Web Conf.}\ }\textbf {\bibinfo {volume} {199}},\ \bibinfo {pages} {02015}
  (\bibinfo {year} {2019}{\natexlab{a}})},\ \Eprint
  {http://arxiv.org/abs/1811.12350} {arXiv:1811.12350 [hep-ph]} \BibitemShut
  {NoStop}%
\bibitem [{\citenamefont {deNiverville}\ \emph {et~al.}(2017)\citenamefont
  {deNiverville}, \citenamefont {Chen}, \citenamefont {Pospelov},\ and\
  \citenamefont {Ritz}}]{deNiverville:2016rqh}%
  \BibitemOpen
  \bibfield  {author} {\bibinfo {author} {\bibfnamefont {P.}~\bibnamefont
  {deNiverville}}, \bibinfo {author} {\bibfnamefont {C.-Y.}\ \bibnamefont
  {Chen}}, \bibinfo {author} {\bibfnamefont {M.}~\bibnamefont {Pospelov}}, \
  and\ \bibinfo {author} {\bibfnamefont {A.}~\bibnamefont {Ritz}},\ }\href
  {\doibase 10.1103/PhysRevD.95.035006} {\bibfield  {journal} {\bibinfo
  {journal} {Phys. Rev.}\ }\textbf {\bibinfo {volume} {D95}},\ \bibinfo {pages}
  {035006} (\bibinfo {year} {2017})},\ \Eprint
  {http://arxiv.org/abs/1609.01770} {arXiv:1609.01770 [hep-ph]} \BibitemShut
  {NoStop}%
\bibitem [{\citenamefont {Burman}\ and\ \citenamefont
  {Smith}(1989)}]{Burman:1989ds}%
  \BibitemOpen
  \bibfield  {author} {\bibinfo {author} {\bibfnamefont {R.~L.}\ \bibnamefont
  {Burman}}\ and\ \bibinfo {author} {\bibfnamefont {E.~S.}\ \bibnamefont
  {Smith}},\ }\href@noop {} {\  (\bibinfo {year} {1989})}\BibitemShut {NoStop}%
\bibitem [{\citenamefont {Aguilar-Arevalo}\ \emph
  {et~al.}(2009{\natexlab{a}})\citenamefont {Aguilar-Arevalo} \emph
  {et~al.}}]{AguilarArevalo:2008yp}%
  \BibitemOpen
  \bibfield  {author} {\bibinfo {author} {\bibfnamefont {A.~A.}\ \bibnamefont
  {Aguilar-Arevalo}} \emph {et~al.} (\bibinfo {collaboration} {MiniBooNE}),\
  }\href {\doibase 10.1103/PhysRevD.79.072002} {\bibfield  {journal} {\bibinfo
  {journal} {Phys. Rev.}\ }\textbf {\bibinfo {volume} {D79}},\ \bibinfo {pages}
  {072002} (\bibinfo {year} {2009}{\natexlab{a}})},\ \Eprint
  {http://arxiv.org/abs/0806.1449} {arXiv:0806.1449 [hep-ex]} \BibitemShut
  {NoStop}%
\bibitem [{\citenamefont {Bonesini}\ \emph {et~al.}(2001)\citenamefont
  {Bonesini}, \citenamefont {Marchionni}, \citenamefont {Pietropaolo},\ and\
  \citenamefont {Tabarelli~de Fatis}}]{Bonesini:2001iz}%
  \BibitemOpen
  \bibfield  {author} {\bibinfo {author} {\bibfnamefont {M.}~\bibnamefont
  {Bonesini}}, \bibinfo {author} {\bibfnamefont {A.}~\bibnamefont
  {Marchionni}}, \bibinfo {author} {\bibfnamefont {F.}~\bibnamefont
  {Pietropaolo}}, \ and\ \bibinfo {author} {\bibfnamefont {T.}~\bibnamefont
  {Tabarelli~de Fatis}},\ }\href {\doibase 10.1007/s100520100656} {\bibfield
  {journal} {\bibinfo  {journal} {Eur. Phys. J.}\ }\textbf {\bibinfo {volume}
  {C20}},\ \bibinfo {pages} {13} (\bibinfo {year} {2001})},\ \Eprint
  {http://arxiv.org/abs/hep-ph/0101163} {arXiv:hep-ph/0101163 [hep-ph]}
  \BibitemShut {NoStop}%
\bibitem [{\citenamefont {Sjostrand}\ \emph {et~al.}(2006)\citenamefont
  {Sjostrand}, \citenamefont {Mrenna},\ and\ \citenamefont
  {Skands}}]{Sjostrand:2006za}%
  \BibitemOpen
  \bibfield  {author} {\bibinfo {author} {\bibfnamefont {T.}~\bibnamefont
  {Sjostrand}}, \bibinfo {author} {\bibfnamefont {S.}~\bibnamefont {Mrenna}}, \
  and\ \bibinfo {author} {\bibfnamefont {P.~Z.}\ \bibnamefont {Skands}},\
  }\href {\doibase 10.1088/1126-6708/2006/05/026} {\bibfield  {journal}
  {\bibinfo  {journal} {JHEP}\ }\textbf {\bibinfo {volume} {05}},\ \bibinfo
  {pages} {026} (\bibinfo {year} {2006})},\ \Eprint
  {http://arxiv.org/abs/hep-ph/0603175} {arXiv:hep-ph/0603175 [hep-ph]}
  \BibitemShut {NoStop}%
\bibitem [{\citenamefont {Sjöstrand}\ \emph {et~al.}(2015)\citenamefont
  {Sjöstrand}, \citenamefont {Ask}, \citenamefont {Christiansen},
  \citenamefont {Corke}, \citenamefont {Desai}, \citenamefont {Ilten},
  \citenamefont {Mrenna}, \citenamefont {Prestel}, \citenamefont {Rasmussen},\
  and\ \citenamefont {Skands}}]{Sjostrand:2014zea}%
  \BibitemOpen
  \bibfield  {author} {\bibinfo {author} {\bibfnamefont {T.}~\bibnamefont
  {Sjöstrand}}, \bibinfo {author} {\bibfnamefont {S.}~\bibnamefont {Ask}},
  \bibinfo {author} {\bibfnamefont {J.~R.}\ \bibnamefont {Christiansen}},
  \bibinfo {author} {\bibfnamefont {R.}~\bibnamefont {Corke}}, \bibinfo
  {author} {\bibfnamefont {N.}~\bibnamefont {Desai}}, \bibinfo {author}
  {\bibfnamefont {P.}~\bibnamefont {Ilten}}, \bibinfo {author} {\bibfnamefont
  {S.}~\bibnamefont {Mrenna}}, \bibinfo {author} {\bibfnamefont
  {S.}~\bibnamefont {Prestel}}, \bibinfo {author} {\bibfnamefont {C.~O.}\
  \bibnamefont {Rasmussen}}, \ and\ \bibinfo {author} {\bibfnamefont {P.~Z.}\
  \bibnamefont {Skands}},\ }\href {\doibase 10.1016/j.cpc.2015.01.024}
  {\bibfield  {journal} {\bibinfo  {journal} {Comput. Phys. Commun.}\ }\textbf
  {\bibinfo {volume} {191}},\ \bibinfo {pages} {159} (\bibinfo {year}
  {2015})},\ \Eprint {http://arxiv.org/abs/1410.3012} {arXiv:1410.3012
  [hep-ph]} \BibitemShut {NoStop}%
\bibitem [{\citenamefont {Darmé}\ \emph {et~al.}(2020)\citenamefont {Darmé},
  \citenamefont {Ellis},\ and\ \citenamefont {You}}]{Darme:2020ral}%
  \BibitemOpen
  \bibfield  {author} {\bibinfo {author} {\bibfnamefont {L.}~\bibnamefont
  {Darmé}}, \bibinfo {author} {\bibfnamefont {S.~A.~R.}\ \bibnamefont
  {Ellis}}, \ and\ \bibinfo {author} {\bibfnamefont {T.}~\bibnamefont {You}},\
  }\href@noop {} {\  (\bibinfo {year} {2020})},\ \Eprint
  {http://arxiv.org/abs/2001.01490} {arXiv:2001.01490 [hep-ph]} \BibitemShut
  {NoStop}%
\bibitem [{\citenamefont {Krusche}(2005)}]{Krusche:2004xz}%
  \BibitemOpen
  \bibfield  {author} {\bibinfo {author} {\bibfnamefont {B.}~\bibnamefont
  {Krusche}},\ }\bibfield  {booktitle} {\emph {\bibinfo {booktitle} {{Lepton
  scattering and the structure of hadrons and nuclei. Proceedings,
  International School of nuclear physics, 26th Course, Erice, Italy, September
  16-24, 2004}}},\ }\href {\doibase 10.1016/j.ppnp.2004.12.002} {\bibfield
  {journal} {\bibinfo  {journal} {Prog. Part. Nucl. Phys.}\ }\textbf {\bibinfo
  {volume} {55}},\ \bibinfo {pages} {46} (\bibinfo {year} {2005})},\ \Eprint
  {http://arxiv.org/abs/nucl-ex/0411033} {arXiv:nucl-ex/0411033 [nucl-ex]}
  \BibitemShut {NoStop}%
\bibitem [{\citenamefont {Batell}\ \emph {et~al.}(2014)\citenamefont {Batell},
  \citenamefont {deNiverville}, \citenamefont {McKeen}, \citenamefont
  {Pospelov},\ and\ \citenamefont {Ritz}}]{Batell:2014yra}%
  \BibitemOpen
  \bibfield  {author} {\bibinfo {author} {\bibfnamefont {B.}~\bibnamefont
  {Batell}}, \bibinfo {author} {\bibfnamefont {P.}~\bibnamefont
  {deNiverville}}, \bibinfo {author} {\bibfnamefont {D.}~\bibnamefont
  {McKeen}}, \bibinfo {author} {\bibfnamefont {M.}~\bibnamefont {Pospelov}}, \
  and\ \bibinfo {author} {\bibfnamefont {A.}~\bibnamefont {Ritz}},\ }\href
  {\doibase 10.1103/PhysRevD.90.115014} {\bibfield  {journal} {\bibinfo
  {journal} {Phys. Rev.}\ }\textbf {\bibinfo {volume} {D90}},\ \bibinfo {pages}
  {115014} (\bibinfo {year} {2014})},\ \Eprint {http://arxiv.org/abs/1405.7049}
  {arXiv:1405.7049 [hep-ph]} \BibitemShut {NoStop}%
\bibitem [{\citenamefont {deNiverville}(8 30)}]{deNiverville:2016dkn}%
  \BibitemOpen
  \bibfield  {author} {\bibinfo {author} {\bibfnamefont {P.}~\bibnamefont
  {deNiverville}},\ }\emph {\bibinfo {title} {{Searching for hidden sector dark
  matter with fixed target neutrino experiments}}},\ \href
  {https://dspace.library.uvic.ca/handle/1828/7502} {Ph.D. thesis},\ \bibinfo
  {school} {U. Victoria (main)} (\bibinfo {year} {2016-08-30})\BibitemShut
  {NoStop}%
\bibitem [{CER(2016)}]{CERN-SHiP-NOTE-2016-004}%
  \BibitemOpen
  \href {https://cds.cern.ch/record/2214092} {\bibfield  {journal} {\bibinfo
  {journal} {SHiP Collaboration}\ } (\bibinfo {year} {2016})},\ \Eprint
  {http://arxiv.org/abs/CERN-SHiP-NOTE-2016-004} {CERN-SHiP-NOTE-2016-004}
  \BibitemShut {NoStop}%
\bibitem [{\citenamefont {deNiverville}\ and\ \citenamefont
  {Frugiuele}(2019)}]{deNiverville:2018dbu}%
  \BibitemOpen
  \bibfield  {author} {\bibinfo {author} {\bibfnamefont {P.}~\bibnamefont
  {deNiverville}}\ and\ \bibinfo {author} {\bibfnamefont {C.}~\bibnamefont
  {Frugiuele}},\ }\href {\doibase 10.1103/PhysRevD.99.051701} {\bibfield
  {journal} {\bibinfo  {journal} {Phys. Rev.}\ }\textbf {\bibinfo {volume}
  {D99}},\ \bibinfo {pages} {051701} (\bibinfo {year} {2019})},\ \Eprint
  {http://arxiv.org/abs/1807.06501} {arXiv:1807.06501 [hep-ph]} \BibitemShut
  {NoStop}%
\bibitem [{\citenamefont {De~Romeri}\ \emph {et~al.}(2019)\citenamefont
  {De~Romeri}, \citenamefont {Kelly},\ and\ \citenamefont
  {Machado}}]{DeRomeri:2019kic}%
  \BibitemOpen
  \bibfield  {author} {\bibinfo {author} {\bibfnamefont {V.}~\bibnamefont
  {De~Romeri}}, \bibinfo {author} {\bibfnamefont {K.~J.}\ \bibnamefont
  {Kelly}}, \ and\ \bibinfo {author} {\bibfnamefont {P.~A.~N.}\ \bibnamefont
  {Machado}},\ }\href {\doibase 10.1103/PhysRevD.100.095010} {\bibfield
  {journal} {\bibinfo  {journal} {Phys. Rev.}\ }\textbf {\bibinfo {volume}
  {D100}},\ \bibinfo {pages} {095010} (\bibinfo {year} {2019})},\ \Eprint
  {http://arxiv.org/abs/1903.10505} {arXiv:1903.10505 [hep-ph]} \BibitemShut
  {NoStop}%
\bibitem [{\citenamefont {Döbrich}\ \emph {et~al.}(2019)\citenamefont
  {Döbrich}, \citenamefont {Jaeckel},\ and\ \citenamefont
  {Spadaro}}]{Dobrich:2019dxc}%
  \BibitemOpen
  \bibfield  {author} {\bibinfo {author} {\bibfnamefont {B.}~\bibnamefont
  {Döbrich}}, \bibinfo {author} {\bibfnamefont {J.}~\bibnamefont {Jaeckel}}, \
  and\ \bibinfo {author} {\bibfnamefont {T.}~\bibnamefont {Spadaro}},\ }\href
  {\doibase 10.1007/JHEP05(2019)213} {\bibfield  {journal} {\bibinfo  {journal}
  {JHEP}\ }\textbf {\bibinfo {volume} {05}},\ \bibinfo {pages} {213} (\bibinfo
  {year} {2019})},\ \Eprint {http://arxiv.org/abs/1904.02091} {arXiv:1904.02091
  [hep-ph]} \BibitemShut {NoStop}%
\bibitem [{\citenamefont {Fermi}(1924)}]{Fermi1924}%
  \BibitemOpen
  \bibfield  {author} {\bibinfo {author} {\bibfnamefont {E.}~\bibnamefont
  {Fermi}},\ }\href@noop {} {\bibfield  {journal} {\bibinfo  {journal} {Z.
  Phys.}\ }\textbf {\bibinfo {volume} {29}},\ \bibinfo {pages} {315} (\bibinfo
  {year} {1924})}\BibitemShut {NoStop}%
\bibitem [{\citenamefont {von Weizsacker}(1934)}]{vonWeizsacker:1934nji}%
  \BibitemOpen
  \bibfield  {author} {\bibinfo {author} {\bibfnamefont {C.~F.}\ \bibnamefont
  {von Weizsacker}},\ }\href {\doibase 10.1007/BF01333110} {\bibfield
  {journal} {\bibinfo  {journal} {Z. Phys.}\ }\textbf {\bibinfo {volume}
  {88}},\ \bibinfo {pages} {612} (\bibinfo {year} {1934})}\BibitemShut
  {NoStop}%
\bibitem [{\citenamefont {Williams}(1934)}]{Williams1934}%
  \BibitemOpen
  \bibfield  {author} {\bibinfo {author} {\bibfnamefont {E.~J.}\ \bibnamefont
  {Williams}},\ }\href@noop {} {\bibfield  {journal} {\bibinfo  {journal}
  {Phys. Rev.}\ }\textbf {\bibinfo {volume} {45}} (\bibinfo {year}
  {1934})}\BibitemShut {NoStop}%
\bibitem [{\citenamefont {Blümlein}\ and\ \citenamefont
  {Brunner}(2014)}]{Blumlein:2013cua}%
  \BibitemOpen
  \bibfield  {author} {\bibinfo {author} {\bibfnamefont {J.}~\bibnamefont
  {Blümlein}}\ and\ \bibinfo {author} {\bibfnamefont {J.}~\bibnamefont
  {Brunner}},\ }\href {\doibase 10.1016/j.physletb.2014.02.029} {\bibfield
  {journal} {\bibinfo  {journal} {Phys. Lett.}\ }\textbf {\bibinfo {volume}
  {B731}},\ \bibinfo {pages} {320} (\bibinfo {year} {2014})},\ \Eprint
  {http://arxiv.org/abs/1311.3870} {arXiv:1311.3870 [hep-ph]} \BibitemShut
  {NoStop}%
\bibitem [{\citenamefont {Feng}\ \emph {et~al.}(2018)\citenamefont {Feng},
  \citenamefont {Galon}, \citenamefont {Kling},\ and\ \citenamefont
  {Trojanowski}}]{Feng:2017uoz}%
  \BibitemOpen
  \bibfield  {author} {\bibinfo {author} {\bibfnamefont {J.~L.}\ \bibnamefont
  {Feng}}, \bibinfo {author} {\bibfnamefont {I.}~\bibnamefont {Galon}},
  \bibinfo {author} {\bibfnamefont {F.}~\bibnamefont {Kling}}, \ and\ \bibinfo
  {author} {\bibfnamefont {S.}~\bibnamefont {Trojanowski}},\ }\href {\doibase
  10.1103/PhysRevD.97.035001} {\bibfield  {journal} {\bibinfo  {journal} {Phys.
  Rev.}\ }\textbf {\bibinfo {volume} {D97}},\ \bibinfo {pages} {035001}
  (\bibinfo {year} {2018})},\ \Eprint {http://arxiv.org/abs/1708.09389}
  {arXiv:1708.09389 [hep-ph]} \BibitemShut {NoStop}%
\bibitem [{\citenamefont {Tsai}\ \emph {et~al.}(2019)\citenamefont {Tsai},
  \citenamefont {deNiverville},\ and\ \citenamefont {Liu}}]{Tsai:2019mtm}%
  \BibitemOpen
  \bibfield  {author} {\bibinfo {author} {\bibfnamefont {Y.-D.}\ \bibnamefont
  {Tsai}}, \bibinfo {author} {\bibfnamefont {P.}~\bibnamefont {deNiverville}},
  \ and\ \bibinfo {author} {\bibfnamefont {M.~X.}\ \bibnamefont {Liu}},\
  }\href@noop {} {\  (\bibinfo {year} {2019})},\ \Eprint
  {http://arxiv.org/abs/1908.07525} {arXiv:1908.07525 [hep-ph]} \BibitemShut
  {NoStop}%
\bibitem [{\citenamefont {Faessler}\ \emph {et~al.}(2010)\citenamefont
  {Faessler}, \citenamefont {Krivoruchenko},\ and\ \citenamefont
  {Martemyanov}}]{Faessler:2009tn}%
  \BibitemOpen
  \bibfield  {author} {\bibinfo {author} {\bibfnamefont {A.}~\bibnamefont
  {Faessler}}, \bibinfo {author} {\bibfnamefont {M.~I.}\ \bibnamefont
  {Krivoruchenko}}, \ and\ \bibinfo {author} {\bibfnamefont {B.~V.}\
  \bibnamefont {Martemyanov}},\ }\href {\doibase 10.1103/PhysRevC.82.038201}
  {\bibfield  {journal} {\bibinfo  {journal} {Phys. Rev.}\ }\textbf {\bibinfo
  {volume} {C82}},\ \bibinfo {pages} {038201} (\bibinfo {year} {2010})},\
  \Eprint {http://arxiv.org/abs/0910.5589} {arXiv:0910.5589 [hep-ph]}
  \BibitemShut {NoStop}%
\bibitem [{\citenamefont {Coloma}\ \emph {et~al.}(2016)\citenamefont {Coloma},
  \citenamefont {Dobrescu}, \citenamefont {Frugiuele},\ and\ \citenamefont
  {Harnik}}]{Coloma:2015pih}%
  \BibitemOpen
  \bibfield  {author} {\bibinfo {author} {\bibfnamefont {P.}~\bibnamefont
  {Coloma}}, \bibinfo {author} {\bibfnamefont {B.~A.}\ \bibnamefont
  {Dobrescu}}, \bibinfo {author} {\bibfnamefont {C.}~\bibnamefont {Frugiuele}},
  \ and\ \bibinfo {author} {\bibfnamefont {R.}~\bibnamefont {Harnik}},\ }\href
  {\doibase 10.1007/JHEP04(2016)047} {\bibfield  {journal} {\bibinfo  {journal}
  {JHEP}\ }\textbf {\bibinfo {volume} {04}},\ \bibinfo {pages} {047} (\bibinfo
  {year} {2016})},\ \Eprint {http://arxiv.org/abs/1512.03852} {arXiv:1512.03852
  [hep-ph]} \BibitemShut {NoStop}%
\bibitem [{\citenamefont {Frugiuele}(2017)}]{Frugiuele:2017zvx}%
  \BibitemOpen
  \bibfield  {author} {\bibinfo {author} {\bibfnamefont {C.}~\bibnamefont
  {Frugiuele}},\ }\href {\doibase 10.1103/PhysRevD.96.015029} {\bibfield
  {journal} {\bibinfo  {journal} {Phys. Rev.}\ }\textbf {\bibinfo {volume}
  {D96}},\ \bibinfo {pages} {015029} (\bibinfo {year} {2017})},\ \Eprint
  {http://arxiv.org/abs/1701.05464} {arXiv:1701.05464 [hep-ph]} \BibitemShut
  {NoStop}%
\bibitem [{\citenamefont {de~Gouvêa}\ \emph {et~al.}(2019)\citenamefont
  {de~Gouvêa}, \citenamefont {Fox}, \citenamefont {Harnik}, \citenamefont
  {Kelly},\ and\ \citenamefont {Zhang}}]{deGouvea:2018cfv}%
  \BibitemOpen
  \bibfield  {author} {\bibinfo {author} {\bibfnamefont {A.}~\bibnamefont
  {de~Gouvêa}}, \bibinfo {author} {\bibfnamefont {P.~J.}\ \bibnamefont {Fox}},
  \bibinfo {author} {\bibfnamefont {R.}~\bibnamefont {Harnik}}, \bibinfo
  {author} {\bibfnamefont {K.~J.}\ \bibnamefont {Kelly}}, \ and\ \bibinfo
  {author} {\bibfnamefont {Y.}~\bibnamefont {Zhang}},\ }\href {\doibase
  10.1007/JHEP01(2019)001} {\bibfield  {journal} {\bibinfo  {journal} {JHEP}\
  }\textbf {\bibinfo {volume} {01}},\ \bibinfo {pages} {001} (\bibinfo {year}
  {2019})},\ \Eprint {http://arxiv.org/abs/1809.06388} {arXiv:1809.06388
  [hep-ph]} \BibitemShut {NoStop}%
\bibitem [{\citenamefont {Hirai}\ \emph {et~al.}(2007)\citenamefont {Hirai},
  \citenamefont {Kumano},\ and\ \citenamefont {Nagai}}]{Hirai:2007sx}%
  \BibitemOpen
  \bibfield  {author} {\bibinfo {author} {\bibfnamefont {M.}~\bibnamefont
  {Hirai}}, \bibinfo {author} {\bibfnamefont {S.}~\bibnamefont {Kumano}}, \
  and\ \bibinfo {author} {\bibfnamefont {T.~H.}\ \bibnamefont {Nagai}},\ }\href
  {\doibase 10.1103/PhysRevC.76.065207} {\bibfield  {journal} {\bibinfo
  {journal} {Phys. Rev.}\ }\textbf {\bibinfo {volume} {C76}},\ \bibinfo {pages}
  {065207} (\bibinfo {year} {2007})},\ \Eprint {http://arxiv.org/abs/0709.3038}
  {arXiv:0709.3038 [hep-ph]} \BibitemShut {NoStop}%
\bibitem [{\citenamefont {Brown}(2018)}]{Brown:2018rcz}%
  \BibitemOpen
  \bibfield  {author} {\bibinfo {author} {\bibfnamefont {G.~R.}\ \bibnamefont
  {Brown}},\ }\emph {\bibinfo {title} {{Sensitivity Study for Low Mass Dark
  Matter Search at DUNE}}},\ \href {\doibase 10.2172/1462086} {Master's
  thesis},\ \bibinfo  {school} {Texas U., Arlington} (\bibinfo {year}
  {2018})\BibitemShut {NoStop}%
\bibitem [{\citenamefont {Hostert}(2019)}]{Hostert:2019iia}%
  \BibitemOpen
  \bibfield  {author} {\bibinfo {author} {\bibfnamefont {M.}~\bibnamefont
  {Hostert}},\ }\emph {\bibinfo {title} {{Hidden Physics at the Neutrino
  Frontier: Tridents, Dark Forces, and Hidden Particles}}},\ \href
  {http://etheses.dur.ac.uk/13289/} {Ph.D. thesis},\ \bibinfo  {school} {Durham
  U.} (\bibinfo {year} {2019})\BibitemShut {NoStop}%
\bibitem [{\citenamefont {Buonocore}\ \emph
  {et~al.}(2019{\natexlab{a}})\citenamefont {Buonocore}, \citenamefont
  {Frugiuele}, \citenamefont {Maltoni}, \citenamefont {Mattelaer},\ and\
  \citenamefont {Tramontano}}]{Buonocore:2018xjk}%
  \BibitemOpen
  \bibfield  {author} {\bibinfo {author} {\bibfnamefont {L.}~\bibnamefont
  {Buonocore}}, \bibinfo {author} {\bibfnamefont {C.}~\bibnamefont
  {Frugiuele}}, \bibinfo {author} {\bibfnamefont {F.}~\bibnamefont {Maltoni}},
  \bibinfo {author} {\bibfnamefont {O.}~\bibnamefont {Mattelaer}}, \ and\
  \bibinfo {author} {\bibfnamefont {F.}~\bibnamefont {Tramontano}},\ }\href
  {\doibase 10.1007/JHEP05(2019)028} {\bibfield  {journal} {\bibinfo  {journal}
  {JHEP}\ }\textbf {\bibinfo {volume} {05}},\ \bibinfo {pages} {028} (\bibinfo
  {year} {2019}{\natexlab{a}})},\ \Eprint {http://arxiv.org/abs/1812.06771}
  {arXiv:1812.06771 [hep-ph]} \BibitemShut {NoStop}%
\bibitem [{\citenamefont {Mills}(2001)}]{Mills:2001wvh}%
  \BibitemOpen
  \bibfield  {author} {\bibinfo {author} {\bibfnamefont {G.~B.}\ \bibnamefont
  {Mills}} (\bibinfo {collaboration} {LSND}),\ }in\ \href@noop {} {\emph
  {\bibinfo {booktitle} {{Proceedings, 34th Rencontres de Moriond on
  Electroweak Interactions and Unified Theories: Les Arcs, France, Mar 13-20,
  1999}}}},\ \bibinfo {organization} {The Gioi}\ (\bibinfo  {publisher} {The
  Gioi},\ \bibinfo {address} {Hanoi},\ \bibinfo {year} {2001})\ pp.\ \bibinfo
  {pages} {41--52}\BibitemShut {NoStop}%
\bibitem [{\citenamefont {Shimizu}\ \emph {et~al.}(1982)\citenamefont
  {Shimizu}, \citenamefont {Kubota}, \citenamefont {Koiso}, \citenamefont
  {Sai}, \citenamefont {Sakamoto},\ and\ \citenamefont
  {Yamamoto}}]{Shimizu:1982dx}%
  \BibitemOpen
  \bibfield  {author} {\bibinfo {author} {\bibfnamefont {F.}~\bibnamefont
  {Shimizu}}, \bibinfo {author} {\bibfnamefont {Y.}~\bibnamefont {Kubota}},
  \bibinfo {author} {\bibfnamefont {H.}~\bibnamefont {Koiso}}, \bibinfo
  {author} {\bibfnamefont {F.}~\bibnamefont {Sai}}, \bibinfo {author}
  {\bibfnamefont {S.}~\bibnamefont {Sakamoto}}, \ and\ \bibinfo {author}
  {\bibfnamefont {S.~S.}\ \bibnamefont {Yamamoto}},\ }\href {\doibase
  10.1016/0375-9474(82)90037-9} {\bibfield  {journal} {\bibinfo  {journal}
  {Nucl. Phys.}\ }\textbf {\bibinfo {volume} {A386}},\ \bibinfo {pages} {571}
  (\bibinfo {year} {1982})}\BibitemShut {NoStop}%
\bibitem [{\citenamefont {Achilli}\ \emph {et~al.}(2011)\citenamefont
  {Achilli}, \citenamefont {Godbole}, \citenamefont {Grau}, \citenamefont
  {Pancheri}, \citenamefont {Shekhovtsova},\ and\ \citenamefont
  {Srivastava}}]{Achilli:2011sw}%
  \BibitemOpen
  \bibfield  {author} {\bibinfo {author} {\bibfnamefont {A.}~\bibnamefont
  {Achilli}}, \bibinfo {author} {\bibfnamefont {R.~M.}\ \bibnamefont
  {Godbole}}, \bibinfo {author} {\bibfnamefont {A.}~\bibnamefont {Grau}},
  \bibinfo {author} {\bibfnamefont {G.}~\bibnamefont {Pancheri}}, \bibinfo
  {author} {\bibfnamefont {O.}~\bibnamefont {Shekhovtsova}}, \ and\ \bibinfo
  {author} {\bibfnamefont {Y.~N.}\ \bibnamefont {Srivastava}},\ }\href
  {\doibase 10.1103/PhysRevD.84.094009} {\bibfield  {journal} {\bibinfo
  {journal} {Phys. Rev.}\ }\textbf {\bibinfo {volume} {D84}},\ \bibinfo {pages}
  {094009} (\bibinfo {year} {2011})},\ \Eprint {http://arxiv.org/abs/1102.1949}
  {arXiv:1102.1949 [hep-ph]} \BibitemShut {NoStop}%
\bibitem [{\citenamefont {Allen}\ \emph {et~al.}(1989)\citenamefont {Allen},
  \citenamefont {Chen}, \citenamefont {Potter}, \citenamefont {Burman},
  \citenamefont {Donahue}, \citenamefont {Krakauer}, \citenamefont {Talaga},
  \citenamefont {Smith},\ and\ \citenamefont {Dodd}}]{Allen:1989dt}%
  \BibitemOpen
  \bibfield  {author} {\bibinfo {author} {\bibfnamefont {R.~C.}\ \bibnamefont
  {Allen}}, \bibinfo {author} {\bibfnamefont {H.~H.}\ \bibnamefont {Chen}},
  \bibinfo {author} {\bibfnamefont {M.~E.}\ \bibnamefont {Potter}}, \bibinfo
  {author} {\bibfnamefont {R.~L.}\ \bibnamefont {Burman}}, \bibinfo {author}
  {\bibfnamefont {J.~B.}\ \bibnamefont {Donahue}}, \bibinfo {author}
  {\bibfnamefont {D.~A.}\ \bibnamefont {Krakauer}}, \bibinfo {author}
  {\bibfnamefont {R.~L.}\ \bibnamefont {Talaga}}, \bibinfo {author}
  {\bibfnamefont {E.~S.}\ \bibnamefont {Smith}}, \ and\ \bibinfo {author}
  {\bibfnamefont {A.~C.}\ \bibnamefont {Dodd}},\ }\href {\doibase
  10.1016/0168-9002(89)90300-8} {\bibfield  {journal} {\bibinfo  {journal}
  {Nucl. Instrum. Meth.}\ }\textbf {\bibinfo {volume} {A284}},\ \bibinfo
  {pages} {347} (\bibinfo {year} {1989})}\BibitemShut {NoStop}%
\bibitem [{\citenamefont {Akimov}\ \emph {et~al.}(2019)\citenamefont {Akimov}
  \emph {et~al.}}]{Akimov:2019xdj}%
  \BibitemOpen
  \bibfield  {author} {\bibinfo {author} {\bibfnamefont {D.}~\bibnamefont
  {Akimov}} \emph {et~al.} (\bibinfo {collaboration} {COHERENT}),\ }\href@noop
  {} {\  (\bibinfo {year} {2019})},\ \Eprint {http://arxiv.org/abs/1911.06422}
  {arXiv:1911.06422 [hep-ex]} \BibitemShut {NoStop}%
\bibitem [{\citenamefont {Auerbach}\ \emph {et~al.}(2001)\citenamefont
  {Auerbach} \emph {et~al.}}]{Auerbach:2001wg}%
  \BibitemOpen
  \bibfield  {author} {\bibinfo {author} {\bibfnamefont {L.~B.}\ \bibnamefont
  {Auerbach}} \emph {et~al.} (\bibinfo {collaboration} {LSND}),\ }\href
  {\doibase 10.1103/PhysRevD.63.112001} {\bibfield  {journal} {\bibinfo
  {journal} {Phys. Rev.}\ }\textbf {\bibinfo {volume} {D63}},\ \bibinfo {pages}
  {112001} (\bibinfo {year} {2001})},\ \Eprint
  {http://arxiv.org/abs/hep-ex/0101039} {arXiv:hep-ex/0101039 [hep-ex]}
  \BibitemShut {NoStop}%
\bibitem [{\citenamefont {Aguilar-Arevalo}\ \emph
  {et~al.}(2009{\natexlab{b}})\citenamefont {Aguilar-Arevalo} \emph
  {et~al.}}]{AguilarArevalo:2008qa}%
  \BibitemOpen
  \bibfield  {author} {\bibinfo {author} {\bibfnamefont {A.~A.}\ \bibnamefont
  {Aguilar-Arevalo}} \emph {et~al.} (\bibinfo {collaboration} {MiniBooNE}),\
  }\href {\doibase 10.1016/j.nima.2008.10.028} {\bibfield  {journal} {\bibinfo
  {journal} {Nucl. Instrum. Meth.}\ }\textbf {\bibinfo {volume} {A599}},\
  \bibinfo {pages} {28} (\bibinfo {year} {2009}{\natexlab{b}})},\ \Eprint
  {http://arxiv.org/abs/0806.4201} {arXiv:0806.4201 [hep-ex]} \BibitemShut
  {NoStop}%
\bibitem [{\citenamefont {Dharmapalan}\ \emph {et~al.}(2012)\citenamefont
  {Dharmapalan} \emph {et~al.}}]{Dharmapalan:2012xp}%
  \BibitemOpen
  \bibfield  {author} {\bibinfo {author} {\bibfnamefont {R.}~\bibnamefont
  {Dharmapalan}} \emph {et~al.} (\bibinfo {collaboration} {MiniBooNE}),\
  }\href@noop {} {\  (\bibinfo {year} {2012})},\ \Eprint
  {http://arxiv.org/abs/1211.2258} {arXiv:1211.2258 [hep-ex]} \BibitemShut
  {NoStop}%
\bibitem [{\citenamefont {collaboration}(2019)}]{Ahdida:2704147}%
  \BibitemOpen
  \bibfield  {author} {\bibinfo {author} {\bibfnamefont {S.}~\bibnamefont
  {collaboration}} (\bibinfo {collaboration} {SHiP Collaboration}),\ }\href
  {https://cds.cern.ch/record/2704147} {\emph {\bibinfo {title} {{SHiP
  Experiment - Comprehensive Design Study report}}}},\ \bibinfo {type} {Tech.
  Rep.}\ \bibinfo {number} {CERN-SPSC-2019-049. SPSC-SR-263}\ (\bibinfo
  {institution} {CERN},\ \bibinfo {address} {Geneva},\ \bibinfo {year}
  {2019})\BibitemShut {NoStop}%
\bibitem [{\citenamefont {Jodłowski}\ \emph {et~al.}(2019)\citenamefont
  {Jodłowski}, \citenamefont {Kling}, \citenamefont {Roszkowski},\ and\
  \citenamefont {Trojanowski}}]{Jodlowski:2019ycu}%
  \BibitemOpen
  \bibfield  {author} {\bibinfo {author} {\bibfnamefont {K.}~\bibnamefont
  {Jodłowski}}, \bibinfo {author} {\bibfnamefont {F.}~\bibnamefont {Kling}},
  \bibinfo {author} {\bibfnamefont {L.}~\bibnamefont {Roszkowski}}, \ and\
  \bibinfo {author} {\bibfnamefont {S.}~\bibnamefont {Trojanowski}},\
  }\href@noop {} {\  (\bibinfo {year} {2019})},\ \Eprint
  {http://arxiv.org/abs/1911.11346} {arXiv:1911.11346 [hep-ph]} \BibitemShut
  {NoStop}%
\bibitem [{\citenamefont {Mohanty}\ and\ \citenamefont
  {Rao}(2015)}]{Mohanty:2015koa}%
  \BibitemOpen
  \bibfield  {author} {\bibinfo {author} {\bibfnamefont {S.}~\bibnamefont
  {Mohanty}}\ and\ \bibinfo {author} {\bibfnamefont {S.}~\bibnamefont {Rao}},\
  }\href@noop {} {\  (\bibinfo {year} {2015})},\ \Eprint
  {http://arxiv.org/abs/1506.06462} {arXiv:1506.06462 [hep-ph]} \BibitemShut
  {NoStop}%
\bibitem [{\citenamefont {Ajimura}\ \emph {et~al.}(2017)\citenamefont {Ajimura}
  \emph {et~al.}}]{Ajimura:2017fld}%
  \BibitemOpen
  \bibfield  {author} {\bibinfo {author} {\bibfnamefont {S.}~\bibnamefont
  {Ajimura}} \emph {et~al.},\ }\href@noop {} {\  (\bibinfo {year} {2017})},\
  \Eprint {http://arxiv.org/abs/1705.08629} {arXiv:1705.08629
  [physics.ins-det]} \BibitemShut {NoStop}%
\bibitem [{\citenamefont {Adamson}\ \emph {et~al.}(2017)\citenamefont {Adamson}
  \emph {et~al.}}]{Adamson:2017qqn}%
  \BibitemOpen
  \bibfield  {author} {\bibinfo {author} {\bibfnamefont {P.}~\bibnamefont
  {Adamson}} \emph {et~al.} (\bibinfo {collaboration} {NOvA}),\ }\href
  {\doibase 10.1103/PhysRevLett.118.151802} {\bibfield  {journal} {\bibinfo
  {journal} {Phys. Rev. Lett.}\ }\textbf {\bibinfo {volume} {118}},\ \bibinfo
  {pages} {151802} (\bibinfo {year} {2017})},\ \Eprint
  {http://arxiv.org/abs/1701.05891} {arXiv:1701.05891 [hep-ex]} \BibitemShut
  {NoStop}%
\bibitem [{\citenamefont {Talebzadeh}\ \emph {et~al.}(1987)\citenamefont
  {Talebzadeh} \emph {et~al.}}]{Talebzadeh:1987rq}%
  \BibitemOpen
  \bibfield  {author} {\bibinfo {author} {\bibfnamefont {M.}~\bibnamefont
  {Talebzadeh}} \emph {et~al.} (\bibinfo {collaboration} {BEBC WA66}),\ }\href
  {\doibase 10.1016/0550-3213(87)90482-2} {\bibfield  {journal} {\bibinfo
  {journal} {Nucl. Phys.}\ }\textbf {\bibinfo {volume} {B291}},\ \bibinfo
  {pages} {503} (\bibinfo {year} {1987})}\BibitemShut {NoStop}%
\bibitem [{\citenamefont {Cooper-Sarkar}\ \emph {et~al.}(1992)\citenamefont
  {Cooper-Sarkar}, \citenamefont {Sarkar}, \citenamefont {Guy}, \citenamefont
  {Venus}, \citenamefont {Hulth},\ and\ \citenamefont
  {Hultqvist}}]{CooperSarkar:1991xz}%
  \BibitemOpen
  \bibfield  {author} {\bibinfo {author} {\bibfnamefont {A.~M.}\ \bibnamefont
  {Cooper-Sarkar}}, \bibinfo {author} {\bibfnamefont {S.}~\bibnamefont
  {Sarkar}}, \bibinfo {author} {\bibfnamefont {J.}~\bibnamefont {Guy}},
  \bibinfo {author} {\bibfnamefont {W.}~\bibnamefont {Venus}}, \bibinfo
  {author} {\bibfnamefont {P.~O.}\ \bibnamefont {Hulth}}, \ and\ \bibinfo
  {author} {\bibfnamefont {K.}~\bibnamefont {Hultqvist}},\ }\href {\doibase
  10.1016/0370-2693(92)90789-7} {\bibfield  {journal} {\bibinfo  {journal}
  {Phys. Lett.}\ }\textbf {\bibinfo {volume} {B280}},\ \bibinfo {pages} {153}
  (\bibinfo {year} {1992})}\BibitemShut {NoStop}%
\bibitem [{\citenamefont {Ge}\ and\ \citenamefont
  {Shoemaker}(2018)}]{Ge:2017mcq}%
  \BibitemOpen
  \bibfield  {author} {\bibinfo {author} {\bibfnamefont {S.-F.}\ \bibnamefont
  {Ge}}\ and\ \bibinfo {author} {\bibfnamefont {I.~M.}\ \bibnamefont
  {Shoemaker}},\ }\href {\doibase 10.1007/JHEP11(2018)066} {\bibfield
  {journal} {\bibinfo  {journal} {JHEP}\ }\textbf {\bibinfo {volume} {11}},\
  \bibinfo {pages} {066} (\bibinfo {year} {2018})},\ \Eprint
  {http://arxiv.org/abs/1710.10889} {arXiv:1710.10889 [hep-ph]} \BibitemShut
  {NoStop}%
\bibitem [{\citenamefont {Jordan}\ \emph {et~al.}(2018)\citenamefont {Jordan},
  \citenamefont {Kahn}, \citenamefont {Krnjaic}, \citenamefont {Moschella},\
  and\ \citenamefont {Spitz}}]{Jordan:2018gcd}%
  \BibitemOpen
  \bibfield  {author} {\bibinfo {author} {\bibfnamefont {J.~R.}\ \bibnamefont
  {Jordan}}, \bibinfo {author} {\bibfnamefont {Y.}~\bibnamefont {Kahn}},
  \bibinfo {author} {\bibfnamefont {G.}~\bibnamefont {Krnjaic}}, \bibinfo
  {author} {\bibfnamefont {M.}~\bibnamefont {Moschella}}, \ and\ \bibinfo
  {author} {\bibfnamefont {J.}~\bibnamefont {Spitz}},\ }\href {\doibase
  10.1103/PhysRevD.98.075020} {\bibfield  {journal} {\bibinfo  {journal} {Phys.
  Rev.}\ }\textbf {\bibinfo {volume} {D98}},\ \bibinfo {pages} {075020}
  (\bibinfo {year} {2018})},\ \Eprint {http://arxiv.org/abs/1806.05185}
  {arXiv:1806.05185 [hep-ph]} \BibitemShut {NoStop}%
\bibitem [{\citenamefont {Buonocore}\ \emph
  {et~al.}(2019{\natexlab{b}})\citenamefont {Buonocore}, \citenamefont
  {deNiverville},\ and\ \citenamefont {Frugiuele}}]{Buonocore:2019esg}%
  \BibitemOpen
  \bibfield  {author} {\bibinfo {author} {\bibfnamefont {L.}~\bibnamefont
  {Buonocore}}, \bibinfo {author} {\bibfnamefont {P.}~\bibnamefont
  {deNiverville}}, \ and\ \bibinfo {author} {\bibfnamefont {C.}~\bibnamefont
  {Frugiuele}},\ }\href@noop {} {\  (\bibinfo {year} {2019}{\natexlab{b}})},\
  \Eprint {http://arxiv.org/abs/1912.09346} {arXiv:1912.09346 [hep-ph]}
  \BibitemShut {NoStop}%
\bibitem [{\citenamefont {Cortina~Gil}\ \emph {et~al.}(2019)\citenamefont
  {Cortina~Gil} \emph {et~al.}}]{CortinaGil:2019nuo}%
  \BibitemOpen
  \bibfield  {author} {\bibinfo {author} {\bibfnamefont {E.}~\bibnamefont
  {Cortina~Gil}} \emph {et~al.} (\bibinfo {collaboration} {NA62}),\ }\href
  {\doibase 10.1007/JHEP05(2019)182} {\bibfield  {journal} {\bibinfo  {journal}
  {JHEP}\ }\textbf {\bibinfo {volume} {05}},\ \bibinfo {pages} {182} (\bibinfo
  {year} {2019})},\ \Eprint {http://arxiv.org/abs/1903.08767} {arXiv:1903.08767
  [hep-ex]} \BibitemShut {NoStop}%
\bibitem [{\citenamefont {Pinfold}(2019)}]{Pinfold:2019nqj}%
  \BibitemOpen
  \bibfield  {author} {\bibinfo {author} {\bibfnamefont {J.~L.}\ \bibnamefont
  {Pinfold}},\ }\bibfield  {booktitle} {\emph {\bibinfo {booktitle}
  {{Proceedings, 7th International Conference on New Frontiers in Physics
  (ICNFP 2018): Kolymbari, Crete, Greece, July 4-12, 2018}}},\ }\href {\doibase
  10.3390/universe5020047} {\bibfield  {journal} {\bibinfo  {journal}
  {Universe}\ }\textbf {\bibinfo {volume} {5}},\ \bibinfo {pages} {47}
  (\bibinfo {year} {2019})}\BibitemShut {NoStop}%
\bibitem [{\citenamefont {Haas}\ \emph {et~al.}(2015)\citenamefont {Haas},
  \citenamefont {Hill}, \citenamefont {Izaguirre},\ and\ \citenamefont
  {Yavin}}]{Haas:2014dda}%
  \BibitemOpen
  \bibfield  {author} {\bibinfo {author} {\bibfnamefont {A.}~\bibnamefont
  {Haas}}, \bibinfo {author} {\bibfnamefont {C.~S.}\ \bibnamefont {Hill}},
  \bibinfo {author} {\bibfnamefont {E.}~\bibnamefont {Izaguirre}}, \ and\
  \bibinfo {author} {\bibfnamefont {I.}~\bibnamefont {Yavin}},\ }\href
  {\doibase 10.1016/j.physletb.2015.04.062} {\bibfield  {journal} {\bibinfo
  {journal} {Phys. Lett.}\ }\textbf {\bibinfo {volume} {B746}},\ \bibinfo
  {pages} {117} (\bibinfo {year} {2015})},\ \Eprint
  {http://arxiv.org/abs/1410.6816} {arXiv:1410.6816 [hep-ph]} \BibitemShut
  {NoStop}%
\bibitem [{\citenamefont {Ball}\ \emph {et~al.}(2016)\citenamefont {Ball} \emph
  {et~al.}}]{Ball:2016zrp}%
  \BibitemOpen
  \bibfield  {author} {\bibinfo {author} {\bibfnamefont {A.}~\bibnamefont
  {Ball}} \emph {et~al.},\ }\href@noop {} {\  (\bibinfo {year} {2016})},\
  \Eprint {http://arxiv.org/abs/1607.04669} {arXiv:1607.04669
  [physics.ins-det]} \BibitemShut {NoStop}%
\bibitem [{\citenamefont {Sher}\ and\ \citenamefont
  {Stevens}(2018)}]{Sher:2017wya}%
  \BibitemOpen
  \bibfield  {author} {\bibinfo {author} {\bibfnamefont {M.}~\bibnamefont
  {Sher}}\ and\ \bibinfo {author} {\bibfnamefont {J.}~\bibnamefont {Stevens}},\
  }\href {\doibase 10.1016/j.physletb.2017.12.022} {\bibfield  {journal}
  {\bibinfo  {journal} {Phys. Lett.}\ }\textbf {\bibinfo {volume} {B777}},\
  \bibinfo {pages} {246} (\bibinfo {year} {2018})},\ \Eprint
  {http://arxiv.org/abs/1710.06894} {arXiv:1710.06894 [hep-ph]} \BibitemShut
  {NoStop}%
\bibitem [{\citenamefont {Frank}\ \emph {et~al.}(2019)\citenamefont {Frank},
  \citenamefont {de~Montigny}, \citenamefont {Ouimet}, \citenamefont {Pinfold},
  \citenamefont {Shaa},\ and\ \citenamefont {Staelens}}]{Frank:2019pgk}%
  \BibitemOpen
  \bibfield  {author} {\bibinfo {author} {\bibfnamefont {M.}~\bibnamefont
  {Frank}}, \bibinfo {author} {\bibfnamefont {M.}~\bibnamefont {de~Montigny}},
  \bibinfo {author} {\bibfnamefont {P.-P.~A.}\ \bibnamefont {Ouimet}}, \bibinfo
  {author} {\bibfnamefont {J.}~\bibnamefont {Pinfold}}, \bibinfo {author}
  {\bibfnamefont {A.}~\bibnamefont {Shaa}}, \ and\ \bibinfo {author}
  {\bibfnamefont {M.}~\bibnamefont {Staelens}},\ }\href@noop {} {\  (\bibinfo
  {year} {2019})},\ \Eprint {http://arxiv.org/abs/1909.05216} {arXiv:1909.05216
  [hep-ph]} \BibitemShut {NoStop}%
\bibitem [{\citenamefont {Kadota}\ and\ \citenamefont
  {Silk}(2014)}]{Kadota:2014mea}%
  \BibitemOpen
  \bibfield  {author} {\bibinfo {author} {\bibfnamefont {K.}~\bibnamefont
  {Kadota}}\ and\ \bibinfo {author} {\bibfnamefont {J.}~\bibnamefont {Silk}},\
  }\href {\doibase 10.1103/PhysRevD.89.103528} {\bibfield  {journal} {\bibinfo
  {journal} {Phys. Rev.}\ }\textbf {\bibinfo {volume} {D89}},\ \bibinfo {pages}
  {103528} (\bibinfo {year} {2014})},\ \Eprint {http://arxiv.org/abs/1402.7295}
  {arXiv:1402.7295 [hep-ph]} \BibitemShut {NoStop}%
\bibitem [{\citenamefont {Primulando}\ \emph {et~al.}(2015)\citenamefont
  {Primulando}, \citenamefont {Salvioni},\ and\ \citenamefont
  {Tsai}}]{Primulando:2015lfa}%
  \BibitemOpen
  \bibfield  {author} {\bibinfo {author} {\bibfnamefont {R.}~\bibnamefont
  {Primulando}}, \bibinfo {author} {\bibfnamefont {E.}~\bibnamefont
  {Salvioni}}, \ and\ \bibinfo {author} {\bibfnamefont {Y.}~\bibnamefont
  {Tsai}},\ }\href {\doibase 10.1007/JHEP07(2015)031} {\bibfield  {journal}
  {\bibinfo  {journal} {JHEP}\ }\textbf {\bibinfo {volume} {07}},\ \bibinfo
  {pages} {031} (\bibinfo {year} {2015})},\ \Eprint
  {http://arxiv.org/abs/1503.04204} {arXiv:1503.04204 [hep-ph]} \BibitemShut
  {NoStop}%
\bibitem [{\citenamefont {Alves}\ \emph {et~al.}(2018)\citenamefont {Alves},
  \citenamefont {Santos},\ and\ \citenamefont {Sinha}}]{Alves:2017uls}%
  \BibitemOpen
  \bibfield  {author} {\bibinfo {author} {\bibfnamefont {A.}~\bibnamefont
  {Alves}}, \bibinfo {author} {\bibfnamefont {A.~C.~O.}\ \bibnamefont
  {Santos}}, \ and\ \bibinfo {author} {\bibfnamefont {K.}~\bibnamefont
  {Sinha}},\ }\href {\doibase 10.1103/PhysRevD.97.055023} {\bibfield  {journal}
  {\bibinfo  {journal} {Phys. Rev.}\ }\textbf {\bibinfo {volume} {D97}},\
  \bibinfo {pages} {055023} (\bibinfo {year} {2018})},\ \Eprint
  {http://arxiv.org/abs/1710.11290} {arXiv:1710.11290 [hep-ph]} \BibitemShut
  {NoStop}%
\bibitem [{\citenamefont {Liu}\ and\ \citenamefont
  {Zhang}(2019)}]{Liu:2018jdi}%
  \BibitemOpen
  \bibfield  {author} {\bibinfo {author} {\bibfnamefont {Z.}~\bibnamefont
  {Liu}}\ and\ \bibinfo {author} {\bibfnamefont {Y.}~\bibnamefont {Zhang}},\
  }\href {\doibase 10.1103/PhysRevD.99.015004} {\bibfield  {journal} {\bibinfo
  {journal} {Phys. Rev.}\ }\textbf {\bibinfo {volume} {D99}},\ \bibinfo {pages}
  {015004} (\bibinfo {year} {2019})},\ \Eprint
  {http://arxiv.org/abs/1808.00983} {arXiv:1808.00983 [hep-ph]} \BibitemShut
  {NoStop}%
\bibitem [{\citenamefont {Gninenko}\ \emph {et~al.}(2019)\citenamefont
  {Gninenko}, \citenamefont {Kirpichnikov},\ and\ \citenamefont
  {Krasnikov}}]{Gninenko:2018ter}%
  \BibitemOpen
  \bibfield  {author} {\bibinfo {author} {\bibfnamefont {S.~N.}\ \bibnamefont
  {Gninenko}}, \bibinfo {author} {\bibfnamefont {D.~V.}\ \bibnamefont
  {Kirpichnikov}}, \ and\ \bibinfo {author} {\bibfnamefont {N.~V.}\
  \bibnamefont {Krasnikov}},\ }\href {\doibase 10.1103/PhysRevD.100.035003}
  {\bibfield  {journal} {\bibinfo  {journal} {Phys. Rev.}\ }\textbf {\bibinfo
  {volume} {D100}},\ \bibinfo {pages} {035003} (\bibinfo {year} {2019})},\
  \Eprint {http://arxiv.org/abs/1810.06856} {arXiv:1810.06856 [hep-ph]}
  \BibitemShut {NoStop}%
\bibitem [{\citenamefont {Acciarri}\ \emph {et~al.}(2019)\citenamefont
  {Acciarri} \emph {et~al.}}]{Acciarri:2019jly}%
  \BibitemOpen
  \bibfield  {author} {\bibinfo {author} {\bibfnamefont {R.}~\bibnamefont
  {Acciarri}} \emph {et~al.} (\bibinfo {collaboration} {ArgoNeuT}),\
  }\href@noop {} {\  (\bibinfo {year} {2019})},\ \Eprint
  {http://arxiv.org/abs/1911.07996} {arXiv:1911.07996 [hep-ex]} \BibitemShut
  {NoStop}%
\bibitem [{\citenamefont {Mertig}\ \emph {et~al.}(1991)\citenamefont {Mertig},
  \citenamefont {Bohm},\ and\ \citenamefont {Denner}}]{Mertig:1990an}%
  \BibitemOpen
  \bibfield  {author} {\bibinfo {author} {\bibfnamefont {R.}~\bibnamefont
  {Mertig}}, \bibinfo {author} {\bibfnamefont {M.}~\bibnamefont {Bohm}}, \ and\
  \bibinfo {author} {\bibfnamefont {A.}~\bibnamefont {Denner}},\ }\href
  {\doibase 10.1016/0010-4655(91)90130-D} {\bibfield  {journal} {\bibinfo
  {journal} {Comput. Phys. Commun.}\ }\textbf {\bibinfo {volume} {64}},\
  \bibinfo {pages} {345} (\bibinfo {year} {1991})}\BibitemShut {NoStop}%
\bibitem [{\citenamefont {Shtabovenko}\ \emph {et~al.}(2016)\citenamefont
  {Shtabovenko}, \citenamefont {Mertig},\ and\ \citenamefont
  {Orellana}}]{Shtabovenko:2016sxi}%
  \BibitemOpen
  \bibfield  {author} {\bibinfo {author} {\bibfnamefont {V.}~\bibnamefont
  {Shtabovenko}}, \bibinfo {author} {\bibfnamefont {R.}~\bibnamefont {Mertig}},
  \ and\ \bibinfo {author} {\bibfnamefont {F.}~\bibnamefont {Orellana}},\
  }\href {\doibase 10.1016/j.cpc.2016.06.008} {\bibfield  {journal} {\bibinfo
  {journal} {Comput. Phys. Commun.}\ }\textbf {\bibinfo {volume} {207}},\
  \bibinfo {pages} {432} (\bibinfo {year} {2016})},\ \Eprint
  {http://arxiv.org/abs/1601.01167} {arXiv:1601.01167 [hep-ph]} \BibitemShut
  {NoStop}%
\bibitem [{\citenamefont {Husek}(2019{\natexlab{b}})}]{Husek:2019axa}%
  \BibitemOpen
  \bibfield  {author} {\bibinfo {author} {\bibfnamefont {T.}~\bibnamefont
  {Husek}},\ }in\ \href@noop {} {\emph {\bibinfo {booktitle} {{22nd High-Energy
  Physics International Conference in Quantum Chromodynamics (QCD 19)
  Montpellier, Languedoc, France, July 2-5, 2019}}}}\ (\bibinfo {year} {2019})\
  \Eprint {http://arxiv.org/abs/1911.06820} {arXiv:1911.06820 [hep-ph]}
  \BibitemShut {NoStop}%
\end{thebibliography}%

\end{document}